\documentclass[12pt,thmsa]{article}
\usepackage{amsmath}
\usepackage{amssymb}
\usepackage{sw20lart}

\input{tcilatex}
\begin{document}

\author{Xijia Miao}
\title{The basic principles to construct a generalized state-locking pulse
field and simulate efficiently the reversible and unitary halting protocol
of a universal quantum computer}
\date{Somerville, Massachusetts. Date: April 2006. }
\maketitle

\begin{abstract}
\end{abstract}

A halting protocol is generally irreversible in the classical computation,
but surprisingly it is usually also irreversible and non-unitary in the
universal quantum computational models. The inherent incompatibility within
the universal quantum computational models between the irreversible and
non-unitary halting protocol and the unitary quantum computational process
that obeys the Schr\"{o}dinger equation in quantum physics has been known
since the early set-up of the universal quantum computational models. The
irreversibility and non-unitarity of the halting protocol is closely related
to the inherent irreversibility and non-unitarity of quantum measurement in
quantum mechanics. The unitary dynamics quantum mechanically implies that
the halting protocol in the universal quantum computational models should be
made reversible and unitary so as to eliminate the inherent incompatibility.
It has been shown in Ref. [24] (Arxiv: quant-ph/0507236) that a universal
quantum computer could be powerful enough to solve efficiently the quantum
search problem in the cyclic-group state subspaces, and the reversible and
unitary halting protocol is the key component to construct the efficient
quantum search processes based on the unitary quantum dynamics, while the
state-locking pulse field is the key component to generate the reversible
and unitary halting protocol. In this paper the reversible and unitary
halting protocol and the generalized state-locking pulse field have been
extensively investigated theoretically. The basic principles to construct
the state-locking pulse field and design the reversible and unitary halting
protocol are described and studied in detail. A generalized state-locking
pulse field is generally dependent upon the time and space variables. It
could be a sequence of time- and space-dependent electromagnetic pulse
fields and could also contain the time- and space-dependent potential
fields. Thus, the reversible and unitary halting protocol built up out of
the state-locking pulse field generally consists of a sequence of time- and
space-dependent unitary evolution processes. It is shown how the quantum
control process is constructed to simulate efficiently the reversible and
unitary halting protocol. An improved subspace-reduction quantum program and
circuit based on the reversible and unitary halting protocol, which is much
simpler than that one in the previous paper [24], is proposed as the key
component to construct further an efficient quantum search process. A simple
atomic physical system which is an atomic ion or a neutral atom in the
double-well potential field is proposed to show how the state-locking pulse
field is generated and how to implement the reversible and unitary halting
protocol. \newline
\newline
\newline
{\large 1. Introduction}

A halting protocol (or a halting operation briefly) of a computational model
such as the Turing machine is one of the key components in computation. It
is well known that the halting operation generally is not a reversible
operation in the classical computation and often is related to measurement
of computational results. Though in the reversible computational model [1]
and probably in the quantum Turing machine [2] the halting protocol could be
made reversible, it is surprised that the halting protocol usually is not
yet reversible and unitary in the universal quantum computational models
including the universal quantum Turing machine [3, 9] and the universal
quantum circuit model [4]. When the universal quantum Turing machine was
proposed in the early day [3], the halting protocol was also introduced as
one of the important components of the universal quantum computational
model. According as the universal quantum Turing machine, in addition to the
computational quantum system that is used to perform quantum computation
there is also an extra quantum bit used to instruct what time a quantum
computational process is halted on the universal quantum Turing machine.
This extra quantum bit is named the halting quantum bit or briefly the
halting qubit. The halting quantum bit should be observed periodically from
the outside in a non-perturbation manner so that the quantum measurement on
the halting quantum bit does not disturb the unitary evolution process of
the quantum system during the quantum computational process. Once the
halting qubit is found to be in the halting state, e.g., the state $%
|1\rangle ,$ the quantum computational process terminates. Therefore, the
halting operation is performed to stop the quantum computational process
only after a sequence of non-perturbation quantum measurements on the
halting qubit were carried out during the quantum computational process.
However, the halting protocol could not be compatible with unitary quantum
computational processes within the universal quantum computational models.
It is well known that a quantum computational process on a universal quantum
computer obeys the quantum physical laws [3, 4], that is, the time evolution
process of the quantum system of the quantum computer obeys the Schr\"{o}%
dinger equation in quantum mechanics during the quantum computational
process. It is also known that the quantum parallel principle [3] that the
quantum computational process is performed on a superposition of the basis
states of the quantum system is the characteristic feature of the universal
quantum computation models. The halting protocol is essentially different
from a unitary quantum computational process in the universal quantum
computational models in that the halting protocol involved in the
non-unitary quantum measurement is generally irreversible and non-unitary.
This basic and inherent incompatibility within the universal quantum
computational models between the facts that the quantum computational
process obeys the unitary quantum dynamics and that the halting protocol is
irreversible and non-unitary due to the non-unitary quantum measurement
really originates from the basic quantum mechanical laws. It is well known
that there is an inherent incompatibility in quantum mechanics [5] between
the non-unitary quantum measurement and the unitary time evolution process
of a closed quantum system that obeys the Schr\"{o}dinger equation.

For many years since the early proposals of the universal quantum
computational models [3, 4] the computational power of the universal quantum
computational models has been investigated extensively and continuously. In
the past two decades it has been shown that there are several possible
candidates to fuel the quantum computational power, which include the
superposition and the quantum parallel principle [3, 4], the quantum
entanglement [29] and the multiple-quantum coherence [17, 20], the quantum
coherence interference [18, 20], and the unitary dynamics quantum
mechanically associated with the symmetry and structure of a quantum system
[17, 18, 20, 22, 23, 24]. A large number of quantum algorithms [13, 14]
including the Shor$^{\prime }s$ prime factorization and discrete logarithm
algorithms [6] and the Grover$^{\prime }s$ quantum search algorithm [7]
which outperform their classical counterparts have been discovered and
developed in the past two decades. Most of these quantum algorithms are
based on the universal quantum circuit model [4]. The computational results
of these quantum algorithms generally are output at the end of the
computational processes by the proper quantum measurement. Thus, these
quantum algorithms are not related to the halting protocol of the universal
quantum computational models. It has been believed extensively that power of
the quantum computation that outperforms the classical computation could
come mainly from the quantum parallel principle [3]. But it was suspected
whether or not an arbitrary recursive function mathematically could be
computed more efficiently on a universal quantum computer than a classical
computer with the help of the quantum parallel principle [8]. The quantum
parallel principle requires that a quantum computational process take a
superposition state as its initial state. However, an incompatibility within
the universal quantum computational models arises when a universal quantum
computer computes a general recursive function in mathematics by starting at
a superposition state. This incompatibility is really due to the conflict
between the halting protocol and the quantum parallel principle [8]. It
results in the question whether or not a universal quantum computer is
capable of computing more efficiently an arbitrary recursive function when
the quantum parallel principle is employed. A quantum computational process
allows its initial state to be a superposition state of the quantum system
of a universal quantum computer [3, 4, 9]. A superposition state may be
expressed as a linear combination of the conventional computational bases of
the quantum system. Each computational base of the superposition state could
be thought of as one input state of a quantum algorithm running on the
universal quantum computer if the superposition state is taken as the
initial state of the quantum computational process of the quantum algorithm.
This implies that the quantum algorithm can be performed in a parallel form
on the universal quantum computer by taking at the same time all these
different computational bases of the superposition state as its input
states. From the point of view of the classical computational model a
different input state corresponds to a different classical computational
process. Then one could imagine that in effect the quantum computational
process really performs simultaneously many different $^{\prime }$classical$%
^{\prime }$ computational processes with different input states when the
initial state of the quantum computational process is a superposition state.
Although these different $^{\prime }$classical$^{\prime }$ computational
processes start at the same initial time, they could end at different times
in computation, respectively. As a typical example, this situation will
occur if the quantum algorithm is used to compute a general recursive
function in mathematics and the end state of the computation is fixed.
Because there is not the same end time for all these different $^{\prime }$%
classical$^{\prime }$ computational processes a conflict arises when the
quantum computer decides what time the quantum computational process is
halted. Thus, from the viewpoint of the mathematical logic principle of
computational model the halting protocol is incompatible with the quantum
parallel principle.

This conflict is apparently between the halting protocol and the quantum
parallel principle, but it really related inherently to the non-unitarity of
quantum measurement which is well known in quantum mechanics [5]. According
as the halting protocol the halting quantum bit usually is set to the
halting state $|n_{h}\rangle =|0\rangle $ at the initial time $t_{0}$ of the
quantum computational process. Suppose that the initial superposition state
of the whole quantum system including the halting qubit of the quantum
computer is given by 
\begin{equation*}
|\Psi (t_{0})\rangle =\underset{j}{\sum }a_{j}(t_{0})|n_{h}\rangle |\varphi
_{j}(t_{0})\rangle =\underset{j}{\sum }a_{j}(t_{0})|0\rangle |\varphi
_{j}(t_{0})\rangle .
\end{equation*}%
According as the halting protocol, when some $^{\prime }$classical$^{\prime
} $ \ computational processes arrive in their end states during the quantum
computational process their halting state $|n_{h}\rangle $ becomes the state 
$|1\rangle .$ Since there is not the same end time for all these different $%
^{\prime }$classical$^{\prime }$ computational processes, there exists some
time $t$ in the quantum computational process such that some $^{\prime }$%
classical$^{\prime }$ computational processes arrive in their end states ($%
\{|\varphi _{l}(t)\rangle \}$) and hence their halting state becomes the
state $|1\rangle $, while the rest have not yet arrived in their end states
and their halting state remains unchanged and is still the state $|0\rangle
. $ Then at the time $t$ the quantum state of the whole quantum system is
written as 
\begin{equation*}
|\Psi (t)\rangle =\underset{k}{\sum }a_{k}(t)|0\rangle |\varphi
_{k}(t)\rangle +\underset{l\neq k}{\sum }a_{l}(t)|1\rangle |\varphi
_{l}(t)\rangle .
\end{equation*}%
The state $|\Psi (t)\rangle $ is clearly a superposition state involved in
the halting state $|n_{h}\rangle $ and the computational states $\{|\varphi
_{j}(t)\rangle \}$ of the quantum system. Again according as the halting
protocol [3], the quantum measurement is carried out on the halting quantum
bit periodically from the outside during the quantum computational process.
Then the quantum measurement will change the state $|\Psi (t)\rangle $ and
spoil the quantum computational process [8]. Actually, according to the
quantum mechanics [5] the quantum-state collapse on the superposition state $%
|\Psi (t)\rangle $ occurs inevitably when the quantum measurement on the
halting qubit is performed on the superposition state $|\Psi (t)\rangle .$
It is well known that a quantum-state collapse process is a non-unitary
process in quantum mechanics [5b]. Thus, the halting protocol generally is
irreversible and non-unitary in the universal quantum computational models
in which the quantum parallel principle is a basic principle.

There are a number of works [3, 8, 9, 10, 11, 12] to discuss in detail the
conflict and to propose schemes to avoid it for the universal quantum
computational models. However, so far there is not any satisfactory and
universal scheme to avoid the conflict when the initial state for a quantum
computational process is a superposition state. On the other hand, the
halting protocol may be made reversible in the reversible computational
model [54] due to that any initial state of computation is generally a
single basis state and hence there is not such a conflict in the reversible
computational model. It has also been shown [3, 8, 9, 10, 11, 12] that this
conflict could be avoided when the quantum computational process consists of
a single $^{\prime }$classical$^{\prime }$ computational process or several
different $^{\prime }$classical$^{\prime }$ computational processes which
arrive in their end states at the same time. This is because in this case
the state $|\Psi (t)\rangle $ above is either the first term with the
halting state $|0\rangle $ or the second term with the halting state $%
|1\rangle $ but not a superposition of the two terms. Since an initial basis
state corresponds one-to-one to a $^{\prime }$classical$^{\prime }$
computational process in the quantum computational process one may also say
that the conflict could be avoided if the initial state of the quantum
computational process is limited to a single basis state. Although from the
viewpoint of the mathematical logic principle of computational model the
halting protocol becomes reasonable if the initial state of the quantum
computational process is limited to a single basis state, it could not be
generally reversible and unitary as the halting protocol contains the
non-unitary quantum measurement. The original halting protocol [3] uses the
quantum measurement to achieve the halting operation, that is, if the
halting qubit is found in the halting state $|1\rangle $ through the
periodic quantum measurement the quantum computational process is stopped by
the brute-force method from the outside. Therefore, even in the case that
the initial state of the quantum computational process is restricted to be a
basis state the original halting protocol generally is not reversible and
unitary. The original halting protocol [3] also implies that there must be a
conditional operation such that when the computational end state appears
during the computational process this conditional operation converts
immediately the initial halting state $|0\rangle $ into the state $|1\rangle
.$ Some improved halting protocols [11, 12] use explicitly the conditional
halting operation to replace the original halting operation. It was also
proposed in the halting protocol [11] that when the quantum computational
process is stopped due to that the halting state $|0\rangle $ is changed to
the state $|1\rangle $ it needs to start an extra unitary evolution process
at the same time so that while both the computational end state and the
halting state $|1\rangle $ are kept unchanged after the halting operation,
the total evolution process can be unitary for the whole quantum system
including the computational quantum system, the halting qubit, and the
auxiliary qubits (the halting protocol [11] needs to use an extra auxiliary
register). In the reversible computational model the reversible halting
protocol could be achieved by executing successively a computational
process, the conditional (logic) halting operation, and the inverse
computational process [54]. However, the quantum measurement must be given
up in these halting protocols [3, 9, 10, 11, 12] if one wants to make these
halting protocols reversible and unitary thoroughly as the quantum
measurement generally could lead to information loss of a quantum system
even when the initial state is a single basis state. Therefore, the
irreversibility and non-unitarity of these halting protocols in the
universal quantum computational models are traced ultimately back to the
irreversibility and non-unitarity of quantum measurement in quantum
mechanics.

The halting protocol usually is not reversible and unitary in the universal
quantum computational models due to the non-unitary quantum measurement, but
for most of the present quantum algorithms based on these universal quantum
computational models the computational results usually are not significantly
affected by the non-unitarity of quantum measurement even though the
non-unitarity could lead to loss of partial information of the computational
results. There are also the incompatibilities within the universal quantum
computational models between the quantum parallel principle and the halting
protocol from viewpoint of the mathematical logic principle and between the
unitary quantum computational process and the irreversible and non-unitary
halting protocol from the viewpoint of the quantum physics, but up to now a
number of powerful quantum algorithms [13, 14] that outperform their
classical counterparts have been found within the universal quantum
computational models and the halting protocol has not yet any significant
effect on these powerful quantum algorithms. Therefore, on one hand, the
quantum parallel principle is paid much attention, on the other hand, the
halting protocol has a negligible influence on the quantum computational
science in the past years. It has been explored in theory to eliminate the
incompatibilities within the universal quantum computational models, but
people are not clear whether or not the incompatibilities could lead to an
essential impact on the power of quantum computation and do not yet know
whether or not the reversibility and unitarity of the halting protocol is
important to discover and develop new and efficient quantum algorithms.
However, these powerful quantum algorithms [13, 14] can only treat
successfully few special mathematical problems. Most of them are based on
the unitary quantum circuit model [4] in which the irreversible and
non-unitary halting protocol neither is used nor has any essential effect on
the computational power of these quantum algorithms. It is also known that
the incompatibilities could lead to that a universal quantum computer could
not be more efficient than a classical computer in computing an arbitrary
recursive function in mathematics. On the other hand, it is well known that
a quantum computational process obeys the unitary quantum dynamics and is
compatible with the mathematical logic principles used by the computational
process. These facts show that it is worthwhile to investigate further how
the halting protocol can be made reversible and unitary so that the
incompatibilities could be eliminated in the universal quantum computation
models.

The unitary dynamical principle of quantum mechanics plays a key role in the
scalable quantum computation in a mixed-state quantum ensemble. This basic
principle states simply that both a closed quantum system and its quantum
ensemble obey the same unitary dynamics quantum mechanically [17, 18, 20,
22, 23, 24]. The unitary quantum dynamics associated with the symmetry and
structure of a quantum system has been used to discover and develop new
quantum search processes in a quantum ensemble [17] and a pure-state quantum
system [23, 24]. Quantum search processes are extremely important in quantum
computation as the unsorted quantum search process has an extensive
application in computational science and can be used to solve the NP-hard
problems. It has been shown that the square speedup for the standard quantum
search algorithm [7] is optimal in a pure-state quantum system [15]. The
first attempt to break through the quantum-searching square speedup limit
was carried out in a nuclear magnetic resonance (NMR)\ spin ensemble [16,
17a]. Though these quantum search processes [16, 17] are not scalable in a
spin ensemble, their speed is really exceed greatly the allowed value of any
quantum search algorithm with the square speedup limit in a range of a few
quantum bits. However, these quantum search processes do not achieve a real
breakthrough of the square speedup limit because their output NMR signal
intensities decay exponentially as the qubit number of the spin system
increases and hence beyond a few quantum bits their quantum-searching speed
falls off rapidly. So far a scalable quantum search algorithm working in a
quantum ensemble has not yet been found. Both the unitary dynamics quantum
mechanically and the quantum coherence interference play a key role in
achieving the fast and scalable quantum computation in a spin ensemble [17,
18, 20]. It has been shown that many oracle-based quantum algorithms
including the parity-determination algorithm are subjected to the polynomial
speedup bounds in a pure-state quantum system [19]. Again the unitary
quantum dynamics has been shown to play an important role in achieving a
much fast computational speed to solve the parity-determination problem in a
spin ensemble which is beyond the polynomial speedup bounds on the
oracle-based quantum algorithms in a range of a few quantum bits [18]. Up to
now, the scalable problem has not yet been solved for the quantum
parity-determination algorithm in a quantum ensemble. These results obtained
in a spin ensemble encourage one to explore further the potential ways to
break through the quantum-searching square speedup limit and the polynomial
speedup bounds upon the oracle-based quantum algorithms. The unitary quantum
dynamics associated with the symmetry and structure of a spin system has
also showed that the prime factorization for a large composite integer may
be implemented in a scalable form in a spin ensemble [20]. The efficient
factoring algorithm was first discovered in a pure-state quantum system [6].
In order to achieve the scalable quantum computation for the prime
factorization in a spin ensemble it is necessary to exploit the symmetric
property and structure of the spin system of the spin ensemble to help the
unitary quantum dynamics to solve the prime factorization problem. Here both
the time-reversal symmetry and the rotation symmetry in spin space [5a] of
the spin system play a key role in the scalable factoring algorithm in a
spin ensemble [20]. The factoring algorithm in the spin ensemble usually is
divided into four time periods [20], which is similar to the conventional
NMR experimental counterparts [21]. The first period is to generate the
multiple-quantum coherences of the spin ensemble. In this period information
of the order of the modular exponential function is loaded on the
multiple-quantum coherences. The second is the time evolution process to
carry out the frequency labeling for these different order multiple-quantum
coherences generated in the first period. The third is the time-reversal
process of the first period. The final is the quantum measurement to output
the NMR multiple-quantum coherence signal that carries the information of
the order of the modular exponential function. Here the output NMR
multiple-quantum spectrum usually is used to determine the order of the
modular exponential function. The multiple-quantum spectrum may be obtained
by Fourier transforming the output NMR\ multiple-quantum coherence signal
that is measured indirectly in experment. The time-reversal symmetry [5a]
ensures that the dephased NMR multiple-quantum coherences in the first
period can be refocused in the third period and hence the multiple-quantum
coherences become inphase so that the multiple-quantum coherence
interference can lead to coherent enhancement of the output NMR
multiple-quantum coherence signal in the factoring algorithm. Therefore, it
becomes possible that the output NMR multiple-quantum coherence signal does
not decrease exponentially as the qubit number of the spin system increases.
The time-reversal symmetry has been used extensively to obtain highly
sensitive and inphase NMR multiple-quantum spectra in high-resolution
nuclear magnetic resonance experiments in solid [21]. On the other hand, the
rotation symmetry in spin space [5a] of a spin system is the basis of the
multiple-quantum operator algebra spaces of the Liouville operator space of
the spin ensemble [22]. Though the total number of independent NMR
multiple-quantum transitions in an $n-$qubit spin system increases
exponentially as the qubit number of the spin system, a large number of the
independent multiple-quantum transitions are really degenerative or nearly
degenerative in transition frequency due to the rotation symmetry in spin
space, and the multiple-quantum transitions with significantly different
transition frequencies are really very few in the spin system. For example,
there may be only $(n+1)$ different order multiple-quantum transitions in
the $n-$qubit spin system and each has its own transition frequency, while
the total number of the independent multiple-quantum transitions of the spin
system is $(4^{n}-2^{n})/2$. In theory the number of the
multiple-quantum-transition spectral lines in the NMR\ multiple-quantum
spectrum may be equal to the number of the independent multiple-quantum
transitions of the spin system. Note that the total spectral intensity of
the NMR multiple-quantum-transition spectrum of the spin system is not more
than the total magnetization ($\thicksim n2^{n}$) of the spin system in the
thermal equilibrium state [55], while the total number $(4^{n}-2^{n})/2$ of
the NMR multiple-quantum-transition spectral lines increases exponentially ($%
\thicksim 4^{n}$) as the qubit number. The intensity of each
multiple-quantum-transition spectral line therefore weakens exponentially ($%
\thicksim n/2^{n}$) as the qubit number increases, although all these
multiple-quantum spectral lines are inphase due to the time-reversal
symmetry. Then due to noise in the detected NMR signal each of these
multiple-quantum-transition spectral lines will become unobservable even
when the qubit number is moderate if most of these
multiple-quantum-transition spectral lines have different resonance
frequencies (i.e., transition frequencies). Fortunately, due to the rotation
symmetry in spin space there are very few multiple-quantum transitions with
significantly different transition frequencies in the spin system and hence
very few observable multiple-quantum spectral lines with significantly
different resonance frequencies in the multiple-quantum spectrum. A large
number of the inphase multiple-quantum spectral lines overlap with each
other in the multiple-quantum spectrum due to that they have the same
resonance frequency. As a result, all these inphase multiple-quantum
spectral lines with the same resonance frequency become really a single
multiple-quantum spectral line, and the intensity of the single
multiple-quantum spectral line is really the sum of all the intensities of
these inphase multiple-quantum spectral lines. This intensity becomes so
large that the noise in the detected NMR signal can not have any significant
effect on the single multiple-quantum spectral line. As shown in Ref. [20],
if now there are only the $(n+1)$ different order multiple-quantum
transitions in the $n-$qubit spin system, each with a significantly
different transition frequency, then each of the $(n+1)$ different order
multiple-quantum transitions is composed of the degenerative
multiple-quantum transitions which have the number $\thicksim $ $%
(4^{n}/2-2^{n-1})/(n+1)$ on average and its multiple-quantum-transition
spectral intensity is proportional to the number $(4^{n}/2-2^{n-1})/(n+1)$
on average. Here the total spectral intensity of the $(4^{n}/2-2^{n-1})$
inphase multiple-quantum-transition spectral lines plus the spectral
intensity of the longitudinal magnetization and spin order components is
really equal to the total magnetization of the spin system [55] which can be
observable in the conventional NMR experiments even for a very large qubit
number. Obviously, the average intensity for each
multiple-quantum-transition spectral line is inversely proportional to the
qubit number and some of these $(n+1)$ multiple-quantum spectral lines do
not weaken exponentially as the qubit number increases. Thus, both the
time-reversal symmetry and the rotation symmetry in spin space ensure that
the output NMR multiple-quantum-transition spectral intensities can be
efficiently detected in the spin ensemble in the factoring algorithm. On the
other hand, so far quantum entanglement has not yet been proven strictly to
play a key role in speeding up a quantum computation, whereas the locally
efficient and scalable factoring algorithm in the spin ensemble [20] shows
that quantum entanglement could not be a key component to make quantum
computation much more powerful than classical computation. Therefore, the
unitary dynamics quantum mechanically associated with the symmetry and
structure of a quantum system could be the key component to power the
quantum computation and hence is a general guidance to discover and develop
new and efficient quantum algorithms in quantum computational science. The
unitary dynamical principle quantum mechanically implies that the
irreversible and non-unitary halting protocol should be modified to be
reversible and unitary so that the incompatibility between the irreversible
and non-unitary halting protocol and the unitary quantum computational
process can be eliminated within the universal quantum computational models.

The unitary dynamics quantum mechanically associated with the symmetry and
structure of a quantum system also plays a key role in discovering the
scalable and efficient quantum algorithms in a pure-state quantum system. It
is well known that there are $2^{n}$ basis states in the Hilbert space of an 
$n-$qubit spin system. These basis states may be chosen as the eigenstates
of the $z-$component operator $J_{z}$ of the total spin angular momentum of
the spin system. But due to the rotation symmetry in spin space of the $n-$%
qubit spin system many of these $2^{n}$ basis states are degenerative and
have the same eigenvalue. Thus, the whole Hilbert space span by these $2^{n}$
eigenstates of the $z-$component operator $J_{z}$ may be divided into $(n+1)$
different state subspaces according to the rotation symmetry in spin space
of the spin system [5a, 5c]. The quantum search space which is just the
whole Hilbert space of the $n-$qubit spin system for the standard quantum
search algorithm therefore is reduced to the largest state subspace among
these $(n+1)$ state subspaces. Hence the standard quantum search algorithm
is improved [23], since the largest state subspace is still much smaller
than the whole Hilbert space. Moreover, it has been shown [23] that any
unknown quantum state can be efficiently transferred to a larger state
subspace from a state subspace in the Hilbert space of the $n-$qubit spin
system, while the inverse process of the state transfer is generally harder
to be carried out. This general rule not only is useful for solving the
quantum search problem but also helpful for understanding deeply
non-equilibrium processes of a quantum ensemble from the viewpoint of the
unitary quantum dynamics instead of the conventional probability theory. As
shown in Ref. [24], this general rule is also closely related to that there
exists a computational-power difference between a unitary evolution process
and its inverse process in a quantum system. Of course, this inverse unitary
process may exist or may not in the quantum system. A direct extension of
the idea that the symmetric property and structure of a quantum system could
help the unitary quantum dynamics to solve the quantum search problem is to
exploit further the symmetric property and structure of a general group such
as a cyclic group in a quantum system to help the unitary quantum dynamics
to discover and develop new and efficient quantum search algorithms [24].
With the help of the reversible and unitary halting protocol based on the
state-locking pulse field and the property and structure of a cyclic group
in a quantum system it has been shown by the unitary quantum dynamics that a
universal quantum computer could be enough powerful to solve efficiently the
quantum search problem in a cyclic-group state space [24]. Here the
important point to arrive at this conclusion is that the halting protocol of
the universal quantum computational models is available and may be made
reversible and unitary for the quantum search process if the initial state
of the quantum search process is limited to a single computational basis
state.

All these conventional halting protocols [3, 9, 10, 11, 12] of the universal
quantum computational models generally can not be used to construct an
efficient quantum search algorithm based on the unitary quantum dynamics
[24]. This is because they are either irreversible and non-unitary or
dependent sensitively upon initial states of the quantum computational
process under study. For example, in the reversible computational model the
step number of computational process in the reversible halting protocol [54]
is dependent sensitively upon the initial state if the output state is fixed
in the reversible halting protocol, while in the reversible and unitary
halting protocol [11] the output state is dependent sensitively upon the
initial state if the step number of computational process is fixed. Thus,
both the reversible and unitary halting protocols are not suitable to
construct an efficient quantum search process based on the unitary quantum
dynamics. Only the specific reversible and unitary halting protocols that
are based on the state-locking pulse field [24] could be useful for solving
efficiently the quantum search problem in the cyclic group state space. This
is because neither the output state nor the step number of computational
process in such a reversible and unitary halting protocol is dependent
sensitively upon any initial state of the computational process and this is
the key point for the reversible and unitary halting protocol to be useful
for solving efficiently the quantum search problem. While the reversible and
unitary halting protocol is the key component of the quantum search process
to solve efficiently the quantum search problem in the cyclic group state
space, the state-locking pulse field plays a key role in constructing such a
reversible and unitary halting protocol. This is because the state-locking
pulse field could make both the output state and the step number of
computational process in the reversible and unitary halting protocol
insensitive to any unknown marked state of the quantum search problem. The
computational complexity for the quantum search process in the cyclic group
state space [24] could be mainly dependent on the performance of the
state-locking pulse field used in the quantum search process. The reversible
and unitary halting protocol not only plays an important role in solving
efficiently a quantum search problem in the cyclic group state space but
also has an extensive application in quantum computation. This is one of the
key components to realize a universal quantum computer to replace fully a
classical computer in future. Therefore, it is necessary to design a
good-performance unitary quantum control unit (or circuit) to simulate
faithfully and efficiently the reversible and unitary halting protocol of
the universal quantum computational models. A unitary quantum control unit
that consists of a trigger pulse, a state-locking pulse field, and a control
state subspace has been proposed to simulate faithfully and efficiently the
reversible and unitary halting protocol [24]. The key component of the
unitary quantum control unit is the state-locking pulse field. A
state-locking pulse field is able to keep a desired state almost unchanged
in a unitary form for a long time in a quantum computational process, and
this is the reason why the output state of the reversible and unitary
halting protocol based on the state-locking pulse field does not depend
sensitively upon any initial states. How to design a good-performance
state-locking pulse field is a challenge and also an important research
subject in quantum computation in future. In general, a general
state-locking pulse field may be dependent upon the time variable and the
space variables and even the quantum field variables. A state-locking pulse
field generally could be a sequence of time- and space-dependent
electromagnetic pulse fields such as the laser pulses and could also contain
any time- and space-dependent potential fields.

In this paper a new quantum program and circuit is constructed explicitly to
reduce the quantum search space which may be generally a cyclic-group state
space or the Hilbert space of an $n-$qubit quantum system. This quantum
program and circuit is much simpler than that one in the previous paper
[24], where in order to show clearly that an ideal universal quantum
computer could be powerful enough to solve efficiently the quantum search
problem in the multiplicative-cyclic-group state space the quantum program
and circuit used to reduce the quantum search space is designed in a more
complex form. It can also be used to construct further a quantum search
process to solve efficiently the unsorted quantum search problem in a
general cyclic group state space or the Hilbert space of an $n-$qubit
quantum system. The quantum program and circuit may be divided into two
almost independent units: the unitary quantum computational unit and the
unitary quantum control unit. The quantum control unit simulates efficiently
the reversible and unitary halting protocol based on the state-locking pulse
field, while the quantum computational unit is responsible for the reduction
of the quantum search space. In the paper the quantum program and circuit is
first analyzed completely. Then the basic properties of an ideal
state-locking pulse field are described in detail, and it is shown how the
quantum control unit simulates faithfully and efficiently the reversible and
unitary halting protocol. The basic principles to construct a general
state-locking pulse field and simulate efficiently the reversible and
unitary halting protocol are suggested and then explained in detail. It is
proposed in the paper that a simple atomic physical system which consists of
an atomic ion or a neutral atom in the double-well potential field is used
to realize the reversible and unitary halting protocol. In the atomic
physical system a generalized state-locking pulse field used to build up the
reversible and unitary halting protocol is also constructed explicitly. 
\newline
\newline
\newline
{\large 2. The reversible and unitary halting protocol and the state-locking
pulse field}

As an important application the reversible and unitary halting protocol may
be used to build up the reversible and unitary quantum program and circuit
to compute some mathematical functions. Here by solving a simple problem
given below one may illustrate how to use the reversible and unitary halting
protocol to solve a general mathematical problem in quantum computation.
Suppose that given a periodic function $f(x)=f(x+x_{T})$ with the period $%
x_{T}$ and the integer variable $x=0,1,...,x_{T}-1,$ there is a
computational circuit $U_{f}$ to compute the functional value $f(x+1)$ from
the functional value $f(x)$ for any integer $x=0,1,...,x_{T}-1$, where the
functional value $f(x)$ is a distinct integer for every distinct integer $x$
for $0\leq x<x_{T}$. The functional operation $U_{f}$ could be expressed as $%
U_{f}f(x)=f(x+1)$ [13, 14] for $x=0,1,...,x_{T}-1.$ Now given an unknown
functional value $f(x_{0})$ with the unknown integer $x_{0}\in
\{0,1,...,x_{T}-1\},$ one wants the unknown functional value $f(x_{0})$ to
be changed to the desired functional value $f(x_{f})$ with the known integer 
$x_{f}\in \{0,1,...,x_{T}-1\},$ e.g., $f(x_{f})=0$ or $1.$ A simple scheme
to solve this simple problem is that one computes one-by-one the functional
value $f(x_{0}+x)$ for the integer $x=0,1,...,x_{T}-1$ by the computational
circuit $U_{f}$ and checks the functional value for each computing step, and
when the functional value $f(x_{0}+x)$ is found to be equal to the desired
functional value $f(x_{f})$ the computational process is halted. Then the
final result of the computational process is clearly the desired functional
value $f(x_{f})$. Though this problem is very simple from the viewpoint of
the computational science, it is surprising that the scheme that can solve
efficiently this simple problem in a reversible and unitary form could also
be used to solve efficiently the quantum search problem. Note that given any
initial value $x_{0}$ and the final value $x_{f}$ there is the unique
integer $x$ within the range $0\leq x<x_{T}$ such that $f(x_{0}+x)=f(x_{f}).$
Now a classical computational program $Q_{cl}$ for the scheme could be
written down as 
\begin{equation*}
n_{h}=0
\end{equation*}%
\begin{equation*}
f(x)=f(x_{0})
\end{equation*}%
\begin{equation*}
\text{For }i=1\text{ to }x_{T}
\end{equation*}%
\begin{equation*}
\text{If }f(x)=f(x_{f})\text{ then }n_{h}=1
\end{equation*}%
\begin{equation*}
\text{while }n_{h}=1,\text{ halting}
\end{equation*}%
\begin{equation*}
\text{else }f(x)\rightarrow f(x+1)\text{ end if}
\end{equation*}%
\begin{equation*}
\text{end for}
\end{equation*}%
where $n_{h}=0$ or $1$ is the halting bit that is used to indicate when the
program terminates. As shown in the program, when the halting bit value $%
n_{h}=1$ the program terminates. The program consists of $x_{T}$ cycles with
the cyclic index $i=1,2,...,x_{T}$. Evidently, this simple program outputs
the desired functional value $f(x_{f})$ no matter what the initial
functional value $f(x_{0})$ is with the possible integer $%
x_{0}=0,1,...,x_{T}-1$. Here the unknown functional value $f(x_{0})$ may be
stored in memory on the computer in advance or is input from outside the
program. The program first checks whether the initial functional value $%
f(x_{0})$ equals $f(x_{f}).$ If $f(x_{f})=f(x_{0})$, the halting bit value $%
n_{h}$ is changed to $1$ from the initial value $0$ and then program
terminates due to $n_{h}=1,$ and the output result is $f(x_{f});$ otherwise
the program computes the functional value $f(x_{0}+1)$ and checks again
whether the functional value $f(x_{0}+1)$ is $f(x_{f})$ or not. This
computing process is repeated until the computed functional value $%
f(x_{0}+i) $ is found to be equal to $f(x_{f}),$ here $(x_{0}+i)\func{mod}%
x_{T}=x_{f}.$ Then the halting bit value $n_{h}=0$ is changed to $1$ so that
the program is halted. Thus, the output functional value for the
computational program $Q_{cl}$ is always $f(x_{f})$ no matter what the
initial functional value $f(x_{0})$ is. This is the important property of
the classical halting protocol. This property is also very important for the
reversible and unitary halting protocol.

If the classical computational program $Q_{cl}$ would be reversible and
unitary, then it could be suitably used to build up an efficient quantum
search process to solve the quantum search problem in a cyclic-group state
space, and such a quantum search process would be much simpler than that one
in the previous paper [24]. For the quantum search problem in the
multiplicative cyclic-group state space $S(C_{p-1})$ [24] the periodic
function $f(x)$ in the program $Q_{cl}$ may be chosen as the modular
exponential function $f_{k}(x)=(g^{M_{k}})^{x}\func{mod}p$ with the order
(or period) $x_{T}=m_{k}$ and the integer variable $x=0,1,...,m_{k}-1$ for $%
k=1,2,...,r$. Then the corresponding reversible functional operation $U_{f}$
may be defined as the unitary cyclic-group operation $U_{g^{M_{k}}}:f_{k}(x)%
\rightarrow f_{k}(x+1)$. Here $p$ is a prime and $g$ the generator of the
multiplicative cyclic group $C_{p-1},$ and the group operation of a
multiplicative cyclic group is the modular multiplication operation. The
multiplicative cyclic group $C_{p-1}$ has the order $%
p-1=p_{1}^{a_{1}}p_{2}^{a_{2}}...p_{r}^{a_{r}}$, where $p_{1},$ $p_{2},$ $%
...,$ $p_{r}$ are distinct primes and the exponents $a_{1},$ $a_{2},$ $...,$ 
$a_{r}>0.$ Here the integer $m_{k}=p_{k}^{a_{k}}$ and $p-1=m_{k}M_{k}$ for $%
k=1,2,...,r$. Any pair of the integers $m_{i}$ and $m_{j}$ are coprime to
each other for $1\leq i<j\leq r,$ and for convenience usually set $%
m_{1}<m_{2}<...<m_{r}$. Obviously, every functional value is distinct for
the modular exponential function $f_{k}(x)$ which satisfies $f_{k}(x)\geq 1$
for $x=0,1,...,m_{k}-1$. The functional states $\{|f_{k}(x)\rangle \}$
really form the multiplicative-cyclic-group state subspace $%
S(m_{k})=\{|(g^{M_{k}})^{s_{k}}\func{mod}p\rangle ,$ $s_{k}=0,1,...,m_{k}-1%
\} $ of the factor cyclic subgroup $C_{p_{k}^{a_{k}}}$ of the multiplicative
cyclic group $C_{p-1}=\{g^{s}\func{mod}p\}=C_{p_{1}^{a_{1}}}\times
C_{p_{2}^{a_{2}}}\times ...\times C_{p_{r}^{a_{r}}},$ whereas the
multiplicative-cyclic-group state space $S(C_{p-1})$ with dimension $p-1$ is
the direct product of the multiplicative-cyclic-group state subspaces $%
\{S(m_{k})\}$ with dimensions $\{m_{k}\}:$%
\begin{equation*}
S(C_{p-1})=S(m_{1})\bigotimes S(m_{2})\bigotimes ...\bigotimes S(m_{r}).
\end{equation*}%
If the quantum search problem is solved in the additive-cyclic-group state
space $S(Z_{p-1})$ with dimension $p-1=m_{1}m_{2}...m_{r},$ then the
periodic function $f(x)$ may be taken as the modular function $f_{k}(x)=x%
\func{mod}m_{k}$ with the period $x_{T}=m_{k}$ and $x=0,1,...,m_{k}-1$ for $%
k=1,2,...,r$. Every modular functional value $f_{k}(x)$ is distinct and $%
f_{k}(x)=0,1,...,m_{k}-1.$ Obviously, the modular functional states $%
\{|f_{k}(x)\rangle \}$ form really the additive-cyclic-group state subspace $%
S(Z_{m_{k}})=\{|s_{k}\func{mod}m_{k}\rangle ,$ $s_{k}=0,1,...,m_{k}-1\}.$
The additive cyclic group $Z_{p-1}=\{0,1,...,p-2\}$ may be decomposed into
the direct sum of the factor additive cyclic subgroups $\{Z_{m_{k}}%
\}:Z_{p-1}=Z_{m_{1}}\bigoplus Z_{m_{2}}\bigoplus ...\bigoplus Z_{m_{r}},$
here the group operation of an additive cyclic group is the modular addition
operation. The additive-cyclic-group state space $S(Z_{p-1})$ then may be
decomposed into the direct product of the factor additive-cyclic-group state
subspaces $\{S(Z_{m_{k}})\}$ with dimensions $\{m_{k}\}:$%
\begin{equation*}
S(Z_{p-1})=S(Z_{m_{1}})\bigotimes S(Z_{m_{2}})\bigotimes ...\bigotimes
S(Z_{m_{r}}).
\end{equation*}%
Though the $2^{n}-$dimensional Hilbert space $S(Z_{2^{n}})$ of an $n-$qubit
quantum system may be thought of as an additive-cyclic-group state space
under the modular addition operation ($\func{mod}2^{n}$), for the Hilbert
space $S(Z_{2^{n}})$ there is not a direct-sum decomposition like $%
S(Z_{p-1}) $ above with $r>1.$ However, the Hilbert space $S(Z_{2^{n}})$ may
be decomposed into the direct product of $n$ $2-$dimensional
additive-cyclic-group state subspaces $\{S(Z_{2})\}$, 
\begin{equation*}
S(Z_{2^{n}})=S(Z_{2})\bigotimes S(Z_{2})\bigotimes ...\bigotimes S(Z_{2}).
\end{equation*}%
If now the periodic function $f(x)$ is chosen as the modular function $%
f_{k}(x)=x\func{mod}2$ with $x=0,1$ for $k=1,2,...,n$, then the quantum
search problem in the $2^{n}-$dimensional Hilbert space of the $n-$qubit
quantum system may also be solved just like that one in the
additive-cyclic-group state space $S(Z_{p-1})$ or in the
multiplicative-cyclic-group state space $S(C_{p-1})$ [24]. The symmetric
properties and structures of both the cyclic groups $C_{p-1}$ and $Z_{p-1}$
have been suggested to help the unitary quantum dynamics to solve the
quantum search problem in the cyclic-group state spaces $S(C_{p-1})$ and $%
S(Z_{p-1}),$ respectively [24].

Obviously, the classical computational program $Q_{cl}$ is not reversible
due to the irreversible operations in the program and especially due to the
fact that the halting operation within the program is not reversible.
However, in order that it can be used to construct a quantum search process
to solve the quantum search problem in the cyclic-group state space the
whole program $Q_{cl}$ including the halting operation must be made
reversible and unitary. Generally, a classical irreversible computational
program may be made reversible in the frame of the reversible computational
model [1] with the help of the reversible mathematical logic gates and
especially the reversible operation of a general function mathematically
[1], the universal quantum gates and especially the conditional quantum
gates [25], and other unitary operations quantum mechanically [26]. One can
also construct directly a unitary quantum circuit equivalent to a classical
computational program on the universal quantum circuit model [4]. It has
been shown that there is an equivalent quantum circuit to a given reversible
quantum program on the universal quantum Turing machine [27]. However, these
conventional methods by which an irreversible classical computational
program is made reversible and unitary could not be always suitable for
transforming the irreversible halting protocol to the reversible and unitary
one in the universal quantum computational models. First, there could not be
a universal halting protocol in the universal quantum computational models.
Second, although there could be a halting protocol in the universal quantum
computational models when the input state of a quantum computational process
is limited to a single basis state, this halting protocol could not be
thought to be reversible in the sense that an information loss occurs in the
halting protocol due to the non-unitary quantum measurement. For example, by
the quantum measurement one could know what time the halting state $%
|0\rangle $ is changed to the state $|1\rangle $ and hence obtains the
information of the instant of time at which the quantum system arrives in
the end state of the quantum computational process, implying that the
quantum computational process loses the information of the instant of time.
However, even if the current quantum measurement could not be non-unitary
according to quantum mechanics [5b] as the measured base now is a single
basis state of the measurement operator and such an information loss could
not have a significant effect on some quantum computational processes, this
information is not available according as the mathematical logic principles
of the quantum search problem and can not be yet allowed to use if the
halting protocol is used in solving the quantum search problem based on the
unitary quantum dynamics. This is because a quantum computational process
obeys the quantum physical laws and is compatible with the used mathematical
logic principles. Therefore, the classical program $Q_{cl}$ may be made
reversible and unitary according as these conventional methods except the
irreversible halting operation within the program. Of course, in the case
that any initial state is restricted to be a single basis state the halting
operation may also be made reversible by the conventional methods [11, 54]
for the conventional computational tasks other than the current tasks. In
the reversible computational model the reversible halting operation in the
program $Q_{cl}$ may be achieved in such a way [54] that after the
computational process from the initial functional value $f(x_{0})$ to the
output functional value $f(x_{f})$ is done the conditional halting operation
is executed and then the inverse computational process is performed from the
functional value $f(x_{f})$ back to the original functional value $f(x_{0}),$
and the cyclic process consisting of the computational process and inverse
computational process repeats incessantly. Evidently, this extra inverse
computational process is dependent upon the cyclic index $i$ of the program
as a different initial functional value $f(x_{0})$ is changed to the
functional value $f(x_{f})$ in a different cycle in the program. One could
also use the halting protocol [11] to achieve the reversible halting
operation for the program. When the program arrives at the output functional
state (corresponding to $f(x_{f})$, see below) the conditional halting
operation is executed and then an extra unitary evolution process starts,
that is, the state of the auxiliary register starts to evolve [11]. If now
the step number of the program $Q_{cl}$ is fixed (for example the step
number may be set to $x_{T})$, then a different initial functional state
(corresponding to $f(x_{0})$) will result in a different output state of the
auxiliary register. Thus, the total output state of the program is really
dependent upon the initial functional state. One therefore concludes that if
any of the two reversible halting protocols [11, 54] is used in the program,
then the program is dependent sensitively upon the initial functional state.
Such a reversible program is not suited as a component of the quantum search
process based on the unitary quantum dynamics. It is necessary to use the
specific method to make the halting operation within the program $Q_{cl}$
reversible and unitary if one wants to use the program to construct a
quantum search process. This specific scheme to make the halting protocol
reversible and unitary may be stated below. As the first point of the
scheme, in order that the halting protocol is reasonable from the viewpoint
of the mathematical logic principle any initial state is limited to a single
basis state for the halting protocol. The cost for this point is that the
quantum parallel principle could become less important. As the second point,
the halting protocol should not contain any quantum measurement so as to
keep away from any irreversible and non-unitary process and avoid any
information loss of the quantum system. Generally, the unitary quantum
dynamics avoids using any non-unitary quantum measurement as its quantum
operation within quantum computational processes. This is quite different
from the early proposals that the quantum measurement may also be used as a
quantum operation to build up quantum circuits and algorithms [13, 28]. If
the halting operation within the classical program $Q_{cl}$ is achieved by
the brute-force method, then the end state (i.e., the output state) of the
classical program is independent of any initial functional value $f(x_{0}).$
But the brute-force method to stop the program from the outside could cause
the whole program irreversible. Therefore, for the third point of the scheme
the brute-force method is replaced with the unitary operation conditionally
depending upon the halting state to stop the program [11, 12, 54]. However,
all these three points above in the scheme can not ensure that the end state
of the program which could include the computational state, halting-qubit
state, and auxiliary state is independent of any initial functional state,
that is, all these three points can not lead to the important property of
the halting operation that the output state is independent of any initial
state. It is of crucial importance to make the end state of the program
independent or almost independent of any initial functional state. This is
because whether or not the quantum search process based on the reversible
and unitary halting protocol is efficient is mainly dependent upon this
point. Thus, as the fourth point, the end state of the program is locked by
the state-locking pulse field so that it is not dependent sensitively upon
any initial functional state. This scheme which consists of the four points
above has been used in the previous paper [24]. However, this scheme is not
intuitive to understand the reversibility and unitarity of the reversible
and unitary halting protocol mainly due to that it is hard to understand how
the state-locking pulse field is capable of keeping a quantum state almost
unchanged in a unitary form for a long time in a quantum computational
process.

It was proposed in the previous paper [24] that a unitary quantum control
unit which consists of a trigger pulse ($P_{t}$), a state-locking pulse
field ($P_{SL}$), and a two-state control state subspace is used to simulate
efficiently the reversible and unitary halting protocol. This quantum
control unit does not use any non-unitary quantum measurement as its
component. The state-locking pulse field which is the key component of the
unitary quantum control unit is used to keep the desired state almost
unchanged in a unitary form for a long time in the quantum computational
process under study. Now by using a similar unitary quantum control unit to
simulate faithfully and efficiently the halting operation within the
classical program $Q_{cl}$ and with the help of the reversible mathematical
logic operations [1], the universal quantum gates and especially the
conditional quantum gates [25], and any other unitary quantum operations, a
reversible and unitary quantum computational program $Q_{c}$ and its
equivalent quantum circuit which correspond to the classical program $Q_{cl}$
can be constructed explicitly in the frames of the reversible computational
model [1] and the universal quantum circuit model [4], respectively, and
they may be represented intuitively by \newline
\begin{equation}
\text{State-Locking Pulse Field }(P_{SL}):ON  \tag{P1}
\end{equation}%
\begin{equation}
|n_{h}\rangle =|0\rangle  \tag{P2}
\end{equation}%
\begin{equation}
|b_{h}\rangle =|0\rangle  \tag{P3}
\end{equation}%
\begin{equation}
|f_{r}(x)\rangle =|f_{r}(x_{0})\rangle  \tag{P4}
\end{equation}%
\begin{equation}
\text{For }i=1\text{ to }m_{r}  \tag{P5}
\end{equation}%
\begin{equation}
\text{If }|f_{r}(x)\rangle =|1\rangle \text{ then }U_{b}^{c}:|b_{h}\rangle
\rightarrow |b_{h}+1\rangle \text{ end if}  \tag{P6}
\end{equation}%
\begin{equation*}
\text{While }|f_{r}(x)\rangle =|1\rangle ,\text{ Do }U_{h}^{c}:|n_{h}\rangle
|f_{r}(x)\rangle =|0\rangle |1\rangle \rightarrow |0\rangle |0\rangle ,
\end{equation*}%
\begin{equation}
\quad \qquad P_{t}:|n_{h}\rangle |f_{r}(x)\rangle =|0\rangle |0\rangle
\rightarrow |c_{1}\rangle |0\rangle ,\text{ }P_{SL}:|c_{1}\rangle
\rightarrow |c_{2}\rangle  \tag{P7}
\end{equation}%
\begin{equation}
\text{If }|b_{h}\rangle =|0\rangle \text{ then }U_{f}:|f_{r}(x)\rangle
\rightarrow |f_{r}(x+1)\rangle \text{ end if}  \tag{P8}
\end{equation}%
\begin{equation}
\text{end for}  \tag{P9}
\end{equation}%
\begin{equation}
\text{State-Locking Pulse Field }(P_{SL}):OFF  \tag{P10}
\end{equation}%
\newline
Note that the halting bit $n_{h}$ and the function $f(x)$ in the classical
program $Q_{cl}$ have already been replaced with the halting state $%
|n_{h}\rangle $ and the functional state $|f(x)\rangle $ in the quantum
program $Q_{c},$ respectively. In the quantum program $Q_{c}$ the functional
state $|f(x)\rangle $ is set to the modular exponentiation state $%
|f_{r}(x)\rangle $ with the period $x_{T}=m_{r}$ of the
multiplicative-cyclic-group state subspace $S(m_{r})=\{|f_{r}(x)\rangle
=|(g^{M_{r}})^{x}\func{mod}p\rangle ;$ $x=0,1,...,m_{r}-1\}$ and the desired
functional state $|f_{r}(x_{f})\rangle $ set to the state $|1\rangle .$ Note
that the multiplicative-cyclic-group state subspace $S(m_{r})$ does not
contain the state $|0\rangle $ because the functional value $f_{r}(x)\geq 1$
for any integer $x=0,1,...,m_{r}-1$. Owing to $f_{r}(x)\neq 0$ here the
state $|f_{r}(x)\rangle =|0\rangle $ means that the state in the register of
the functional state $|f_{r}(x)\rangle $ takes the state $|0\rangle $ rather
than that the functional value $f_{r}(x)=0.$ The unitary functional
operation $U_{f}$ is defined by $U_{f}|f_{r}(x)\rangle =|f_{r}(x+1)\rangle $
for $|f_{r}(x)\rangle \in S(m_{r}),$ here $U_{f}$ is really the cyclic-group
unitary operation $U_{g^{M_{r}}}$ of the factor cyclic subgroup $%
C_{p_{r}^{a_{r}}}$ of the multiplicative cyclic group $C_{p-1}$. If the
state $|f_{r}(x)\rangle =|0\rangle ,$ then $U_{f}|f_{r}(x)\rangle
=U_{f}|0\rangle =|0\rangle $ as the state $|0\rangle $ does not belong to
the multiplicative-cyclic-group state subspace $S(m_{r})$. More generally
the functional operation $U_{f}$ satisfies $U_{f}|g(x)\rangle =|g(x)\rangle $
if the state $|g(x)\rangle $ is not in the state subspace $S(m_{r})$. Though
here the quantum program $Q_{c}$ and its equivalent quantum circuit are
designed for the multiplicative-cyclic-group state subspaces $\{S(m_{k}),$ $%
k=1,2,...,r\}$ (the index $r$ may be replaced with $k$ in the quantum
program $Q_{c}$), similar quantum programs and circuits can also be
constructed for the additive-cyclic-group state subspace $%
S(Z_{m_{k}})=\{|f_{k}(x)\rangle =|x\func{mod}m_{k}\rangle ;$ $%
x=0,1,...,m_{k}-1\}$ for $k=1,2,...,r$ and for a general periodic-function
state space. The quantum program $Q_{c}$ contains mainly $m_{r}$ cycles with
the cyclic index $i=1$ to $m_{r}$ and the state-locking pulse field ($P_{SL}$%
).

The quantum program $Q_{c}$ consists of ten statements, which are denoted
conveniently as the statement P1, statement P2, ..., statement P10,
respectively. In particular, the statement P7 consists of the conditional
unitary operation $U_{h}^{c}$, the trigger pulse $P_{t}$, and the
state-locking pulse field $P_{SL}.$ The quantum program $Q_{c}$ is more
complex than the classical one $Q_{cl}$ mainly due to the statement P7. The
statement P7 contains the unitary quantum control unit that simulates
efficiently the reversible and unitary halting protocol. The unitary quantum
control unit consists of the trigger pulse $P_{t},$ the state-locking pulse
field $P_{SL},$ and the control state subspace $S(C)=\{|c_{1}\rangle
,|c_{2}\rangle \}$. Any one of these two states $|c_{1}\rangle $ and $%
|c_{2}\rangle $ of the control state subspace is different from the initial
halting state $|n_{h}\rangle =|0\rangle $. Actually the three states, i.e.,
the halting state $|n_{h}\rangle =|0\rangle $ and these two states $%
|c_{1}\rangle $ and $|c_{2}\rangle ,$ should be orthogonal to each other and
belong to the same register which is named the halting register here. These
three states are also called the halting-register states or briefly the
halting states. It will be seen in next section that the halting register
may be generated from the simple physical system of an atomic ion or a
neutral atom in the double-well potential field. The physical system of the
halting register may also be called the quantum control system whose Hilbert
space contains the control state subspace $S(C)$ and probably other relevant
states. Both the trigger pulse $P_{t}$ and the state-locking pulse field $%
P_{SL}$ are applied only to the physical system of the halting register.
Actually, in the quantum program $Q_{c}$ the state-locking pulse field $%
P_{SL}$ is applied only to the control state subspace $S(C).$ The trigger
pulse $P_{t}$ acts on the initial halting state $|n_{h}\rangle =|0\rangle $
to convert conditionally it into the state $|c_{1}\rangle $ of the control
state subspace only if the state $|f_{r}(x)\rangle $ in the register of the
functional state is the state $|0\rangle .$ Thus, here the unitary
transformation $U_{t}$ during the trigger pulse $P_{t}$ could be defined
simply by 
\begin{equation*}
U_{t}:|n_{h}\rangle |f_{r}(x)\rangle =|0\rangle |0\rangle \leftrightarrow
|c_{1}\rangle |0\rangle .
\end{equation*}%
Both the trigger pulse and the state-locking pulse field are very important
in the unitary quantum control unit and will be discussed separately in
detail later. In order that the halting quantum bit $\{|n_{h}\rangle
,n_{h}=0,1\}$ is separated from other qubits of the computational state
subspace in the quantum program an extra quantum bit named the
branch-control quantum bit $\{|b_{h}\rangle ,b_{h}=0,1\}$ is used to control
directly the functional operation of the function $f_{r}(x)$ in place of the
halting quantum bit. Therefore, there is the unitary conditional functional
operation $U_{f}^{c}$ for the functional state $|f_{r}(x)\rangle $ defined
by 
\begin{equation*}
U_{f}^{c}|b_{h}\rangle |f_{r}(x)\rangle =\left\{ 
\begin{array}{c}
|0\rangle |f_{r}(x+1)\rangle \text{ if }b_{h}=0 \\ 
|b_{h}\rangle |f_{r}(x)\rangle \text{ if }b_{h}\neq 0\quad \ 
\end{array}%
\right.
\end{equation*}%
This definition shows that the conditional functional operation $U_{f}^{c}$
is applied only to both the functional state $|f_{r}(x)\rangle $ and the
branch-control state $|b_{h}\rangle .$ If the branch-control state $%
|b_{h}\rangle $ is the state $|0\rangle $ the functional operation $U_{f}$
changes the functional state $|f_{r}(x)\rangle $ to another functional state 
$|f_{r}(x+1)\rangle ,$ otherwise the operation $U_{f}$ does not change the
functional state. Therefore, the branch-control state $|b_{h}\rangle $ can
control conditionally the action of the functional operation $U_{f}$ upon
the functional state. When the branch-control state $|b_{h}\rangle $ is
changed from the state $|0\rangle $ to the state $|1\rangle $ the functional
operation $U_{f}$ is halted to act on the functional state $|f_{r}(x)\rangle 
$ even though the conditional functional operation $U_{f}^{c}$ continues to
apply to the functional state. In the quantum program the halting quantum
bit $\{|n_{h}\rangle \}$ is designed to control the branch-control quantum
bit $\{|b_{h}\rangle \}$ indirectly and hence controls ultimately the action
of the functional operation $U_{f}$ on the functional state. The conditional
unitary transformation $U_{b}^{c}$ in the quantum program could be defined
simply by 
\begin{equation*}
U_{b}^{c}|b_{h}\rangle |f_{r}(x)\rangle =\left\{ 
\begin{array}{c}
|(b_{h}+1)\func{mod}2\rangle |f_{r}(x)\rangle ,\text{ if }|f_{r}(x)\rangle
=|1\rangle \\ 
|b_{h}\rangle |f_{r}(x)\rangle ,\text{ if }|f_{r}(x)\rangle \neq |1\rangle
\qquad \qquad \quad \ 
\end{array}%
\right.
\end{equation*}%
and the conditional unitary transformation $U_{h}^{c}:|n_{h}\rangle
|f_{r}(x)\rangle =|0\rangle |1\rangle $ $\leftrightarrow |0\rangle |0\rangle 
$ is defined explicitly by 
\begin{equation*}
U_{h}^{c}|n_{h}\rangle |f_{r}(x)\rangle =\left\{ 
\begin{array}{c}
|0\rangle |0\rangle ,\text{ if }|f_{r}(x)\rangle =|1\rangle \text{ and }%
|n_{h}\rangle =|0\rangle \\ 
|0\rangle |1\rangle ,\text{ if }|f_{r}(x)\rangle =|0\rangle \text{ and }%
|n_{h}\rangle =|0\rangle \\ 
|n_{h}\rangle |f_{r}(x)\rangle ,\text{ otherwise\qquad \qquad \qquad\ \ }%
\end{array}%
\right.
\end{equation*}

The detailed analysis for the quantum program $Q_{c}$ is given below. The
state-locking pulse field $P_{SL}$ is first switched on at the beginning of
the quantum program (see the statement P1 in the program)\ and could be
switched off (or partly switched off)\ after the quantum program finished
(the statement P10). It is mainly used to manipulate the states $%
|c_{1}\rangle $ and $|c_{2}\rangle $ and lock the state $|c_{2}\rangle $ of
the control state subspace $S(C)$. Both the initial halting state $%
|n_{h}\rangle $ (statement P2) and the initial branch-control state $%
|b_{h}\rangle $ (the statement P3) are simply set to the state $|0\rangle .$
The initial functional state $|f_{r}(x_{0})\rangle $ (the statement P4)
could be unknown and may be stored in the memory of the quantum computer in
advance or is input from outside the quantum program. There are $m_{r}$
possible initial functional states $\{|f_{r}(x_{0})\rangle ,$ $%
x_{0}=0,1,...,m_{r}-1\}$ at most. The quantum program first checks whether
or not the initial functional state $|f_{r}(x_{0})\rangle $ is the desired
functional state $|f_{r}(x_{f})\rangle =|1\rangle $ (the statement P6). Then
there are two possible cases to be considered, that is, either $%
|f_{r}(x_{0})\rangle =|f_{r}(x_{f})\rangle $ or $|f_{r}(x_{0})\rangle \neq
|f_{r}(x_{f})\rangle $. Consider the first case that $|f_{r}(x_{0})\rangle
=|f_{r}(x_{f})\rangle =|1\rangle $. Since the state $|f_{r}(x_{0})\rangle
=|f_{r}(x_{f})\rangle $ the branch-control state $|b_{h}\rangle =|0\rangle $
is first changed to the state $|1\rangle $ by the conditional unitary
operation $U_{b}^{c}$ (the statement P6). Then the desired state $%
|f_{r}(x_{f})\rangle $ is changed conditionally to the state $|0\rangle $
(the statement P7)\ by the conditional unitary operation $U_{h}^{c}$ due to
that the halting state $|n_{h}\rangle $ now is the state $|0\rangle ,$ the
initial halting state $|n_{h}\rangle =|0\rangle $ then is changed to the
state $|c_{1}\rangle $ of the control state subspace by the trigger pulse $%
P_{t}$ and then the state $|c_{1}\rangle $ to another orthogonal state $%
|c_{2}\rangle $ of the control state subspace, since then the state $%
|c_{2}\rangle $ is kept unchanged by the state-locking pulse field $P_{SL}$
(the statement P7). When the quantum program executes the statement P8 the
conditional functional operation $U_{f}^{c}$ will not have a net effect on
the functional state $|f_{r}(x)\rangle $ according as the definition of the
operation $U_{f}^{c}$ because the branch-control state $|b_{h}\rangle $ now
is the state $|1\rangle $ and the state $|f_{r}(x)\rangle =|0\rangle ,$
although now the operation $U_{f}^{c}$ is still applied to the whole quantum
system of the quantum computer. This shows that the functional operation $%
U_{f}$ acting on the functional state $|f_{r}(x)\rangle $ is really halted
after the initial branch-control state $|0\rangle $ is changed to the state $%
|1\rangle $. Now the quantum program finished the first cycle with the index 
$i=1$. It then returns and executes the statement P5 of the second cycle
with the index $i=2$. When the quantum program executes the statement P6 in
the second cycle, the current branch-control state $|b_{h}\rangle =|1\rangle 
$ keeps unchanged under the action of the operation $U_{b}^{c}$ as the
current state $|f_{r}(x)\rangle =|0\rangle .$ Since the current state $%
|n_{h}\rangle |f_{r}(x)\rangle =|c_{2}\rangle |0\rangle ,$ that is, the
state $|n_{h}\rangle \neq |0\rangle ,$ the unitary operation $U_{h}^{c}$
does not have a real effect on the quantum system when the quantum program
executes the statement P7, and since now the halting-register state $%
|n_{h}\rangle $ is the state $|c_{2}\rangle ,$ that is, the state $%
|n_{h}\rangle $ is neither $|0\rangle $ nor $|c_{1}\rangle ,$ the trigger
pulse $P_{t}$ does not yet have a real effect on the quantum system. The key
point in the quantum program is that the state $|c_{2}\rangle $ of the
control state subspace has been locked by the state-locking pulse field $%
P_{SL}$ since the quantum program executes the statement P7 in the first
cycle $(i=1).$ Hence the state $|c_{2}\rangle $ is not yet changed when the
statement P7 is executed in the second cycle. Actually, the state $%
|c_{2}\rangle $ may be kept unchanged till the end of the quantum program
after the statement P7 was executed in the first cycle. If now the program
continues to execute the rest statements and even run till the end of the
program, then all these states $|n_{h}\rangle =|c_{2}\rangle ,$ $%
|b_{h}\rangle =|1\rangle ,$ and $|f_{r}(x)\rangle =|0\rangle $ of the whole
quantum system of the quantum computer are still kept unchanged due to the
fact that the halting-register state $|n_{h}\rangle $ is kept in the state $%
|c_{2}\rangle $ by the state-locking pulse field. Note that the conditional
functional operation $U_{f}^{c}$ is applied continuously to the quantum
system of the quantum computer even after the branch-control state $%
|b_{h}\rangle =|0\rangle $ is changed to the state $|1\rangle ,$ which leads
to that the functional operation $U_{f}$ acting on the functional state $%
|f_{r}(x)\rangle $ is halted. Of course, in this case the conditional
functional operation $U_{f}^{c}$ has not a net effect on the functional
state $|f_{r}(x)\rangle .$ This process is repeated from the second cycle ($%
i=2$) to the end ($i=m_{r}$) of the quantum program. The analysis above
shows that if the functional state $|f_{r}(x)\rangle $ is the desired state $%
|f(x_{f})\rangle $, then when the quantum program executes the statement P6
the branch-control state $|b_{h}\rangle =|0\rangle $ is changed to the state 
$|1\rangle ,$ and then on the statement P7 the initial halting state $%
|n_{h}\rangle =|0\rangle $ is changed to the state $|c_{1}\rangle $ and then
further to the state $|c_{2}\rangle $ which is ultimately kept unchanged by
the state-locking pulse field, since then the state $|n_{h}\rangle
|b_{h}\rangle |f_{r}(x)\rangle =$ $|c_{2}\rangle |1\rangle |0\rangle $ of
the whole quantum system is kept unchanged till the end of the program.
Therefore, the quantum program outputs the final state $|n_{h}\rangle
|b_{h}\rangle |f_{r}(x)\rangle =$ $|c_{2}\rangle |1\rangle |0\rangle $ (the
state $|f_{r}(x)\rangle =$ $|0\rangle $ is easily changed to the desired
state $|f_{r}(x_{f})\rangle =|1\rangle ).$ Next consider the second case:
the state $|f_{r}(x_{0})\rangle \neq |f_{r}(x_{f})\rangle .$ If the initial
functional state $|f_{r}(x_{0})\rangle \neq |f_{r}(x_{f})\rangle =|1\rangle
, $ then the initial branch-control state $|b_{h}\rangle =|0\rangle $ is not
changed to the state $|1\rangle $ by the unitary operation $U_{b}^{c}$ when
the program executes the statement P6. Again due to that the state $%
|f_{r}(x_{0})\rangle \neq |1\rangle $ and $|0\rangle $ the unitary operation 
$U_{h}^{c}$ does not really act on the state $|n_{h}=0\rangle
|f_{r}(x_{0})\rangle $ and both the trigger pulse $P_{t}$ and the
state-locking pulse field $P_{SL}$ do not yet act on the initial halting
state $|n_{h}\rangle =|0\rangle $ when the program executes the statement
P7. Since the branch-control state $|b_{h}\rangle =|0\rangle $ the
functional state $|f_{r}(x_{0})\rangle $ is changed to the state $%
|f_{r}(x_{0}+1)\rangle $ by the conditional functional operation $U_{f}^{c}$
after executing the statement P8. Now the quantum program returns to execute
the statement P6 of the second cycle after the cyclic index $i=1$ is changed
to $i=2$ on the statement P5. Again the program first checks whether or not
the functional state $|f_{r}(x_{0}+1)\rangle $ is the desired state $%
|f_{r}(x_{f})\rangle $. Just like before there are also two possible cases,
the first case is $|f_{r}(x_{0}+1)\rangle =|f_{r}(x_{f})\rangle $ and the
second $|f_{r}(x_{0}+1)\rangle \neq |f_{r}(x_{f})\rangle $. As shown above,
for the first case $|f_{r}(x_{0}+1)\rangle =|f_{r}(x_{f})\rangle $ the
program will output the final state $|n_{h}\rangle |b_{h}\rangle
|f_{r}(x)\rangle =$ $|c_{2}\rangle |1\rangle |0\rangle .$ For the second
case $|f_{r}(x_{0}+1)\rangle \neq |f_{r}(x_{f})\rangle $ the functional
state $|f_{r}(x_{0}+1)\rangle $ will be changed to the state $%
|f_{r}(x_{0}+2)\rangle $ at the end of the second cycle ($i=2$) of the
program (the statement P8). This process is repeated till the $k-$th cycle ($%
m_{r}>k\geq 1$) when the functional state $|f_{r}(x_{0}+k)\rangle
=|f_{r}(x_{f})\rangle $ at the end of the $k-$th cycle$.$ Here the index $k$
is unique for $0\leq k<m_{r}$ and $k=0$ corresponds to the earlier case $%
|f_{r}(x_{0})\rangle =|f_{r}(x_{f})\rangle .$ Now for the $(k+1)-$th cycle
the initial branch-control state $|b_{h}\rangle =|0\rangle $ is first
changed to the state $|1\rangle $ (the statement P6), following the
statement P6 the functional state $|f_{r}(x_{0}+k)\rangle
=|f_{r}(x_{f})\rangle $ is changed to the state $|0\rangle ,$ then the
initial halting state $|n_{h}\rangle =|0\rangle $ to the state $%
|c_{1}\rangle $ by the trigger pulse $P_{t}$ and further to the state $%
|c_{2}\rangle $ by the state-locking pulse field (the statement P7), and
since then the state $|c_{2}\rangle $ is kept unchanged by the state-locking
pulse field till the end of the program. Therefore, the quantum program
outputs finally the state $|n_{h}\rangle |b_{h}\rangle |f_{r}(x)\rangle =$ $%
|c_{2}\rangle |1\rangle |0\rangle $. This shows that after executing quantum
program $Q_{c}$ the initial state $|n_{h}\rangle |b_{h}\rangle
|f_{r}(x)\rangle =$ $|0\rangle |0\rangle |f_{r}(x_{0})\rangle $ is always
transferred to the output state $|c_{2}\rangle |1\rangle |0\rangle $ no
matter what the initial functional state $|f_{r}(x_{0})\rangle $ is with $%
x_{0}=0,1,...,m_{r}-1.$ Note that the quantum program $Q_{c}$ is reversible
and unitary because all these operations of the quantum program are
reversible and unitary. One therefore concludes that by the unitary quantum
program $Q_{c}$ different initial states $\{|0\rangle |0\rangle
|f_{r}(x_{0})\rangle \}$ are transferred to the same output state $%
|c_{2}\rangle |1\rangle |0\rangle $ and hence the output state of the
quantum program is not dependent sensitively upon any initial states.
However, the first part of the conclusion is apparently in conflict with the
fact that different input states can not be completely transferred to the
same output state by a given unitary transformation. Therefore, the first
part of the conclusion is expressed exactly as that different initial states 
$\{|0\rangle |0\rangle |f_{r}(x_{0})\rangle \}$ are transferred to the same
output state $|c_{2}\rangle |1\rangle |0\rangle $ in probabilities
approaching infinitely 100\% in theory by the unitary quantum program $Q_{c}$%
. In theory there is only one initial state $|0\rangle |0\rangle
|f_{r}(x_{0})\rangle $ that may be completely transferred to the output
state $|c_{2}\rangle |1\rangle |0\rangle $ by the unitary quantum program.
As shown in next sections, due to the fact that the quantum program $Q_{c}$
is reversible and unitary the output state $|c_{2}\rangle |1\rangle
|0\rangle $ can be really obtained from the initial state $|0\rangle
|0\rangle |f_{r}(x_{0})\rangle $ only in a probability close to 100\% rather
than in the probability 100\% for any initial functional state $%
|f_{r}(x_{0})\rangle $ in a real physical system.

The state-locking pulse field $P_{SL}$ plays a key role in the quantum
control process that simulates efficiently the reversible and unitary
halting protocol in the quantum program $Q_{c}$. It is the state-locking
pulse field that keeps the state $|c_{2}\rangle $ of the control state
subspace unchanged till the end of the quantum program after the functional
state $|f_{r}(x)\rangle $ is changed to the desired functional state $%
|f_{r}(x_{f})\rangle $, while keeping the state $|c_{2}\rangle $ unchanged
for a long time is the key step to achieve the reversible and unitary
halting protocol. As pointed out before, the statement P7 of the quantum
program which is involved in the state-locking pulse field $P_{SL}$ is
mainly used to simulate the reversible and unitary halting protocol. The
statement P7 really forms a unitary quantum control process (or unit). This
process (or unit) is almost independent of the quantum computational process
(or unit) to compute the desired functional state $|f_{r}(x_{f})\rangle $ in
the quantum program, but it really controls the quantum computational
process (or unit). A quantum program and its quantum circuit may be
generally divided into two parts, one part is the quantum computational unit
(or process) and another the quantum control unit (or process). As an
example, for the quantum program $Q_{c}$ the quantum computational unit is
used to compute the desired functional state $|f_{r}(x_{f})\rangle ,$ while
the quantum control unit is used to perform the reversible and unitary
halting protocol. As pointed out before, the unitary quantum control unit
consists of the trigger pulse $P_{t}$, the state-locking pulse field $P_{SL}$%
, and the control state subspace $S(C)$. Generally, the control state
subspace $S(C)$ is different from the computational state subspace such as
the multiplicative-cyclic-group state subspace $S(m_{r})$ in the quantum
program, but they all belong to the Hilbert space of the whole quantum
system of the quantum computer. The simplest control state subspace is a
two-state subspace such as $S(C)=\{|c_{1}\rangle ,|c_{2}\rangle \},$ but
generally a control state subspace is not restricted to be a two-state
subspace in the quantum program. The trigger pulse $P_{t}$ could be used for
communication between the control state subspace and the computational state
subspace. It generally triggers a time- and space-dependent unitary
evolution process in the quantum control system with the control state
subspace $S(C)$. The unitary transformation for the trigger pulse may be
defined explicitly. For example, in the quantum program $Q_{c}$ the initial
halting state $|n_{h}\rangle =|0\rangle $ is converted into the state $%
|c_{1}\rangle $ and vice versa by the trigger pulse $P_{t}$ only if the
state $|f(x)\rangle $ is the state $|0\rangle $. Then the unitary
transformation for the trigger pulse $P_{t}$ is defined simply by $U_{t},$
as shown before. Here the state $|f_{r}(x)\rangle =|0\rangle $ in the
register of the functional state could be thought to be related to the
computational state subspace, while the state $|c_{1}\rangle $ is of the
control state subspace. A different definition for the unitary
transformation during the trigger pulse $P_{t}$ can be seen in next section.
The unitary quantum control process that simulates the reversible and
unitary halting protocol may be simply described below. Here it is first
pointed out that the state-locking pulse field $P_{SL}$ is applied only to
the quantum control system with the control state subspace $S(C)$ in the
quantum program. The state-locking pulse field is first switched on to apply
to the quantum control system of the quantum computer at the beginning of
the quantum program (see the statement P1)$,$ but it usually does not take a
real action on the quantum control system at the beginning time. However,
when the functional state $|f_{r}(x)\rangle $ is changed to the desired
state $|f_{r}(x_{f})\rangle $ and then the initial halting state $%
|n_{h}\rangle =|0\rangle $ changed conditionally to the state $|c_{1}\rangle 
$, the state-locking pulse field $P_{SL}$ which has been applying to the
control state subspace since the beginning of the quantum program starts to
take a real action on the states of the control state subspace. The state $%
|c_{1}\rangle $ is first sent to the state $|c_{2}\rangle $ in the control
state subspace under the state-locking pulse field. This process usually is
a time- and space-dependent unitary evolution process. Then the state $%
|c_{2}\rangle $ is kept unchanged by the state-locking pulse field to the
end of the quantum program and circuit. Because now the branch-control state 
$|b_{h}\rangle $ leaves the initial one and is kept in the state $|1\rangle $
unchanged due to that the state $|c_{2}\rangle $ is kept unchanged by the
state-locking pulse field the computational process is halted conditionally
and the reversible and unitary halting operation therefore is achieved.
According to this picture that the quantum control process simulates the
reversible and unitary halting protocol the state-locking pulse field $%
P_{SL} $ is applied continuously to the quantum control system from the
beginning to the end of the quantum program$.$ If the quantum system of the
quantum computer which includes the quantum control system now is acted on
by a unitary operation such as one of the unitary operations $U_{b}^{c},$ $%
U_{h}^{c},$ $U_{f}^{c},$ and the trigger pulse $P_{t}$ of the quantum
program, then actually it is acted on by both the state-locking pulse field
and the unitary operation simultaneously. Then the state-locking pulse field
could be designed in such a way that the effect of the state-locking pulse
field on the quantum system can be negligible during the period of the
unitary operation applying to the quantum system. For example, the unitary
transformation on the quantum system is approximately equal to the single
unitary transformation of the functional operation $U_{f}^{c}$ when both the
functional operation $U_{f}^{c}$ and the state-locking pulse field $P_{SL}$
are applied to the quantum system simultaneously in the quantum program.
This is because in this case the state-locking pulse field has a negligible
effect on the quantum system. However, as shown in next section, there may
also be another case that the unitary transformation during the trigger
pulse $P_{t}$ could be really generated by both the state-locking pulse
field and the trigger pulse. Then in this case the contribution of the
state-locking pulse field is not negligible. These general properties of an
ideal state-locking pulse field could be used to measure the performance of
a real state-locking pulse field used in the reversible and unitary halting
protocol.

The quantum control process that simulates the reversible and unitary
halting protocol could be a single time-dependent unitary evolution process,
but generally it may be a time- and space-dependent unitary evolution
process of the quantum control system. However, if the quantum control
process is restricted to be dependent only upon a single time variable,
there could be a large drawback for such a quantum control process with a
two-state control state subspace $S(C)$. This can be explained in detail
below. Suppose that the state $|c_{1}\rangle $ of the control state subspace
is generated completely from the initial halting state $|n_{h}\rangle
=|0\rangle $ by the trigger pulse $P_{t}$ at the instant of time $t_{0i}$ in
the $i-$th cycle of the quantum program (see statement P7). The instant of
time $t_{0i}$ is special in that the state-locking pulse field really starts
to act on the control state subspace $S(C)$ at the instant of time $t_{0i}$
in the quantum program $Q_{c}$. Here suppose that the state-locking pulse
field has a negligible effect on the state $|c_{1}\rangle $ during the
trigger pulse $P_{t}$. Evidently, a different initial functional state $%
|f_{r}(x_{0})\rangle $ corresponds to a different time $t_{0i}$, and there
are $m_{r}$ possible and different times $\{t_{0i},$ $i=1,2,...,m_{r}\}$ at
most in the quantum program because there are $m_{r}$ different initial
functional states $\{|f_{r}(x_{0})\rangle ,$ $x_{0}=0,1,...,m_{r}-1\}$ of
the cyclic-group state subspace $S(m_{r}).$ Suppose that $\Delta T_{i}$ $%
(1\leq i\leq m_{r})$ is the time period of the $i-$th cycle of the quantum
program. Obviously, $t_{0i}=t_{0(i-1)}+\Delta T_{i}$ for $i=1,2,...,m_{r},$
where $t_{00}$ may be defined as $t_{00}=t_{01}-\Delta T_{1}$. For
convenience, here the time period $\Delta T_{i}$ is set to the same one $%
\Delta T_{c}$ for every cycle $i=1,2,...,m_{r}.$ Suppose that in the quantum
program $Q_{c}$ the duration of the trigger pulse $P_{t}$ is $\delta t_{r}$
and the duration is denoted as $\Delta t_{0}$ during which the state $%
|c_{1}\rangle $ is converted completely into the state $|c_{2}\rangle $ in
the control state subspace by the state-locking pulse field. In order to
show that a quantum control process that is restricted to be dependent only
upon a single time variable is not a good one to simulate efficiently the
reversible and unitary halting operation there are two possible situations
to be investigated in the quantum control process. Consider the first
situation. Note that in the $i-$th cycle of the quantum program the initial
halting state $|n_{h}\rangle =|0\rangle $ is changed completely to the state 
$|c_{1}\rangle $ by the trigger pulse $P_{t}$ in the time interval from the
time $t_{0i}-\delta t_{r}$ to the time $t_{0i}.$ This means that the
functional state $|f_{r}(x)\rangle $ is changed to the desired state $%
|f_{r}(x_{f})\rangle $ at the end of the $(i-1)-$th cycle, and during the
period from the time $t_{0i}$ to the time $t_{0i}+\Delta t_{0}$ the state $%
|c_{1}\rangle $ is converted into the state $|c_{2}\rangle $, and from the
time $t_{0i}+\Delta t_{0}$ on, the state $|c_{2}\rangle $ is kept unchanged
to the end of the quantum program by the state-locking pulse field $P_{SL}$.
Evidently, before the instant of time $t_{0(i+1)}-\delta t_{r}$ in the $%
(i+1)-$th cycle the state $|c_{1}\rangle $ must be completely converted into
the state $|c_{2}\rangle $ and since then the state $|c_{2}\rangle $ is kept
unchanged by the state-locking pulse field, otherwise it is possible that
the residual state $|c_{1}\rangle $ could be changed back to the initial
halting state $|n_{h}\rangle =|0\rangle $ by the trigger pulse $P_{t}$
during the period from the time $t_{0(i+1)}-\delta t_{r}$ to the time $%
t_{0(i+1)}$ in the $(i+1)-$th cycle. Therefore, for the first situation the
quantum control process requires that the quantum control system be
completely in the state $|c_{2}\rangle $ in the time interval between $%
t_{0(i+1)}-\delta t_{r}$ and $t_{0(i+1)}$ in the $(i+1)-$th cycle of the
quantum program. Now consider the second situation. There is also another
possibility that unlike the first situation the functional state $%
|f_{r}(x)\rangle $ is converted into the desired state $|f_{r}(x_{f})\rangle 
$ in the $i-$th cycle instead of the $(i-1)-$th cycle, because there are
different initial functional states $\{|f_{r}(x_{0})\rangle \}$ in the
quantum program$.$ Then during the period from the time $t_{0(i+1)}-\delta
t_{r}$ to the time $t_{0(i+1)}$ in the $(i+1)-$th cycle the initial halting
state $|n_{h}\rangle =|0\rangle $ is changed completely to the state $%
|c_{1}\rangle $ by the trigger pulse $P_{t}.$ This shows that for the second
situation the quantum control system is in the state $|c_{1}\rangle $ at the
time $t_{0(i+1)}$ in the $(i+1)-$th cycle. Then it can be seen from the two
possible situations that at the instant of time $t_{0(i+1)}$ in the $(i+1)-$%
th cycle of the quantum program the quantum control system is either
completely in the state $|c_{1}\rangle $ for the second situation or
completely in the state $|c_{2}\rangle $ for the first situation. Note that
these two states $|c_{1}\rangle $ and $|c_{2}\rangle $ of the control state
subspace are orthogonal to one another. Now one considers the $(i+2)-$th
cycle of the quantum program. The quantum program requires for any one of
the two possible situations that the quantum control system be completely in
the state $|c_{2}\rangle $ so as to avoid any real effect of the trigger
pulse on the quantum system during the period of the trigger pulse between
the time $t_{0(i+2)}-\delta t_{r}$ and the time $t_{0(i+2)}$ in the $(i+2)-$%
th cycle. Evidently, given a state-locking pulse field and a trigger pulse
as well as other unitary operations in the quantum program there is the same
time evolution propagator $U(t_{0(i+2)}-\delta t_{r},t_{0(i+1)})$ of the
quantum control system during the period from the time $t_{0(i+1)}$ in the $%
(i+1)-$th cycle to the time $t_{0(i+2)}-\delta t_{r}$ in the $(i+2)-$th
cycle no matter that the quantum control system is in the state $%
|c_{1}\rangle $ for the second situation or in the state $|c_{2}\rangle $
for the first situation at the instant of time $t_{0(i+1)}.$ Thus, there are
two possibilities to be considered. The first one is that the state of the
control state subspace is the state $|c_{2}\rangle $ at the instant of time $%
t_{0(i+1)}$ in the $(i+1)-$th cycle, which corresponds to the first
situation above. Since during the period between the time $t_{0(i+2)}-\delta
t_{r}$ and the time $t_{0(i+2)}$ in the $(i+2)-$th cycle the quantum control
system must be in the state $|c_{2}\rangle $ as required by the quantum
program and circuit, one has the unitary transformation for the state $%
|n_{h}\rangle |b_{h}\rangle |f_{r}(x)\rangle $ of the whole quantum system: 
\begin{equation*}
U(t_{0(i+2)}-\delta t_{r},t_{0(i+1)})|c_{2}\rangle |1\rangle |0\rangle
=|c_{2}\rangle |1\rangle |0\rangle
\end{equation*}%
where the state of the whole quantum system is $|c_{2}\rangle |1\rangle
|0\rangle $ at the time $t_{0(i+1)}$ in the $(i+1)-$th cycle and also $%
|c_{2}\rangle |1\rangle |0\rangle $ at the time $t_{0(i+2)}-\delta t_{r}$ in
the $(i+2)-$th cycle. The second one is that the state of the control state
subspace is the state $|c_{1}\rangle $ at the instant of time $t_{0(i+1)}$
in the $(i+1)-$th cycle, which corresponds to the second situation above.
Then for this situation the unitary state transformation is given by 
\begin{equation*}
U(t_{0(i+2)}-\delta t_{r},t_{0(i+1)})|c_{1}\rangle |1\rangle |0\rangle
=|c_{2}\rangle |1\rangle |0\rangle
\end{equation*}%
where the state of the whole quantum system is also $|c_{2}\rangle |1\rangle
|0\rangle $ at the time $t_{0(i+2)}-\delta t_{r}$, as required by the
quantum program and circuit. One sees that these two orthogonal states $%
|c_{1}\rangle $ and $|c_{2}\rangle $ are changed completely to the same
state $|c_{2}\rangle $ in the control state subspace by the same unitary
transformation $U(t_{0(i+2)}-\delta t_{0},t_{0(i+1)})$ during the period
from the time $t_{0(i+1)}$ to the time $t_{0(i+2)}-\delta t_{r}$. Obviously,
this is impossible and there is a conflict between the unitarity of the
propagator $U(t_{0(i+2)}-\delta t_{0},t_{0(i+1)})$ and the requirement of
the quantum program and circuit that the quantum control system be in the
state $|c_{2}\rangle $ at the time $t_{0(i+2)}-\delta t_{r}$ in the $(i+2)-$%
th cycle, because the requirement leads to non-unitarity of the propagator $%
U(t_{0(i+2)}-\delta t_{0},t_{0(i+1)})$. Therefore, the quantum control
process with the two-state control state subspace $S(C)$ could not simulate
faithfully and efficiently the reversible and unitary halting protocol and
could fail to control the quantum computational process in the quantum
program $Q_{c}$ if it is constrained to be a single time-dependent evolution
process. This conflict could be related to the square speedup limit of the
quantum search algorithm if the quantum program $Q_{c}$ is used to construct
the quantum search algorithm. Here it should be pointed out that this
conflict could be avoided in a larger control state subspace rather than the
two-state control state subspace, but this could lead to that the output
state of the quantum program is still dependent sensitively upon initial
states so that the quantum program $Q_{c}$ becomes unvalued for building up
an efficient quantum search process. Of course, it is usually better to use
a simpler control state subspace to control the quantum computational
process.

It is well known that a quantum system with a time-independent Hamiltonian
satisfies the time-displacement symmetry (or invariance) [5a]. The time
evolution process of such a quantum system is independent of any initial
time but depends upon the time difference between the end time and the
initial time of the process. Therefore, one possible scheme to make the
output state of the quantum program $Q_{c}$ independent of any initial state
could be that the Hamiltonian that governs the quantum control process is
restricted to be time-independent. As shown before, the times $\{t_{0i}\}$
could be thought of as the starting times for the state-locking pulse field
to really act on the control state subspace in the quantum control process.
Actually, the quantum control process starts to work after the trigger pulse
is applied at the instant of time $t_{0i}-\delta t_{r},$ and it may be
stated simply as that the initial halting state $|n_{h}\rangle =|0\rangle $
is changed completely to the state $|c_{1}\rangle $ by the trigger pulse $%
P_{t}$ in the time interval from the time $t_{0i}-\delta t_{r}$ to the time $%
t_{0i},$ then the state $|c_{1}\rangle $ is changed to the state $%
|c_{2}\rangle $ by the state-locking pulse field in the time interval $%
\Delta t_{0}$ from the time $t_{0i}$ to the time $t_{0i}+\Delta t_{0},$ and
from the time $t_{0i}+\Delta t_{0}$ on, the state $|c_{2}\rangle $ is locked
by the state-locking pulse field. According to this picture the quantum
control process could be expressed conveniently in terms of a sequence of
unitary transformations on the state $|n_{h}\rangle |b_{h}\rangle
|f_{r}(x)\rangle $ of the whole quantum system, 
\begin{equation}
\{P_{SL}(\{\varphi _{k}\},t_{0i},t_{0i}-\delta t_{r}),P_{t}\}|0\rangle
|1\rangle |0\rangle =|c_{1}\rangle |1\rangle |0\rangle ,\qquad \quad \quad
\quad  \label{1}
\end{equation}%
\begin{equation*}
P_{SL}(\{\varphi _{k}\},t,t_{0i})|c_{1}\rangle |1\rangle |0\rangle \quad 
\hspace{1in}\hspace{1in}\qquad
\end{equation*}%
\begin{equation}
=\{\varepsilon (t,t_{0i})|c_{1}\rangle +e^{-i\gamma (t,t_{0i})}\sqrt{%
1-|\varepsilon (t,t_{0i})|^{2}}|c_{2}\rangle \}|1\rangle |0\rangle ,\text{ }%
t_{0i}\leq t.  \label{2}
\end{equation}%
Here $P_{SL}(\{\varphi _{k}\},t,t_{0i})$ is the unitary propagator of the
state-locking pulse field $P_{SL}$ applying separately to the quantum system
during the period from the initial time $t_{0i}$ to the time $t$ and $%
\{P_{SL}(\{\varphi _{k}\},t_{0i},t_{0i}-\delta t_{r}),P_{t}\}$ represents
the unitary propagator when the trigger pulse $P_{t}$ and the state-locking
pulse field $P_{SL}$ are applied to the quantum system simultaneously during
the pulse duration $\delta t_{r}$ of the trigger pulse from the time $%
t_{0i}-\delta t_{r}$ to the time $t_{0i}.$ The parameters $\{\varphi _{k}\}$
in the unitary propagator $P_{SL}(\{\varphi _{k}\},t,t_{0})$ are the control
parameters of the state-locking pulse field which may be generally dependent
on the time variable, the spatial variables, or even the quantum field
variables. Here the unitary propagator $\{P_{SL}(\{\varphi
_{k}\},t_{0i},t_{0i}-\delta t_{r}),P_{t}\}$ is really equal to the unitary
transformation $U_{t}$ during the trigger pulse $P_{t}$, indicating that the
unitary transformation $U_{t}$ is really generated by both the state-locking
pulse field and the trigger pulse. As supposed before, the state-locking
pulse field is negligible during the trigger pulse in the quantum program $%
Q_{c}$, and hence the unitary transformation $U_{t}$ is really generated
approximately by the single trigger pulse. In the unitary transformation (2) 
$\gamma (t,t_{0i})$ is the phase angle of the amplitude of the state $%
|c_{2}\rangle $ and $\varepsilon (t,t_{0i})$ the residual amplitude of the
state $|c_{1}\rangle $ at the time $t$ after the unitary transformation $%
P_{SL}(\{\varphi _{k}\},t,t_{0i})$ acts on the state $|c_{1}\rangle $. As
required by the quantum program $Q_{c}$, the state $|c_{1}\rangle $ must be
converted completely into the state $|c_{2}\rangle $ in the control state
subspace by the unitary transformation $P_{SL}(\{\varphi _{k}\},t,t_{0i})$
within the period $\Delta t_{0}$ from the time $t_{0i}$ to the time $%
t=t_{0i}+\Delta t_{0}.$ Here the time interval $\Delta t_{0}$ is shorter
than the cyclic period $\Delta T_{c}$ minus the duration $\delta t_{r}$ of
the trigger pulse, that is, $\Delta t_{0}<\Delta T_{c}-\delta t_{r}.$
Evidently, when the time $t\geq t_{0i}+\Delta t_{0}$ the absolute amplitude
value $|\varepsilon (t,t_{0i})|$ should approach infinitely zero in theory
but can not equal exactly zero for every time $t_{0i}$ for $i=1,2,...,m_{r}$
even for an ideal state-locking pulse field, and generally it should be
close to zero for a real state-locking pulse field. One may say that the
amplitude $\varepsilon (t,t_{0i})$ equals zero in theory if it approaches
infinitely zero. The amplitude value $|\varepsilon (t,t_{0i})|$ measures the
performance of a real state-locking pulse field, that is, the smaller the
amplitude value $|\varepsilon (t,t_{0i})|$ is for $t\geq t_{0i}+\Delta
t_{0}, $ the better the performance is for a real state-locking pulse field $%
P_{SL}$. If now the quantum control process is governed by a
time-independent Hamiltonian during the state-locking pulse field, then the
time evolution process of Eq. (2) from the state $|c_{1}\rangle $ to the
state $|c_{2}\rangle $ does not depend directly upon any initial time $%
t_{0i} $ or the end time $t$ but it is dependent upon the time difference $%
\Delta t_{i}=t-t_{0i}$, that is, $P_{SL}(\{\varphi
_{k}\},t,t_{0i})=P_{SL}(\{\varphi _{k}\},\Delta t_{i})$. This also means
that both the amplitude $\varepsilon (t,t_{0i})$ of the state $|c_{1}\rangle 
$ and $e^{-i\gamma (t,t_{0i})}\sqrt{1-|\varepsilon (t,t_{0i})|^{2}}$ of the
state $|c_{2}\rangle $ in Eq. (2) are dependent upon the time difference $%
\Delta t_{i},$ that is, $\varepsilon (t,t_{0i})=\varepsilon (\Delta t_{i})$
and $\gamma (t,t_{0i})=\gamma (\Delta t_{i}).$ It was pointed out before
that there are $m_{r}$ different times $\{t_{0i},$ $i=1,2,...,m_{r}\}$ at
most for the quantum control process in the quantum program. If the
state-locking pulse field $P_{SL}$ is designed in such a way that the
Hamiltonian that governs the quantum control process of Eq. (2) is
time-independent, then the time difference $\Delta t_{i}=t-t_{0i}$ for $%
i=1,2,...,m_{r}$ replaces the time variable $t$ to become a new time
variable of the quantum control process of Eq. (2), and consequently it is
not necessary to deal with directly the time variable $t$ in the quantum
control process of Eq. (2). The unitary transformations (1) and (2) indicate
that by the quantum program $Q_{c}$ with the quantum control process of Eq.
(2)\ governed by a time-independent Hamiltonian these $m_{r}$ different
initial functional states $\{|f_{r}(x_{0})\rangle \}$ (as well as the
initial halting state and branch-control state) can be transferred
one-by-one to the $m_{r}$ states on the right-hand side of Eq. (2) whose
phases and amplitudes are dependent upon the time differences $\{\Delta
t_{i}\}$ for $i=1,2,...,m_{r}$, respectively. That the Hamiltonian that
governs the quantum control process of Eq. (2) under the state-locking pulse
field is constrained to be time-independent could be helpful for designing a
state-locking pulse field with a good performance to control the quantum
computational process in the quantum program $Q_{c}$.

However, the time-displacement symmetry is not sufficient to solve
thoroughly the conflict mentioned above and it can not yet figure out
completely a state-locking pulse field. This is because although the
amplitude $\varepsilon (\Delta t_{i})$ of the state $|c_{1}\rangle $ in Eq.
(2) is independent of any single initial time $t_{0i}$, it is still
dependent upon the time difference $\Delta t_{i}$. Obviously, a different
time difference $\Delta t_{i}$ generally leads to a different residual
amplitude $\varepsilon (\Delta t_{i})$ of the state $|c_{1}\rangle $ in Eq.
(2). Then there could be still a problem that the amplitude $|\varepsilon
(\Delta t_{i})|$ could fail to be close to zero during the trigger pulse $%
P_{t}$ from the time $t_{0k}-\delta t_{r}$ to time $t_{0k},$ here the time
difference $(t_{0k}-\delta t_{r}-t_{0i})>\Delta t_{0}$ for $%
k=i+1,i+2,...,m_{r}.$ As a result, the residual state $|c_{1}\rangle $ with
a large amplitude $|\varepsilon (\Delta t_{i})|$ may be changed back to the
initial halting state $|n_{h}\rangle =|0\rangle $ by the trigger pulse $%
P_{t} $ again. Obviously, this result is not consistent with the requirement
of quantum program that any residual amplitude $|\varepsilon (\Delta t_{i})|$
of the state $|c_{1}\rangle $ in Eq. (2) equal zero in theory\ when $\Delta
t_{i}\geq \Delta t_{0}$ for $i=1,2,...,m_{r}$. Now suppose that these two
states $|c_{1}\rangle $ and $|c_{2}\rangle $ of the control state subspace
are degenerative eigenstates of the time-independent Hamiltonian $H$ and the
unitary propagator $P_{SL}(\{\varphi _{k}\},t,t_{0i})=\exp (-iH\Delta t_{i})$
($\hslash =1$)$.$ Then these two eigenstates have the same eigenvalue and
the eigen-equations are written as $H|c_{1}\rangle =\lambda |c_{1}\rangle $
and $H|c_{2}\rangle =\lambda |c_{2}\rangle $ with the common eigenvalue $%
\lambda ,$ respectively. Thus, the time evolution process similar to Eq. (2)
with the initial superposition state $a|c_{1}\rangle +b|c_{2}\rangle $ of
the control state subspace $S(C),$ which is drived by the time-independent
Hamiltonian $H$, may be generally expressed by 
\begin{equation*}
P_{SL}(\{\varphi _{k}\},\Delta t_{i})(a|c_{1}\rangle +b|c_{2}\rangle
)|1\rangle |0\rangle =\exp (-i\lambda \Delta t_{i})(a|c_{1}\rangle
+b|c_{2}\rangle )|1\rangle |0\rangle .
\end{equation*}%
Evidently, only the global phase factor of the state $a|c_{1}\rangle
+b|c_{2}\rangle $ is changed by the unitary propagator $P_{SL}(\{\varphi
_{k}\},\Delta t_{i})$. The global phase factor is dependent upon the time
difference $\Delta t_{i},$ but the absolute amplitude of the state does not
change as the time difference. Therefore, the time-independent Hamiltonian
that drives the time evolution process of Eq. (2) in the two-state control
state subspace is very useful for keeping the amplitude of any state of the
control state subspace unchanged for a long time. However, such a
Hamiltonian could not be suitable for transferring the state $|c_{1}\rangle $
to the state $|c_{2}\rangle $ in the control state subspace, while the state
transfer is necessary for a quantum control process such as Eq. (2) to
simulate the reversible and unitary halting protocol.

It is shown above that it is not sufficient to build up the state-locking
pulse field with a good performance by constraining the Hamiltonian of the
quantum control system with the two-state subspace $S(C)$ under the
state-locking pulse field to be time-independent. A more suitable
Hamiltonian that governs the quantum control process of Eq. (2) may be
dependent upon the spatial variables and/or the quantum field variables but
independent of the time variable so that the propagator $P_{SL}(\{\varphi
_{k}\},\Delta t)$ is still dependent upon the time difference $\Delta t$. In
a quantum computer architecture different quantum bits of the quantum system
of the quantum computer must be addressed spatially or distinguished from
each other by some properties of the quantum system such as the
spectroscopic properties so that they can be manipulated at will. However,
such spatial-dependent properties of the quantum system are static and
different from a space-dependent evolution process. A quantum computational
process generally is considered as a unitary time evolution process of a
quantum system [3] which may be generally dependent upon both time and
space. According as quantum mechanics [5], time- and space-dependent
evolution processes of a quantum system such as the conventional quantum
scattering process, the quantum tunneling process, the quantum collision
process, the molecular chemical dissociation process, and so on, obey the
Schr\"{o}dinger equation as well and hence they are also governed by the
unitary quantum dynamics in time and space. The force to drive a time- and
space-dependent evolution process such as a quantum scattering process
usually could be the motional momentum of a particle or an electromagnetic
field and so on. The time- and space-dependent unitary evolution processes
could also be used to build up quantum computational processes just like the
conventional quantum gates [25], although space-dependent unitary evolution
processes usually are more complicated and difficult to be manipulated at
will than those space-independent ones. A large advantage of a time- and
space-dependent unitary evolution process over a space-independent one for
building up a quantum computational process could be that a time- and
space-dependent unitary evolution process may be manipulated separately
either in the time dimension or in the space dimensions or in both the time
and space dimensions. While a time-dependent and space-independent unitary
evolution process could be inadequate as the quantum control process of Eq.
(2)\ for the two-state control system, a time- and space-dependent unitary
evolution process could be better to act as the quantum control process.
Thus, a time- and space-dependent unitary evolution process could be very
useful for some specific purposes in quantum computation, although a quantum
computational process usually is simply designed to be a space-independent
unitary evolution process of a quantum system and any space-dependent
evolution processes of the quantum system are suppressed so that in
algorithm the quantum computational process may be constructed as simply as
possible. As shown before, if the quantum control process is purely
time-dependent in the two-state subspace $S(C),$ there is a conflict between
the unitarity of the quantum control process and the performance of the
quantum control process that the state $|c_{2}\rangle $ in the control state
subspace $S(C)$ is kept unchanged for a long time by the state-locking pulse
field. Then it could be better to use a time- and space-dependent unitary
evolution process such as a quantum scattering process to realize the
quantum control process of Eq. (2), meanwhile the quantum computational
process may be set to a single time-dependent unitary evolution process of
the quantum system of the quantum computer. This is because both the single
time-dependent quantum computational process and the time- and
space-dependent quantum control process may be manipulated separately in
time and space and hence they become almost independent upon each other. If
the Hamiltonian to drive the time- and space-dependent quantum control
process of Eq. (2) is space-dependent and time-independent, then the energy
of the quantum control system is conservative during the quantum control
process of Eq. (2) and hence both these two states $|c_{1}\rangle $ and $%
|c_{2}\rangle $ of the control state subspace have the same energy. Then the
quantum control process of Eq. (2) in the quantum control system is directly
dependent upon the time interval $\Delta t_{i}=t-t_{0i}$ rather than the
initial time $t_{0i}$ or the time variable $t$ separately. Since now these
two states $|c_{1}\rangle $ and $|c_{2}\rangle $ are degenerate in energy
only their global phase is dependent upon the time difference $\Delta t_{i}$
but their amplitudes are not during the quantum control process of Eq. (2),
and hence the state $|c_{2}\rangle $ may be kept unchanged for a long time
under the state-locking pulse field. However, here the state transfer from
the state $|c_{1}\rangle $ to the state $|c_{2}\rangle $ in the control
state subspace $S(C)$ could be achieved by the time- and space-dependent
unitary evolution process such as the quantum scattering process if the
time-independent and space-dependent Hamiltonian of the quantum control
system is chosen suitably. Obviously, the quantum scattering process should
be designed suitably according to the properties of an ideal state-locking
pulse field.\newline
\newline
\newline
{\large 3. An atomic physical model to simulate efficiently the reversible
and unitary halting protocol}

The detailed analysis in the former sections for the quantum control process
shows that a quantum control process that simulates faithfully and
efficiently the reversible and unitary halting protocol should contain a
time- and space-dependent unitary evolution process such as a quantum
scattering process and the control state subspace $S(C)$ should not be
restricted to be only the smallest two-state subspace. Generally, the
quantum control process to simulate efficiently the reversible and unitary
halting protocol in the quantum program $Q_{c}$ may be implemented in a real
quantum physical system. A trapped atomic-ion system has been proposed as a
real physical system to implement quantum computation [33]. Here a simple
quantum physical system of an atomic ion or a neutral atom in
one-dimensional double-well potential field is proposed as the quantum
control system of the quantum program and circuit $Q_{c}$ with the control
state subspace $S(C)$ which is larger than the two-state control state
subspace. Now the simple atomic physical system will be used to illustrate
how the quantum control process simulates really the reversible and unitary
halting protocol and how to construct explicitly the state-locking pulse
field. In this simple physical system the atomic ion or the neutral atom in
the double-well potential field is called the halting-quantum-bit atom or
the halting-qubit atom briefly. Hereafter the halting-qubit atom is referred
to as the atomic ion or the neutral atom in the double-well potential field
unless otherwise stated. In the double-well potential field the left-hand
potential well could be approximately a conventional harmonic potential
well, while the right-hand potential well could be simply a square potential
well. Here also suppose that the right-hand square potential well is
sufficiently wide such that the halting-qubit atom motions almost freely in
one-dimensional space in the square potential well. The intermediate part
between these two potential wells is a square potential barrier which is
used to block free transportation of the halting-qubit atom from one
potential well to another in the double-well potential field. Here the
maximum height of the right potential wall of the left-hand potential well
is just equal to the height of the intermediate square potential barrier.
The left and right potential walls of the double-well potential field may be
infinitely high in theory, but the intermediate square potential barrier is
finitely high and wide. Actually the left-hand potential well should be an
asymmetric harmonic potential well with an infinitely high left potential
wall and a finitely high right potential wall, respectively. If the
intermediate square potential barrier is infinitely high and finitely wide
then any two states of the halting-qubit atom in the left- and right-hand
potential wells respectively are completely orthogonal to one another, and
if the square potential barrier is high enough and finitely wide then the
two states are also considered to be orthogonal to one another
approximately. The time-independent double-well potential field in the
atomic physical system could be generated by the external electromagnetic
field [30]. More generally the double-well potential field could be thought
of as an effective potential field of the halting-qubit atom, that is, this
potential field could be generated effectively by the interaction between
the halting-qubit atom and the external electromagnetic field, the
interactions between the halting-qubit atom and those atoms of the
computational state subspace in the quantum system of the quantum computer,
and those interactions between the halting-qubit atom and its environment.
The halting-qubit atom in the double-well potential field could be coupled
to those quantum-bit atoms of the computational state subspace either
through the Coulomb interactions between charged atomic ions [31] or through
the atomic dipole-dipole interactions between atomic ions in an atomic-ion
physical system [32b], while the halting-qubit atom could also be coupled to
other atoms by the dipole-dipole interactions of neutral atoms in a neutral
atomic physical system [32a, 32c]. These interactions may be used to set up
two-qubit quantum gates between the halting-qubit atom and those quantum-bit
atoms of the computational state subspace in the quantum system. With these
two-qubit quantum gates and one-qubit quantum gates one can build up those
efficient unitary operations that act on both the halting qubit and those
qubits of the computational state subspace such as the unitary operation $%
U_{h}^{c}$ in the quantum program $Q_{c}$ and $V_{h}^{c}$ in the quantum
control unit $Q_{h}$ (see below). It is required by the quantum control
process that these interactions be available only when the halting-qubit
atom is in the left-hand harmonic potential well, while they are negligible
when the halting-qubit atom is in the right-hand potential well due to that
the halting-qubit atom in the right-hand potential well is much farther from
those atoms of the computational state subspace, and both the Coulomb
interaction and the dipole-dipole interaction may become very weak as the
distance between the interacting atoms become large [31, 32]. Therefore,
when the halting-qubit atom enters the right-hand potential well from the
left-hand one these interactions between the halting-qubit atom and those
atoms of the computational state subspace should be decoupled and can be
negligible so that the two-qubit quantum gates between the halting-qubit
atom in the right-hand potential well and those atoms of the computational
state subspace can not be built up effectively. More generally any quantum
operations involved in the halting-qubit atom in the left-hand potential
well could be really hung up when the halting-qubit atom leaves the
left-hand potential well. Therefore, the halting operation could be achieved
due to that these quantum operations are hung up when the halting-qubit atom
enters into the right-hand potential well from the left-hand one.

Generally, an atom has both the internal electronic states and the
center-of-mass motional states. Here a center-of-mass motional state of an
atom may be a wave-packet motional state, and for a heavy particle the
wave-packet picture in quantum mechanics is close to the classical particle
picture [5a]. The internal electronic states of an atom are generally
quantized bound states, but the center-of-mass motional states may be either
the quantized bound states in a potential field or the continuous states in
free space [5a, 5c]. A quantum bit of an atom generally may be chosen
suitably as a pair of the internal electronic ground states of the atom, but
sometime the quantized motional states of an atom in a harmonic potential
field are also taken as the quantum bits in quantum computation [30, 33].
Thus, the halting-qubit atom has the internal electronic states and also the
center-of-mass motional states in the double-well potential field. The
halting quantum bit generally may be chosen as a pair of the specific
internal electronic ground states of the halting-qubit atom. The
time-independent double-well potential field generally affects the
center-of-mass motional states of the halting-qubit atom [30], but it could
have a negligible effect on the internal electronic states of the
halting-qubit atom so that these internal electronic states could keep
unchanged when the halting-qubit atom moves from one potential well to
another [34]. Actually, the internal electronic states of the halting-qubit
atom are determined mainly by the internal interactions of the halting-qubit
atom itself, although the complete quantum structure of the halting-qubit
atom in the double-well potential field is determined by both the external
double-well potential field and the internal interactions of the
halting-qubit atom itself. As shown in the previous section, the
implementation of the reversible and unitary halting protocol in the atomic
physical system is involved in the time- and space-dependent unitary
evolution process that the halting-qubit atom moves from one potential well
to another in the double-well potential field. Then the center-of-mass
motional states and especially the wave-packet motional states of the
halting-qubit atom in the double-well potential field will play an important
role in implementing the reversible and unitary halting protocol in the
atomic physical system.

If now the atomic physical system of the halting-qubit atom in the
one-dimensional double-well potential field is considered to be a quantum
control system, then one must define explicitly the initial halting state $%
|n_{h}\rangle $ and the states $|c_{1}\rangle $ and $|c_{2}\rangle $ of the
control state subspace $S(C)$ in the quantum program $Q_{c}$. First of all,
the total wavefunction of the halting-qubit atom in the one-dimensional
double-well potential field may be generally written as 
\begin{equation*}
|n_{h}^{\prime },CM,R\rangle =|n_{h}^{\prime }\rangle |CM,R\rangle .
\end{equation*}%
Here the states $|n_{h}^{\prime }\rangle $ and $|CM,R\rangle $ represent the
internal electronic state and the center-of-mass motional state of the
halting-qubit atom, respectively. The integer $n_{h}^{\prime }$ and the
index $CM$ are the quantum numbers of the internal state $|n_{h}^{\prime
}\rangle $ and the motional state $|CM,R\rangle ,$ respectively, and $R$ is
the spatial coordinate of the center of mass of the halting-qubit atom with
the motional state $|CM,R\rangle $ in the double-well potential field. The
center-of-mass spatial position $R$ is generally time-dependent, i.e., $%
R=R(t)$. Actually, $CM$ is also used to represent the expectation value or
eigenvalue of the motional energy (or momentum) of the halting-qubit atom in
the double-well potential field particularly when the halting-qubit atom is
in a wave-packet motional state. Before the quantum program $Q_{c}$ is
performed the halting-qubit atom is located in the left-hand harmonic
potential well of the double-well potential field by the ionic or atomic
trapping techniques [30, 35]. For convenience the left-hand potential well
may be prepared temporarily as a conventional harmonic potential well before
the quantum program starts to work. This can be achieved easily by setting
the height of the right-hand wall of the left-hand potential well to be
sufficiently large, since the left-hand potential well can be thought of
approximately as a conventional harmonic potential well when the right-hand
potential wall is sufficiently high (note that the left-hand potential wall
is infinitely high in theory). Thus, before the quantum program starts the
internal and motional states of the halting-qubit atom may be really
prepared to be the ground internal state and the ground motional state of
the conventional harmonic oscillator by the laser cooling techniques [30,
36], respectively. Now the global ground state of the halting-qubit atom in
the left-hand harmonic potential well may be written as $|0,CM0,R_{0}\rangle
=|0\rangle |CM0,R_{0}\rangle ,$ which is the product state of the ground
internal state $|n_{h}^{\prime }\rangle =|0\rangle $ and the ground motional
state $|CM0,R_{0}\rangle $ of the atom in the harmonic potential well. Note
that the ground motional state $|CM0,R_{0}\rangle $ of the atom in the
conventional one-dimensional harmonic potential well is a Gaussian
wave-packet motional state [5a]. After the total ground state $%
|0,CM0,R_{0}\rangle $ is prepared the left-hand harmonic potential well is
suddenly changed back to the original double-well potential field at the
starting time of the quantum program. Actually, this process changes merely
the sufficiently high right-hand wall of the harmonic potential well to the
finitely high one of the left-hand potential well of the double-well
potential field. According as quantum mechanics [5a] that a wavefunction of
a quantum system must be continuous in time, the state $|0,CM0,R_{0}\rangle $
still remains unchanged at the starting time of the quantum program when the
harmonic potential field is suddenly changed back to the double-well
potential field. Then at the starting time of the quantum program the
motional state for the halting-qubit atom in the left-hand potential well of
the double-well potential field is just $|CM0,R_{0}\rangle $ and hence still
a one-dimensional Gaussian wave-packet motional state. The wave-packet state 
$|0,CM0,R_{0}\rangle $ could not be exactly the global ground state of the
halting-qubit atom in the double-well potential field. However, the energy
of the wave-packet state $|0,CM0,R_{0}\rangle $ may be very close to that
one of the global ground state of the halting-qubit atom in the double-well
potential field if the double-well potential field is designed suitably.
Therefore, the initial halting state $|n_{h}\rangle =|0\rangle $ in the
quantum program $Q_{c}$ may be set to the wave-packet state $%
|0,CM0,R_{0}\rangle $. Then the halting quantum bit may be chosen as the two
ground internal states $\{|n_{h}^{\prime }\rangle ,$ $n_{h}^{\prime }=0,1\}$
of the wave-packet states $\{|n_{h}^{\prime },CM0,R_{0}\rangle \}$. The
state $|c_{1}\rangle $ of the control state subspace in the quantum program $%
Q_{c}$ could be taken as the wave-packet state $|1,CM1(t_{0i}),R_{1}(t_{0i})%
\rangle $ with the internal state $|n_{h}^{\prime }\rangle =|1\rangle $ and
the wave-packet motional state $|CM1(t_{0i}),R_{1}(t_{0i})\rangle $ of the
halting-qubit atom in the left-hand potential well. Here the internal state $%
|n_{h}^{\prime }\rangle =|1\rangle $ could be chosen as another hyperfine
ground internal electronic state of the halting-qubit atom different from
the hyperfine ground internal electronic state $|n_{h}^{\prime }\rangle
=|0\rangle ,$ and the wave-packet spatial position $R_{1}(t_{0i})$ is within
the left-hand potential well. The wave-packet motional state $%
|CM1(t_{0i}),R_{1}(t_{0i})\rangle $ is generated from the wave-packet
motional state $|CM0,R_{0}\rangle $ by the trigger pulse $P_{t}$ (see
below). The mean motional energy ($CM1$) of the wave-packet motional state $%
|CM1(t_{0i}),R_{1}(t_{0i})\rangle $ is much higher than that one ($CM0$) of
the motional state $|CM0,R_{0}\rangle $ and also the height of the
intermediate potential barrier in the double-well potential field. Thus, the
wave-packet state $|c_{1}\rangle =|1,CM1(t_{0i}),R_{1}(t_{0i})\rangle $ is
an unstable state. When the halting-qubit atom in the left-hand potential
well is in the unstable wave-packet state $|c_{1}\rangle $, its motional
energy ($CM1$) is much higher than the height of the intermediate potential
barrier so that the halting-qubit atom is driven by the high motional
momentum ($CM1$) of the atom to pass the intermediate potential barrier to
enter into the right-hand potential well. This is a time- and
space-dependent quantum scattering process for the halting-qubit atom in the
double-well potential field. This quantum scattering process will be taken
as the quantum control process of Eq. (2). Note that the quantum scattering
process starts at the initial time $t_{0i}$. Now suppose that the
halting-qubit atom enters completely into the right-hand potential well at
the time $t_{mi}=t_{0i}+\Delta t_{0}$ in the quantum scattering process and
at the time $t_{mi}$ the wave-packet state of the halting-qubit atom in the
right-hand potential well is denoted as $|1,CM2(t_{mi}),R_{2}(t_{mi})\rangle 
$. Here the wave-packet spatial position $R_{2}(t_{mi})$ is within the
right-hand potential well. Since the quantum scattering process for the
halting-qubit atom from one potential well to another in the double-well
potential field does not change the ground internal states of the atom [34]
both the wave-packet states $|c_{1}\rangle
=|1,CM1(t_{0i}),R_{1}(t_{0i})\rangle $ and $|c_{2}^{\prime }\rangle
=|1,CM2(t_{mi}),R_{2}(t_{mi})\rangle $ have the same internal state with $%
n_{h}^{\prime }=1$. Now another state $|c_{2}\rangle $ of the control state
subspace in the quantum program $Q_{c}$ could be temporarily set to the
wave-packet state $|c_{2}^{\prime }\rangle $. Actually, the state $%
|c_{2}\rangle $ in the quantum program $Q_{c}$ will correspond to any
wave-packet state of the halting-qubit atom in the right-hand potential
well, as can be seen later, for the control state subspace of the atomic
physical system is not a two-state subspace. Due to the motional energy
conservation during the quantum scattering process the motional energy ($CM1$%
) of the wave-packet state $|1,CM1(t_{0i}),R_{1}(t_{0i})\rangle $ of the
halting-qubit atom in the left-hand potential well is really equal to that
one ($CM2$) of the wave-packet state $|1,CM2(t_{mi}),R_{2}(t_{mi})\rangle $
of the atom in the right-hand potential well. Thus, the wave-packet state $%
|1,CM2(t_{mi}),R_{2}(t_{mi})\rangle $ is also an unstable state just like
the wave-packet state $|1,CM1(t_{0i}),R_{1}(t_{0i})\rangle $. The spatial
spreads of these two wave-packet motional states $|CM1(t_{0i}),R_{1}(t_{0i})%
\rangle $ and $|CM2(t_{mi}),R_{2}(t_{mi})\rangle $ could be generally
different from each other, but these two motional states could be
approximately Gaussian in coordinate space [5a, 5c] just like the motional
state $|CM0,R_{0}\rangle $. Though these two wave-packet motional states
could not be exactly orthogonal to one another, they could be almost
completely orthogonal to one another if the wave-packet spatial distance $%
|R_{2}(t_{mi})-$ $R_{1}(t_{0i})|$ between their spatial positions $%
R_{1}(t_{0i})$ and $R_{2}(t_{mi})$ is large enough, because the overlapping
integral between these two wave-packet motional states decays exponentially
as the wave-packet spatial distance increases. Therefore, the width of the
intermediate potential barrier of the double-well potential field must be
large enough such that any wave-packet motional state of the halting-qubit
atom within one potential well is almost completely orthogonal to that one
within another potential well in the double-well potential field.

In the atomic physical system the time-independent double-well potential
field could be considered as one of the components of the state-locking
pulse field $P_{SL}$. More generally, a complete state-locking pulse field
in the atomic physical system could consist of three parts below. The first
part is the double-well potential field itself. The second is the sequences
of the time- and space-dependent electromagnetic pulse fields which are
applied only to the right-hand potential well. As can be seen below, these
sequences include mainly the unitary decelerating sequence and the unitary
accelerating sequence. The unitary decelerating sequence (the unitary
accelerating sequence) is used to decelerate (accelerate) the halting-qubit
atom in motion in the right-hand potential well. In this part the sequences
of the electromagnetic pulse fields are used to manipulate coherently the
wave-packet motional state of the halting-qubit atom in the right-hand
potential well so that the halting-qubit atom can stay in the right-hand
potential well for a long time required by the quantum computational process
in the quantum program. The third part is the sequences of the time- and
space-dependent electromagnetic pulse fields and includes also the
time-dependent potential fields which are applied mainly to the left-hand
potential well after the quantum program $Q_{c}$ terminates. In this part
the electromagnetic pulse fields associated with the unitary accelerating
sequence in the second part are used to drive the halting-qubit atom in the
right-hand potential well to go back to the left-hand potential well after
the quantum computational process finished, and at the same time transfer
each of possible wave-packet states of the halting-qubit atom in the
right-hand potential well to some given wave-packet state of the atom in the
left-hand potential well so that the final state of the halting-qubit atom
in the left-hand potential well is not dependent sensitively upon these
possible wave-packet states. Therefore, in both the second and third parts
these electromagnetic pulse fields should have a negligible effect on any
motional state of the halting-qubit atom if the atom locates within the
left-hand potential well during the quantum computational process. In this
and next sections it is discussed how the double-well potential field
affects the wave-packet state of the halting-qubit atom before the atom
leaves the left-hand potential well in the quantum computational process,
while in the section 5 it will be discussed how the sequences of the time-
and space-dependent electromagnetic pulse fields acts on the halting-qubit
atom to keep the atom in the right-hand potential well for a long time and
how by the electromagnetic pulse fields and the time-dependent potential
fields each of the possible wave-packet states of the halting-qubit atom in
the right-hand potential well is changed to some given wave-packet state of
the atom in the left-hand potential well after the quantum computational
process finished. The time-independent double-well potential field must be
designed properly. Before the halting-qubit atom leaves the left-hand
potential well during the quantum computational process the atom may have
its zero-point oscillatory motion in the left-hand harmonic potential well,
and it could also penetrate through the intermediate potential barrier and
enter into the right-hand potential well due to the quantum tunneling effect
[5a] even if the motional energy of the halting-qubit atom in the left-hand
potential well is lower than the height of the intermediate potential
barrier. The zero-point oscillatory motion is allowed normally in quantum
computation [30], but the quantum control process that simulates efficiently
the reversible and unitary halting protocol could become degraded if the
halting-qubit atom enters into the right-hand potential well in a
non-negligible probability due to the quantum tunneling effect before the
halting operation is performed according as the quantum program. Thus, the
double-well potential field should be designed in such a way that the
probability for the halting-qubit atom going from the left-hand potential
well to the right-hand one due to the quantum tunneling effect should be
minimized and can be neglected for the quantum computational process. The
height and width of the intermediate potential barrier in the double-well
potential field may control the quantum tunneling effect, that is, the
higher and wider the intermediate potential barrier, the lower the
penetrating probability for the halting-qubit atom. For example, consider a
simple physical model that a free particle with motional energy $E_{h}$ and
mass $m_{h}$ hits a square potential barrier with height $V_{0}$ and width $%
a $ [5a]$.$ The particle will be reflected and/or transmitted by the square
potential barrier. If the motional energy $E_{h}$ of the particle is much
less than the potential barrier height $V_{0}$ such that $\beta a>>1$ with $%
\beta \hslash =\sqrt{2m_{h}(V_{0}-E_{h})},$ then the transmission
coefficient of the particle is approximately proportional to the exponential
factor $\exp (-2\beta a)$ [5a]$.$ Therefore, the probability for the free
particle to penetrate through the potential barrier is also proportional to
the exponential factor $\exp (-2\beta a)$. When the potential barrier height 
$V_{0}$ and width $a$ are large enough, this probability falls off rapidly.
Now a bound particle with the same motional energy $E_{h}<<V_{0}$ like the
halting-qubit atom in the left-hand potential well is more difficult to
penetrate through the same square potential barrier than a free particle. An
atom has a much heavier mass $m_{h}$ than an electron. The probability for
the halting-qubit atom with the motional energy $E_{h}<<V_{0}$ in the
left-hand potential well to penetrate through the intermediate square
potential barrier with the height $V_{0}$ and width $a$ decays rapidly in an
exponential form as the height $V_{0},$ the width $a$, and the atomic mass $%
m_{h}$ increase. Evidently, if the intermediate potential barrier is high
and wide enough such that the ground-state motional energy ($CM0$) of the
halting-qubit atom is much lower than the barrier height, then the quantum
tunneling effect could have a negligible effect on the initial halting state 
$|n_{h}\rangle =|0,CM0,R_{0}\rangle $ and also the state $|n_{h}\rangle
=|1,CM0,R_{0}\rangle $ of the halting quantum bit, and hence the wave-packet
states $\{|n_{h}^{\prime },CM0,R_{0}\rangle \}$ with $n_{h}^{\prime }=0,1$
of the halting qubit atom in the left-hand potential well may keep almost
unchanged during the quantum computational process before the halting-qubit
atom leaves the left-hand potential well due to the halting operation in the
quantum program.

The coherent stimulated Raman adiabatic passage ($STIRAP$) method has been
used to prepare the $^{\prime }$Schr\"{o}dinger Cat$^{\prime }$
superposition state of a trapped atom [37]. Here the coherent $STIRAP$
method [37--41] may also be used to transfer selectively the halting state $%
|n_{h}\rangle =|1,CM0,R_{0}\rangle $ of the halting-qubit atom in the
left-hand potential well to the unstable state $|c_{1}\rangle
=|1,CM1(t_{0i}),$ $R_{1}(t_{0i})\rangle $ of the control state subspace.
Therefore, the state-dependent trigger pulse $P_{t}$ in the quantum program $%
Q_{c}$ could be chosen as the coherent Raman adiabatic laser pulse. The
spatial action zone of the coherent Raman adiabatic laser pulse $P_{t}$ is
confined within the left-hand potential well. A coherent Raman adiabatic
laser pulse $P_{t}$ consists of a pair of the coherent adiabatic laser
beams. Here denote these two adiabatic laser beams as $A$ and $B$,
respectively. When the coherent Raman adiabatic laser pulse $P_{t}$ is
applied to the halting-qubit atom in the left-hand potential well, one of
these two adiabatic laser beams (e.g., the beam $A$) first excites
selectively the transition of the halting-qubit atom from the wave-packet
state $|n_{h}^{\prime },CM0,R_{0}\rangle $ with the internal state $%
|n_{h}^{\prime }\rangle =|1\rangle $ to some specific excited state $%
|n_{e},CM_{e},R_{e}\rangle ,$ meanwhile the halting-qubit atom in the
excited state $|n_{e},CM_{e},R_{e}\rangle $ is stimulated by another
adiabatic laser beam $B$ to jump to the state $|c_{1}\rangle
=|1,CM1(t_{0i}),R_{1}(t_{0i})\rangle $. Here the excited internal state $%
|n_{e}\rangle $ $(n_{e}\neq 0,1)$ of the excited state $|n_{e},CM_{e},R_{e}%
\rangle $ has a higher internal energy than the ground internal states $%
\{|n_{h}^{\prime }\rangle ,$ $n_{h}^{\prime }=0,1\}$ and the halting-qubit
atom with the wave-packet motional state $|CM_{e},R_{e}\rangle $ is still
within the left-hand potential well. For example, if the halting-qubit atom
is chosen as a single $^{9}B_{e}^{+}$ ion, then these two internal states $%
\{|n_{h}^{\prime }\rangle ,$ $n_{h}^{\prime }=0,1\}$ could be taken as the
ionic hyperfine ground states $^{2}S_{1/2}$ $(F=2,m_{F}=-2)$ and $%
^{2}S_{1/2} $ $(F=1,m_{F}=-1),$ respectively, while the excited internal
state $|n_{e}\rangle $ could be the excited electronic state $^{2}P_{1/2}$ $%
(F=2,m_{F}=-2)$ of the ion [37a]. This state-dependent excitation process
under the coherent Raman adiabatic laser trigger pulse $P_{t}$ may be simply
expressed as 
\begin{equation*}
|1,CM0,R_{0}\rangle \overset{A}{\leftrightarrow }|n_{e},CM_{e},R_{e}\rangle 
\overset{B}{\leftrightarrow }|1,CM1(t_{0i}),R_{1}(t_{0i})\rangle .
\end{equation*}%
In order that only the two desired transitions: $|1,$ $CM0,R_{0}\rangle
\leftrightarrow |n_{e},CM_{e},R_{e}\rangle $ and $|n_{e},CM_{e},R_{e}\rangle
\leftrightarrow |1,CM1(t_{0i}),R_{1}(t_{0i})\rangle $ are excited
selectively by the coherent Raman adiabatic laser pulse $P_{t}$ the
frequencies of both the laser beams $A$ and $B$ must be close to the
resonance frequencies of these two desired transitions, respectively, and
they are also much far from resonance frequencies of any other transitions
including the transition: $|0,CM0,R_{0}\rangle \leftrightarrow
|n_{e},CM_{e},R_{e}\rangle $. Any wave-packet state $|n_{h}^{\prime
},CM,R\rangle $ of the halting-qubit atom with the internal state $%
|n_{h}^{\prime }\rangle \neq |1\rangle $ will not be affected effectively by
any one of these two adiabatic laser beams. Thus, the coherent Raman
adiabatic laser pulse $P_{t}$ does not act on any wave-packet state with the
internal state $|n_{h}^{\prime }\rangle \neq |1\rangle $ such as the initial
halting state $|0,CM0,R_{0}\rangle $. On the other hand, in order to
suppress irreversible spontaneous emission processes both the laser beams $A$
and $B$ are detuned properly from the excitation state $|n_{e},CM_{e},R_{e}%
\rangle $. It might be better that the wave vector difference of these two
laser beams $A$ and $B$ is set to point to the left-hand potential wall of
the double-well potential field as the left-hand potential wall may be
sufficiently high in practice. This means that the halting-qubit atom moves
along $-x$ axis toward the left-hand potential wall under the action of the
coherent Raman adiabatic laser pulse $P_{t}$. Obviously, in the
state-dependent excitation process the halting state $|1,CM0,R_{0}\rangle $
is excited to the higher motional energy state $|c_{1}\rangle
=|1,CM1(t_{0i}),R_{1}(t_{0i})\rangle $ by the coherent Raman adiabatic laser
pulse $P_{t}$. In particular, the mean motional energy of the motional state 
$|CM1(t_{0i}),R_{1}(t_{0i})\rangle $ of the halting-qubit atom must be much
higher than the height of the intermediate potential barrier so that the
unitary quantum scattering process\ can take place automatically for the
halting-qubit atom from the left-hand potential well to the right-hand one.

The quantum control process in the quantum program $Q_{c}$ usually should be
modified properly when it is implemented in a real quantum physical system
such as the simple atomic physical system mentioned above. For the quantum
control system of the atomic physical system the initial halting state $%
|n_{h}\rangle $ and these two states $|c_{1}\rangle $ and $|c_{2}\rangle $
of the control state subspace $S(C)$ in the quantum program $Q_{c}$ are
defined as the wave-packet states of the halting-qubit atom in the left-hand
potential well: $|n_{h}\rangle =|0,CM0,R_{0}\rangle ,$ $|c_{1}\rangle
=|1,CM1(t_{0i}),R_{1}(t_{0i})\rangle ,$ and $|c_{2}\rangle =|c_{2}^{\prime
}\rangle =|1,CM2(t_{mi}),R_{2}(t_{mi})\rangle ,$ respectively. The
state-dependent trigger pulse $P_{t}$ is taken as a suitable coherent Raman
adiabatic laser pulse that is applied to within the left-hand potential well
in space. The state-locking pulse field in the atomic physical system
consists of the double-well potential field and the sequences of the time-
and space-dependent electromagnetic pulse fields and the time-dependent
potential fields, as mentioned in the previous paragraphs. Now a quantum
control process (or unit) $Q_{h}$ of the atomic physical system which
replaces the statement P7 of the quantum program $Q_{c}$ is designed to
simulate efficiently the reversible and unitary halting protocol. The
quantum control process (or unit) $Q_{h}$ is written as 
\begin{equation*}
\text{While }|f_{r}(x)\rangle =|f_{r}(x_{f})\rangle ,\text{ \qquad \qquad
\qquad \qquad \qquad \qquad \qquad \quad }
\end{equation*}%
\begin{equation*}
\quad \text{Do }U_{h}^{c}:|n_{h}\rangle |f_{r}(x_{f})\rangle
=|0,CM0,R_{0}\rangle |1\rangle \rightarrow |0,CM0,R_{0}\rangle |0\rangle ,
\end{equation*}%
\begin{equation*}
\quad \quad V_{h}^{c}:|n_{h}\rangle |f_{r}(x)\rangle =|0,CM0,R_{0}\rangle
|0\rangle \rightarrow |1,CM0,R_{0}\rangle |0\rangle
\end{equation*}%
\begin{equation*}
\text{State-dependent excitation process }(P_{t}):\qquad \ \quad \quad
\qquad \qquad
\end{equation*}%
\begin{equation*}
|n_{h}\rangle =|1,CM0,R_{0}\rangle \rightarrow |c_{1}\rangle
=|1,CM1(t_{0i}),R_{1}(t_{0i})\rangle
\end{equation*}%
\begin{equation*}
\text{Quantum scattering process in time and space }(P_{SL}):\qquad \quad
\end{equation*}%
\begin{equation*}
\qquad \qquad |c_{1}\rangle =|1,CM1(t_{0i}),R_{1}(t_{0i})\rangle \rightarrow
|c_{2}^{\prime }\rangle =|1,CM2(t_{mi}),R_{2}(t_{mi})\rangle
\end{equation*}%
\newline
Here the desired functional state is $|f_{r}(x_{f})\rangle =|1\rangle $ if
the quantum program $Q_{c}$ works in the multiplicative-cyclic-group state
subspace $S(m_{r})$ and $|f_{r}(x_{f})\rangle =|0\rangle $ if the quantum
program $Q_{c}$ works in the additive-cyclic-group state subspace $%
S(Z_{m_{r}})$ and in later case the unitary operation $U_{h}^{c}$ may be
omitted from the quantum control unit $Q_{h}$. In the quantum control unit $%
Q_{h}$ the unitary operation $U_{h}^{c}$ is the same as the original one in
the quantum program $Q_{c}$: 
\begin{equation*}
U_{h}^{c}:|0,CM0,R_{0}\rangle |b_{h}\rangle |1\rangle \leftrightarrow
|0,CM0,R_{0}\rangle |b_{h}\rangle |0\rangle ,
\end{equation*}%
and the new conditional unitary operation $V_{h}^{c}$ is defined as 
\begin{equation*}
V_{h}^{c}:|0,CM0,R_{0}\rangle |b_{h}\rangle |0\rangle \leftrightarrow
|1,CM0,R_{0}\rangle |b_{h}\rangle |0\rangle
\end{equation*}%
with $b_{h}=0,1$, and the conditional unitary operation $U_{tr}^{c}$ during
the state-dependent trigger pulse $P_{t}$ is simply defined by 
\begin{equation*}
U_{tr}^{c}:|1,CM0,R_{0}\rangle |b_{h}\rangle |f_{r}(x)\rangle
\leftrightarrow |1,CM1(t_{0i}),R_{1}(t_{0i})\rangle |b_{h}\rangle
|f_{r}(x)\rangle .
\end{equation*}%
Here the unitary operation $U_{tr}^{c}$ corresponds to the original one $%
U_{t}$ in the quantum program $Q_{c}$. The conditional unitary operations $%
U_{h}^{c}$ and $V_{h}^{c}$ generally are dependent upon both the functional
states and the internal states of the halting-qubit atom but may be
independent of any motional states of the atom. They may be built up
efficiently out of the interactions between the halting-qubit atom in the
left-hand potential well and those atoms of the computational state
subspace. It follows from the quantum control unit $Q_{h}$ that only when
the functional state $|f_{r}(x)\rangle $ becomes the desired state $%
|f_{r}(x_{f})\rangle $ can the unitary operations $U_{h}^{c}$ and $V_{h}^{c}$
take a real action on the quantum system of the quantum computer. One of the
important processes in the quantum control unit $Q_{h}$ is the
state-dependent excitation process involved in the trigger pulse $P_{t}$
with the pulse duration $\delta t_{r}$. Though the coherent Raman adiabatic
laser trigger pulse $P_{t}$ is applied to the halting-qubit atom in the
left-hand potential well at the starting time $t_{0i}-\delta t_{r}$ for
every cycle of the quantum program $Q_{c}$ with the cyclic index $%
i=1,2,...,m_{r}$, it can only take a real action on those wave-packet states
of the atom with the internal state $|n_{h}^{\prime }\rangle =|1\rangle $
such as the halting state $|1,CM0,R_{0}\rangle $. Therefore, only when the
initial halting state $|n_{h}\rangle =|0,CM0,R_{0}\rangle $ is changed to
the halting state $|n_{h}\rangle =|1,CM0,R_{0}\rangle $ by the conditional
unitary operation $V_{h}^{c}$ in the quantum program can the trigger pulse $%
P_{t}$ excite the state $|n_{h}\rangle =|1,CM0,R_{0}\rangle $ with the
ground motional energy ($CM0$) to the unstable wave-packet state $%
|c_{1}\rangle =|1,CM1(t_{0i}),R_{1}(t_{0i})\rangle $ with a much higher
motional energy ($CM1$) and a different spatial position $R_{1}(t_{0i})\neq
R_{0}$ in the left-hand potential well. Since the mean motional energy ($CM1$%
) of the wave-packet state $|1,CM1(t_{0i}),R_{1}(t_{0i})\rangle $ is much
higher than the height of the intermediate potential barrier the
halting-qubit atom in the left-hand potential well with the wave-packet
state $|1,CM1(t_{0i}),R_{1}(t_{0i})\rangle $ can easily pass the
intermediate potential barrier to enter into the right-hand potential well.
This process is a quantum scattering process in space and time: $%
|1,CM1(t_{0i}),R_{1}(t_{0i})\rangle \rightarrow
|1,CM2(t_{mi}),R_{2}(t_{mi})\rangle ,$ that is, a time- and space-dependent
unitary evolution process which is driven by the motional momentum of the
halting-qubit atom in the unstable state $|1,CM1(t_{0i}),R_{1}(t_{0i})%
\rangle $. This is a key process for the quantum control process $Q_{h}$ to
achieve the reversible and unitary halting operation. The halting operation
will take place when the halting-qubit atom leaves the left-hand potential
well because the coherent Raman adiabatic laser trigger pulse $P_{t}$ does
not have a real action on any wave-packet state of the halting-qubit atom if
the atom is not in the left-hand potential well. On the other hand, if the
halting-qubit atom enters into the right-hand potential well, then any
two-qubit quantum gates become unavailable between the halting-qubit atom
and those atoms of the computational state subspace due to that the
effective interactions vanish between the halting-qubit atom and those
atoms. Then any one of the unitary operations $U_{h}^{c}$ and $V_{h}^{c}$ in
the quantum control unit $Q_{h}$ becomes yet unavailable and consequently
the halting operation is achieved too.\newline
\newline
\newline
{\large 4. The unitary evolution processes for the quantum program and
circuit\ in the atomic physical system}

Now the original quantum control unit (the statement P7) of the quantum
program $Q_{c}$ is replaced with the quantum control unit $Q_{h}$ of the
atomic physical system. The time evolution process of the atomic physical
system of the quantum computer under the quantum program and circuit $Q_{c}$
is investigated in detail below. This investigation will be helpful for
understanding more clearly and deeply the general properties of a
state-locking pulse field and especially the properties of the unitary
transformations related to the state-locking pulse field. First of all, the
quantum program and circuit $Q_{c}$ with the quantum control unit $Q_{h}$
may be written in the simple form 
\begin{equation*}
Q_{c}=\{P_{SL}:OFF\}\{U_{f}^{c}P_{SL}(\{\varphi _{i}\},\Delta
t_{0})U_{tr}^{c}V_{h}^{c}U_{h}^{c}U_{b}^{c}\}^{m_{r}}\{P_{SL}:ON\}.
\end{equation*}%
Here the initial state of the quantum circuit $Q_{c}$ is set to the basis
state $|n_{h}\rangle |b_{h}\rangle |f_{r}(x)\rangle =|0,CM0,R_{0}\rangle
|0\rangle |f_{r}(x_{0})\rangle .$ The state-locking pulse field $P_{SL}$ is
applied continuously to the quantum system from the beginning to the end of
the quantum circuit. During the periods of these unitary operations $%
U_{f}^{c},$ $U_{tr}^{c},$ $U_{h}^{c},$ $V_{h}^{c},$ and $U_{b}^{c}$ the
quantum system is really acted upon simultaneously by both the state-locking
pulse field and these unitary operations. The unitary transformation $%
P_{SL}(\{\varphi _{i}\},\Delta t_{0})$ represents the quantum scattering
process during the period from the time $t_{0i}$ to the time $%
t_{mi}=t_{0i}+\Delta t_{0}.$ Obviously, if a unitary operation is applied
only to the computational state subspace it commutes with the unitary
propagator $P_{SL}(\{\varphi _{k}\},t,t_{0})$ of the state-locking pulse
field as the state-locking pulse field is applied only to the halting-qubit
atom in the double-well potential field. Now the unitary functional
operation $U_{f}^{c}$ and the unitary operation $U_{b}^{c}$ are applied only
to the computational state subspace in the quantum circuit. Thus, the
unitary propagator $P_{SL}(\{\varphi _{k}\},t,t_{0})$ always commutes with
these conditional unitary operations$:$ 
\begin{equation}
P_{SL}(\{\varphi _{k}\},t,t_{0})U_{b}^{c}\equiv U_{b}^{c}P_{SL}(\{\varphi
_{k}\},t,t_{0}),  \label{3}
\end{equation}%
\begin{equation}
P_{SL}(\{\varphi _{k}\},t,t_{0})U_{f}^{c}\equiv U_{f}^{c}P_{SL}(\{\varphi
_{k}\},t,t_{0}).  \label{4}
\end{equation}%
Though the conditional unitary operations $U_{h}^{c}$ and $V_{h}^{c}$ may be
independent of any atomic motional states according as their definitions,
these two unitary operations may require that the wave-packet motional state 
$|CM0,R_{0}\rangle $ of the halting states \{$|n_{h}^{\prime
},CM0,R_{0}\rangle ,n_{h}^{\prime }=0,1$\} of the halting-qubit atom in the
left-hand potential well keep unchanged up to a global phase factor when
these two unitary operations are applied to the quantum system in the period
from the initial time $t_{0}$ of the quantum circuit to the time $%
t_{0i}-\delta t_{r}$ before the state $|1,CM0,R_{0}\rangle $ is changed to
the state $|c_{1}\rangle =|1,CM1(t_{0i}),R_{1}(t_{0i})\rangle $ by the
state-dependent trigger pulse $P_{t}.$ Here the quantum circuit $Q_{c}$
starts to run the first cycle at the initial time $t_{0}$. The requirement
may be necessary when both the unitary operations are built up out of the
dipole-dipole interactions or the Coulomb interactions which are dependent
on the interdistances of atoms. Since the double-well potential field is
considered as one of the components of the state-locking pulse field $P_{SL}$
in the atomic physical system, any motional state of the halting-qubit atom
in the double-well potential field is affected inevitably by the
state-locking pulse field. However, as pointed out before, the intermediate
potential barrier is so high and wide that up to a global phase factor the
wave-packet motional state $|CM0,R_{0}\rangle $ of the halting states $%
\{|n_{h}^{\prime },CM0,R_{0}\rangle \}$ is almost unchanged in the period
from the initial time $t_{0}$ to the time $t_{0i}-\delta t_{r}$ before the
state $|1,CM0,R_{0}\rangle $ is changed to the unstable state $|c_{1}\rangle
.$ This property of the state-locking pulse field could be simply expressed
by the unitary transformation: 
\begin{equation*}
P_{SL}(\{\varphi _{k}\},t,t_{0})|n_{h}^{\prime },CM0,R_{0}\rangle \qquad
\qquad \qquad \qquad \qquad \qquad
\end{equation*}%
\begin{equation}
=\exp [-iE_{0}(t-t_{0})/\hslash ]|n_{h}^{\prime },CM0,R_{0}\rangle ,\text{ }%
t_{0}\leq t\leq t_{0i}-\delta t_{r}.  \label{5}
\end{equation}%
where $n_{h}^{\prime }=0,$ $1$ and $E_{0}$ is the motional energy of the
ground motional state $|CM0,R_{0}\rangle .$ Hereafter for convenience the
global phase factor such as $\exp [-iE_{0}(t-t_{0})/\hslash ]$ in Eq. (5) is
omitted without confusion. Therefore, both the unitary operations $U_{h}^{c}$
and $V_{h}^{c}$ commute approximately with the unitary propagator $%
P_{SL}(\{\varphi _{k}\},t,t_{0})$ ($t_{0}\leq t\leq t_{0i}-\delta t_{r}$) of
the state-locking pulse field, 
\begin{equation}
P_{SL}(\{\varphi _{k}\},t,t_{0})U_{h}^{c}=U_{h}^{c}P_{SL}(\{\varphi
_{k}\},t,t_{0}),  \label{6}
\end{equation}%
\begin{equation}
P_{SL}(\{\varphi _{k}\},t,t_{0})V_{h}^{c}=V_{h}^{c}P_{SL}(\{\varphi
_{k}\},t,t_{0}).  \label{7}
\end{equation}%
Moreover, the commutation relations (6) and (7)\ still hold when the
halting-qubit atom enters into the right-hand potential well from the
left-hand one, because in this case these two unitary operations $U_{h}^{c}$
and $V_{h}^{c}$ have not any real effect on the halting-qubit atom and are
reduced theoretically to the unity operation. Therefore, the commutation
relations (6) and (7)\ hold for the whole quantum circuit. The commutation
relations (5), (6), and (7) lead to that there hold the state unitary
transformations, 
\begin{equation*}
\{P_{SL}(\{\varphi _{k}\},t,t_{0}),U_{h}^{c}\}|n_{h}^{\prime
},CM0,R_{0}\rangle |b_{h}\rangle |f_{r}(x)\rangle
\end{equation*}%
\begin{equation}
\hspace{1in}=U_{h}^{c}|n_{h}^{\prime },CM0,R_{0}\rangle |b_{h}\rangle
|f_{r}(x)\rangle ,\text{ }t_{0}\leq t\leq t_{0i}-\delta t_{r},  \label{8}
\end{equation}%
\begin{equation*}
\{P_{SL}(\{\varphi _{k}\},t,t_{0}),V_{h}^{c}\}|n_{h}^{\prime
},CM0,R_{0}\rangle |b_{h}\rangle |f_{r}(x)\rangle
\end{equation*}%
\begin{equation}
\hspace{1in}=V_{h}^{c}|n_{h}^{\prime },CM0,R_{0}\rangle |b_{h}\rangle
|f_{r}(x)\rangle ,\text{ }t_{0}\leq t\leq t_{0i}-\delta t_{r}.  \label{9}
\end{equation}%
Evidently, the unitary propagator $P_{SL}(\{\varphi _{k}\},t,t_{0})$
generally could not commute with the unitary operation $U_{tr}^{c}$ of the
trigger pulse $P_{t}$ in the atomic physical system, 
\begin{equation*}
P_{SL}(\{\varphi _{k}\},t,t_{0})U_{tr}^{c}\neq U_{tr}^{c}P_{SL}(\{\varphi
_{k}\},t,t_{0}).
\end{equation*}%
This is because in the atomic physical system the state-dependent excitation
process of the trigger pulse $P_{t}$ from the state $|1,CM0,R_{0}\rangle $
to the unstable state $|1,CM1(t_{0i}),R_{1}(t_{0i})\rangle $ is affected
inevitably by the left-hand harmonic potential field. Then the coherent
Raman adiabatic laser trigger pulse $P_{t}$ must be designed suitably such
that when both the trigger pulse $P_{t}$ and the state-locking pulse field $%
P_{SL}$ are applied simultaneously to the halting-qubit atom in the
left-hand potential well the unitary propagator $\{P_{SL}(\{\varphi
_{k}\},t_{0i},t_{0i}-\delta t_{r}),P_{t}\}$ satisfies the relation: 
\begin{equation*}
\{P_{SL}(\{\varphi _{k}\},t_{0i},t_{0i}-\delta
t_{r}),P_{t}\}|1,CM0,R_{0}\rangle |b_{h}\rangle |f_{r}(x)\rangle \qquad
\qquad \qquad \quad
\end{equation*}%
\begin{equation}
\equiv U_{tr}^{c}|1,CM0,R_{0}\rangle |b_{h}\rangle |f_{r}(x)\rangle
=|1,CM1(t_{0i}),R_{1}(t_{0i})\rangle |b_{h}\rangle |f_{r}(x)\rangle .
\label{10}
\end{equation}%
The approximate calculation in theory for the unitary propagator $%
U_{tr}^{c}\equiv \{P_{SL}(\{\varphi _{k}\},t_{0i},$ $t_{0i}-\delta
t_{r}),P_{t}\}$ could be carried out on the simple physical model of the
time-dependent forced harmonic oscillator [5c, 37, 50g, 50h, 50i, 51b].

Suppose again that the six unitary operations $U_{f}^{c},$ $P_{SL}(\{\varphi
_{k}\},\Delta t_{0}),$ $U_{tr}^{c},$ $V_{h}^{c},$ $U_{h}^{c},$ and $U_{b}^{c}
$ in the quantum circuit $Q_{c}$ have the durations $\delta t_{f},$ $\Delta
t_{0},$ $\delta t_{r},$ $\delta t_{h}^{\prime },$ $\delta t_{h},$ and $%
\delta t_{b},$ respectively, and the period of each cycle of the quantum
circuit is $\Delta T=\delta t_{b}+\delta t_{h}+\delta t_{h}^{\prime }+\delta
t_{r}+\Delta t_{0}+\delta t_{f}.$ As stated before, the unstable state $%
|c_{1}\rangle =|1,CM1(t_{0i}),R_{1}(t_{0i})\rangle $ of the control state
subspace is generated completely at the instant of time $t_{0i}$ in the $i-$%
th cycle of the quantum circuit$.$ Then it follows that the functional state 
$|f_{r}(x)\rangle $ is converted into the desired functional state $%
|f_{r}(x_{f})\rangle $ during the period $\delta t_{f}$ from the time $%
(t_{0i}-\delta t_{r}-\delta t_{h}^{\prime }-\delta t_{h}-\delta
t_{b})-\delta t_{f}$ to the time $(t_{0i}-\delta t_{r}-\delta t_{h}^{\prime
}-\delta t_{h}-\delta t_{b})$ in the $(i-1)-$th cycle. Note that the initial
state of the quantum circuit is the wave-packet state $|0,CM0,R_{0}\rangle
|0\rangle |f_{r}(x_{0})\rangle $ and at the initial time $t_{0}$ the
halting-qubit atom is in the state $|0,CM0,R_{0}\rangle $ and in the
left-hand potential well. Then the desired functional state $%
|f_{r}(x_{f})\rangle $ takes the integer $x_{f}=(x_{0}+i-1)\func{mod}m_{r}$.
Before the $i-$th cycle the initial halting state $|n_{h}\rangle
=|0,CM0,R_{0}\rangle $ and the initial branch-control state $|b_{h}\rangle
=|0\rangle $ keep unchanged but only the initial functional state $%
|f_{r}(x_{0})\rangle $ is consecutively changed to other functional state $%
|f_{r}(x_{0}+j)\rangle $ for $j=0,1,2,...,i-1$ in the quantum circuit$,$
where the last functional state $|f_{r}(x_{0}+i-1)\rangle $ is the desired
functional state $|f_{r}(x_{f})\rangle $. Obviously, before the $i-$th cycle
the time evolution process of the whole quantum system of the quantum
computer with the initial state $|0,CM0,R_{0}\rangle |0\rangle
|f_{r}(x_{0})\rangle $ can be expressed as 
\begin{equation*}
\{P_{SL}(\{\varphi _{k}\},t_{j},t_{0}),(U_{T})^{j}\}|0,CM0,R_{0}\rangle
|0\rangle |f_{r}(x_{0})\rangle \qquad \quad \quad 
\end{equation*}%
\begin{equation*}
=(U_{T})^{j}|0,CM0,R_{0}\rangle |0\rangle |f_{r}(x_{0})\rangle \ \ \qquad
\qquad \hspace{1in}\qquad \qquad 
\end{equation*}%
\begin{equation}
=|0,CM0,R_{0}\rangle |0\rangle |f_{r}(x_{0}+j)\rangle ,\text{ }%
t_{j}=t_{0}+j\Delta T,\text{ }0\leq j\leq i-1,  \label{11}
\end{equation}%
where the unitary operator $U_{T}$ is denoted as the unitary operation
sequence $U_{f}^{c}P_{SL}(\{\varphi _{k}\},\Delta
t_{0})U_{tr}^{c}V_{h}^{c}U_{h}^{c}U_{b}^{c}$ of the quantum circuit $Q_{c}.$
In the unitary transformation (11) the first equation shows that up to a
global phase factor the unitary propagator of the quantum system under both
the state-locking pulse field $P_{SL}$\ and the unitary operation sequence $%
(U_{T})^{j}$ for $0\leq j\leq i-1$ acting on the initial state is really
equal to the single unitary operation sequence $(U_{T})^{j}$ acting on the
same initial state before the $i-$th cycle of the quantum circuit. Here the
integer $j$ is also used as the cyclic index of the quantum circuit.

For example, there are some cases to be considered for the unitary
transformation (11). For the first case $(i)$ the initial functional state $%
|f_{r}(x_{0})\rangle $ happens to be just the desired functional state $%
|f_{r}(x_{f})\rangle .$ In this case the index $i=1$. Then $x_{f}=x_{0}$ and
the cyclic index takes $j=0$ in the unitary transformation (11). Therefore, $%
t_{j}=t_{0}$ for $j=0.$ Here $t_{01}=t_{0}+\delta t_{b}+\delta t_{h}+\delta
t_{h}^{\prime }+\delta t_{r}.$ Both the unitary propagator $P_{SL}(\{\varphi
_{k}\},t_{j},t_{0})$ and the unitary operation sequence $(U_{T})^{j}$ become
the unity operator in effect as the cyclic index $j=0$ and the unitary
transformation (11) is a state identity. For the second case $(ii)$ the
initial functional state $|f_{r}(x_{0})\rangle $ could be changed to the
desired functional state $|f_{r}(x_{f})\rangle $ at the end of the first
cycle of the quantum circuit. In this case the index $i=2$. Then $%
x_{f}=(x_{0}+1)\func{mod}m_{r}$ and the cyclic index takes $j=0$ and $1$ in
the unitary transformation (11). Note that the end time of the first cycle
is just equal to the beginning time of the second cycle in the quantum
circuit. Thus, $t_{0}=t_{0},$ $t_{1}=t_{0}+\Delta T=t_{01}+\Delta
t_{0}+\delta t_{f}=t_{02}-\delta t_{r}-\delta t_{h}^{\prime }-\delta
t_{h}-\delta t_{b}.$ Here $t_{01}=t_{0}+\delta t_{b}+\delta t_{h}+\delta
t_{h}^{\prime }+\delta t_{r}$ and $t_{02}=t_{01}+\Delta T.$ Generally, for
the third case $(iii)$ the initial functional state $|f_{r}(x_{0})\rangle $
could be changed to the desired functional state $|f_{r}(x_{f})\rangle $ at
the end of the $(i-1)-$th cycle of the quantum circuit. In this case the
cyclic index $i>1$. Then $x_{f}=(x_{0}+i-1)\func{mod}m_{r}$ and the cyclic
index takes $j=0,1,...,(i-1)$ in the unitary transformation (11). Here the
end time of the $(i-1)-$th cycle is just the beginning time of the $i-$th
cycle in the quantum circuit. Thus, $t_{0}=t_{0}$ and $t_{j}=t_{0}+j\Delta
T=t_{0j}+\Delta t_{0}+\delta t_{f}=t_{0(j+1)}-\delta t_{r}-\delta
t_{h}^{\prime }-\delta t_{h}-\delta t_{b}$ for $j=1,...,(i-1).$ Here $%
t_{01}=t_{0}+\delta t_{b}+\delta t_{h}+\delta t_{h}^{\prime }+\delta t_{r}$
and $t_{0(j+1)}=t_{01}+j\Delta T$ for $j=1,...,(i-1).$

The $i-$th cycle ($i\geq 1$) of the quantum circuit will be analyzed
separately as follows. In the $i-$th cycle of the quantum circuit the
starting time is $t_{(i-1)}=t_{0}+(i-1)\Delta T=t_{0i}-\delta t_{r}-\delta
t_{h}^{\prime }-\delta t_{h}-\delta t_{b}$ and the starting state $%
|0,CM0,R_{0}\rangle |0\rangle |f_{r}(x_{f})\rangle $ with $x_{f}=(x_{0}+i-1)%
\func{mod}m_{r}.$ Suppose that now the quantum circuit starts to execute the 
$i-$th cycle. First, the unitary transformation $U_{b}^{c}$ changes the
initial branch-control state $|b_{h}\rangle =|0\rangle $ to the state $%
|1\rangle ,$%
\begin{equation*}
\{P_{SL}(\{\varphi _{k}\},t_{(i-1)}+\delta
t_{b},t_{(i-1)}),U_{b}^{c}\}|0,CM0,R_{0}\rangle |0\rangle
|f_{r}(x_{f})\rangle 
\end{equation*}%
\begin{equation*}
\equiv P_{SL}(\{\varphi _{k}\},t_{(i-1)}+\delta
t_{b},t_{(i-1)})U_{b}^{c}|0,CM0,R_{0}\rangle |0\rangle |f_{r}(x_{f})\rangle 
\end{equation*}%
\begin{equation}
=|0,CM0,R_{0}\rangle |1\rangle |f_{r}(x_{f})\rangle \hspace{1.5in}\qquad
\quad \quad .  \label{12}
\end{equation}%
Here the unitary propagator $\{P_{SL}(\{\varphi _{k}\},t_{(i-1)}+\delta
t_{b},t_{(i-1)}),U_{b}^{c}\}$ is identical to the unitary operation $%
P_{SL}(\{\varphi _{k}\},t_{(i-1)}+\delta t_{b},t_{(i-1)})U_{b}^{c}$ due to
the fact that both the unitary transformation $P_{SL}(\{\varphi
_{k}\},t_{(i-1)}+\delta t_{b},t_{(i-1)})$ of the state-locking pulse field
and the unitary operation $U_{b}^{c}$ commute with each other, as shown in
Eq. (3). The second equation holds due to the state unitary transformation $%
P_{SL}(\{\varphi _{k}\},t_{(i-1)}+\delta t_{b},t_{(i-1)})|0,CM0,R_{0}\rangle
=|0,CM0,R_{0}\rangle ,$ as shown in the unitary transformation (5). Then,
the unitary operation $U_{h}^{c}$ converts the desired functional state $%
|f_{r}(x_{f})\rangle =|1\rangle $ into the state $|0\rangle ,$%
\begin{equation*}
\{P_{SL}(\{\varphi _{k}\},t,t_{(i-1)}+\delta
t_{b}),U_{h}^{c}\}|0,CM0,R_{0}\rangle |1\rangle |f_{r}(x_{f})\rangle 
\end{equation*}%
\begin{equation*}
=P_{SL}(\{\varphi _{k}\},t,t_{(i-1)}+\delta
t_{b})U_{h}^{c}|0,CM0,R_{0}\rangle |1\rangle |f_{r}(x_{f})\rangle 
\end{equation*}%
\begin{equation}
=|0,CM0,R_{0}\rangle |1\rangle |0\rangle ,\qquad \text{ }t=t_{(i-1)}+\delta
t_{b}+\delta t_{h},\qquad   \label{13}
\end{equation}%
and the unitary operation $V_{h}^{c}$ further changes the initial halting
state $|n_{h}\rangle =|0,CM0,R_{0}\rangle $ to the state $%
|1,CM0,R_{0}\rangle $, 
\begin{equation*}
\{P_{SL}(\{\varphi _{k}\},t,t_{(i-1)}+\delta t_{b}+\delta
t_{h}),V_{h}^{c}\}|0,CM0,R_{0}\rangle |1\rangle |0\rangle 
\end{equation*}%
\begin{equation*}
=P_{SL}(\{\varphi _{k}\},t,t_{(i-1)}+\delta t_{b}+\delta
t_{h})V_{h}^{c}|0,CM0,R_{0}\rangle |1\rangle |0\rangle 
\end{equation*}%
\begin{equation}
=|1,CM0,R_{0}\rangle |1\rangle |0\rangle ,\ \ \hspace{1.5in}\qquad \qquad 
\label{14}
\end{equation}%
where $t=t_{(i-1)}+\delta t_{b}+\delta t_{h}+\delta t_{h}^{\prime
}=t_{0i}-\delta t_{r}.$ These two unitary transformations (13) and (14) may
be obtained by the relations (8) and (9), respectively, and here equation
(5) has also been used. Now the state-dependent trigger pulse $P_{t}$ starts
to act on the halting state $|n_{h}\rangle =|1,CM0,R_{0}\rangle $ at the
time $t_{0i}-\delta t_{r}$ due to that the halting state has the internal
state with $n_{h}^{\prime }=1$ and the halting-qubit atom now is in the
left-hand potential well. With the help of the relation (10) the
state-dependent excitation process of the trigger pulse $P_{t}$ from the
halting state to the unstable state $|c_{1}\rangle $ of the control state
subspace may be expressed as 
\begin{equation*}
\{P_{SL}(\{\varphi _{k}\},t_{0i},t_{0i}-\delta
t_{r}),U_{tr}^{c}\}|1,CM0,R_{0}\rangle |1\rangle |0\rangle 
\end{equation*}%
\begin{equation}
=|1,CM1(t_{0i}),R_{1}(t_{0i})\rangle |1\rangle |0\rangle   \label{15}
\end{equation}%
where $t_{0i}=t_{(i-1)}+\delta t_{b}+\delta t_{h}+\delta t_{h}^{\prime
}+\delta t_{r}$ and $\{P_{SL}(\{\varphi _{k}\},t_{0i},t_{0i}-\delta
t_{r}),U_{tr}^{c}\}$ of the quantum circuit $Q_{c}$ is just defined as the
unitary operation $U_{tr}^{c}$ (see Eq. (10)). This excitation process
increases the motional energy of the halting-qubit atom in the left-hand
potential well so that the following unitary quantum scattering process for
the atom can take place automatically.

From the instant of time $t_{0i}$ on, the quantum circuit $Q_{c}$ really
starts to execute simultaneously its own two almost independent processes:
the quantum control process and the quantum computational process. The
quantum control process could be really thought to start at the time $t_{0i}$
and at the initial state $|c_{1}\rangle =|1,CM1(t_{0i}),R_{1}(t_{0i})\rangle 
$ for the physical system of the halting-qubit atom. Here for convenience
the state-dependent excitation process of the trigger pulse is not included
in the quantum control process. In the quantum control process the
halting-qubit atom with the state $|1,CM1(t_{0i}),R_{1}(t_{0i})\rangle $
first carries out a unitary quantum scattering process in space and time.
During the period $\Delta t_{0}$ of the quantum scattering process the
halting-qubit atom is driven by its own motional momentum to leave the
left-hand potential well and pass the intermediate potential barrier to
enter into the right-hand potential well. This quantum scattering process
may be expressed formally by the unitary transformation: 
\begin{equation*}
\{P_{SL}(\{\varphi _{k}\},t_{mi},t_{0i}),P_{SL}(\{\varphi _{k}\},\Delta
t_{0})\}|1,CM1(t_{0i}),R_{1}(t_{0i})\rangle |1\rangle |0\rangle \newline
\newline
\end{equation*}%
\begin{equation*}
\equiv P_{SL}(\{\varphi _{k}\},\Delta
t_{0})|1,CM1(t_{0i}),R_{1}(t_{0i})\rangle |1\rangle |0\rangle \qquad
\end{equation*}%
\begin{equation}
=|1,CM2(t_{mi}),R_{2}(t_{mi})\rangle |1\rangle |0\rangle ,\text{ }%
t_{mi}=t_{0i}+\Delta t_{0},\ \   \label{16}
\end{equation}%
where $\{P_{SL}(\{\varphi _{k}\},t_{mi},t_{0i}),P_{SL}(\{\varphi
_{k}\},\Delta t_{0})\}$ in the quantum circuit $Q_{c}$ is defined as $%
P_{SL}(\{\varphi _{k}\},\Delta t_{0}),$ and the duration $\Delta t_{0}$ must
be long enough so that by the quantum scattering process the wave-packet
state $|1,CM1(t_{0i}),R_{1}(t_{0i})\rangle $ of the halting-qubit atom in
the left-hand potential well can be almost completely transferred to the
wave-packet state $|1,CM2(t_{mi}),R_{2}(t_{mi})\rangle $ of the atom in the
right-hand potential well. Actually, the duration $\Delta t_{0}$ must ensure
that the wave-packet spatial distance $|R_{2}(t_{mi})-R_{0}|$ is large
enough such that both the wave-packet state $|1,CM0,R_{0}\rangle $ and $%
|1,CM2(t_{mi}),R_{2}(t_{mi})\rangle $ are almost orthogonal to each other.
Note that the distance $|R_{2}(t_{mi})-R_{1}(t_{0i})|$ is longer than $%
|R_{2}(t_{mi})-R_{0}|.$ This means that such a duration $\Delta t_{0}$
ensures that both the wave-packet states $|1,CM1(t_{0i}),R_{1}(t_{0i})%
\rangle $ and $|1,CM2(t_{mi}),R_{2}(t_{mi})\rangle $ are also almost
orthogonal to each other. The quantum scattering process (16) is dependent
only upon the time difference $\Delta t_{0}$ rather than the instant of time 
$t_{mi}$ or $t_{0i},$ that is, $P_{SL}(\{\varphi _{k}\},t_{mi},t_{0i})\equiv
P_{SL}(\{\varphi _{k}\},\Delta t_{0})$, and it is also an energy
conservative process that the motional energy of the final state $%
|c_{2}^{\prime }\rangle =|1,CM2(t_{mi}),R_{2}(t_{mi})\rangle $ is really
equal to that of the initial state $|c_{1}\rangle
=|1,CM1(t_{0i}),R_{1}(t_{0i})\rangle .$ After the quantum scattering process
the quantum computational process continues to execute the functional
unitary operation $U_{f}^{c}$ as usual, but from the time $t_{mi}$ on, the
quantum computational process is really halted in effect and meanwhile the
halting-qubit atom also motions continuously in the right-hand potential
well under the state-locking pulse field. This process may be expressed in
terms of the unitary transformation: 
\begin{equation*}
\{P_{SL}(\{\varphi _{k}\},t_{mi}+\delta
t_{f},t_{mi}),U_{f}^{c}\}|1,CM2(t_{mi}),R_{2}(t_{mi})\rangle |1\rangle
|0\rangle
\end{equation*}%
\begin{equation*}
\equiv P_{SL}(\{\varphi _{k}\},t_{mi}+\delta
t_{f},t_{mi})U_{f}^{c}|1,CM2(t_{mi}),R_{2}(t_{mi})\rangle |1\rangle |0\rangle
\end{equation*}%
\begin{equation}
=|n_{h}^{\prime }(t_{mi}+\delta t_{f}),CM2(t_{mi}+\delta
t_{f}),R_{2}(t_{mi}+\delta t_{f})\rangle |1\rangle |0\rangle ,\quad
\label{17}
\end{equation}%
where $|n_{h}^{\prime }(t_{mi}+\delta t_{f})\rangle $ is the internal state
of the halting-qubit atom at the time $t_{mi}+\delta t_{f}.$ The functional
operation $U_{f}^{c}$ does not have a real effect on the state $%
|f_{r}(x)\rangle =|0\rangle $ also due to the branch-control state $%
|b_{h}\rangle =|1\rangle .$ Here it must be ensured that the halting-qubit
atom is in the right-hand potential well during the period from the time $%
t_{mi}$ to the time $t_{mi}+\delta t_{f}$ so that the wave-packet state $%
|n_{h}^{\prime }(t),CM2(t),R_{2}(t)\rangle $ with $t_{mi}\leq t\leq
t_{mi}+\delta t_{f}$ is almost orthogonal to any one of the three
wave-packet states $|n_{h}^{\prime },CM0,R_{0}\rangle $ ($n_{h}^{\prime
}=0,1 $) and $|1,CM1(t_{0i}),R_{1}(t_{0i})\rangle .$ Obviously, here the
width of the intermediate potential barrier should be large enough so that a
wave-packet state of the halting-qubit atom in one potential well is almost
orthogonal to any wave-packet state of the atom in another potential well in
the double-well potential field.

From the $(i+1)-$th cycle to the end of the quantum circuit the conditional
unitary operations $U_{b}^{c},$ $U_{h}^{c},$ $V_{h}^{c},$ $U_{tr}^{c}$, and $%
U_{f}^{c}$ do not really affect the quantum system, although according as
the quantum circuit these unitary operations are still applied continuously
to the quantum system of the quantum computer, and only the state-locking
pulse field takes a real action on the halting-qubit atom in the right-hand
potential well. Then the time evolution process of the quantum system from
the time $t_{i}=t_{mi}+\delta t_{f}$ to the end of the quantum computational
process could be generally written as 
\begin{equation*}
\{P_{SL}(\{\varphi _{k}\},t_{i+j},t_{i}),(U_{T})^{j}\}|n_{h}^{\prime
}(t_{i}),CM2(t_{i}),R_{2}(t_{i})\rangle |1\rangle |0\rangle
\end{equation*}%
\begin{equation*}
=P_{SL}(\{\varphi _{k}\},t_{i+j},t_{i})|n_{h}^{\prime
}(t_{i}),CM2(t_{i}),R_{2}(t_{i})\rangle |1\rangle |0\rangle
\end{equation*}%
\begin{equation}
=|n_{h}^{\prime }(t_{i+j}),CM2(t_{i+j}),R_{2}(t_{i+j})\rangle |1\rangle
|0\rangle .\text{ }  \label{18}
\end{equation}%
Here the time $t_{i+j}=t_{i}+j\Delta T$ for $0\leq j\leq m_{r}-i$ and the
end time of the computational process is given by $t_{m_{r}}=t_{0}+m_{r}%
\Delta T$. From Eq. (11) and Eq. (18) one sees that before the initial
halting state $|0,CM0,R_{0}\rangle $ is changed to the state $%
|1,CM0,R_{0}\rangle $ the unitary propagator $\{P_{SL}(\{\varphi
_{k}\},t_{j},t_{0}),(U_{T})^{j}\}$ acting on the initial state $%
|0,CM0,R_{0}\rangle |0\rangle |f_{r}(x_{0})\rangle $ is really equal to the
single unitary operation sequence $(U_{T})^{j}$ acting on the same initial
state (see Eq. (11)), but after the halting-qubit atom enters into the
right-hand potential well the unitary propagator $\{P_{SL}(\{\varphi
_{k}\},t_{i+j},t_{i}),(U_{T})^{j}\}$ acting on the starting state $%
|n_{h}^{\prime }(t_{i}),CM2(t_{i}),R_{2}(t_{i})\rangle |1\rangle |0\rangle $
is really equal to the single unitary propagator $P_{SL}(\{\varphi
_{k}\},t_{i+j},t_{i})$ of the state-locking pulse field acting on the same
state (see Eq. (18)). Here any wave-packet state $|n_{h}^{\prime
}(t),CM2(t),R_{2}(t)\rangle $ for $t_{i}\leq t\leq t_{m_{r}}$ of the
halting-qubit atom in the right-hand potential well must be orthogonal or
almost orthogonal to any one of the three wave-packet states $|n_{h}^{\prime
},CM0,R_{0}\rangle $ ($n_{h}^{\prime }=0,1$) and $%
|1,CM1(t_{0i}),R_{1}(t_{0i})\rangle $ of the atom in the left-hand potential
well. Therefore, the quantum program and circuit requires that the
halting-qubit atom be in the right-hand potential well from the time $t_{mi}$
to the end of the quantum computational process.

Though the halting-qubit atom must be in the right-hand potential well at
the end of the computational process no matter what the initial functional
state $|f_{r}(x_{0})\rangle $ is with $x_{0}=0,1,...,m_{r}-1$, each possible
wave-packet state of the halting-qubit atom $|n_{h}^{\prime
}(t_{m_{r}}),CM2(t_{m_{r}}),R_{2}(t_{m_{r}})\rangle $ with $%
t_{m_{r}}=t_{i}+(m_{r}-i)\Delta T$ for $i=0,1,...,m_{r}-1$ could be
different in spatial position in the right-hand potential well at the end of
the computational process. This is because a different initial functional
state $|f_{r}(x_{0})\rangle $ corresponds to a different wave-packet state $%
|n_{h}^{\prime }(t_{m_{r}}),CM2(t_{m_{r}}),R_{2}(t_{i}+(m_{r}-i)\Delta
T)\rangle $ which is located at a different spatial position $%
R_{2}(t_{i}+(m_{r}-i)\Delta T)$ in the right-hand potential well. Here $%
n_{h}^{\prime }(t_{m_{r}})$ and $CM2(t_{m_{r}})$ with $%
t_{m_{r}}=t_{i}+(m_{r}-i)\Delta T$ each takes a same value for all different
index values $i=0,1,...,m_{r}-1,$ respectively$.$ Actually, similar to the
situations in the conventional halting protocol [11, 54], all these $m_{r}$
possible wave-packet states $\{|n_{h}^{\prime
}(t_{m_{r}}),CM2(t_{m_{r}}),R_{2}(t_{i}+(m_{r}-i)\Delta T)\rangle \}$ for $%
i=0,1,...,m_{r}-1$ should be almost orthogonal to one another. Therefore,
the output wave-packet states of the quantum circuit at the end of the
computational process are dependent sensitively upon the initial functional
states $\{|f_{r}(x_{0})\rangle \}.$ However, as pointed out before, the
quantum circuit $Q_{c}$ will not be a suitable component of the quantum
search processes based on the unitary quantum dynamics [24] if its output
state is dependent sensitively upon any initial functional state $%
|f_{r}(x_{0})\rangle .$ Evidently, if each of these $m_{r}$ possible
wave-packet states $\{|n_{h}^{\prime
}(t_{m_{r}}),CM2(t_{m_{r}}),R_{2}(t_{i}+(m_{r}-i)\Delta T)\rangle \}$ at the
end of the computational process can be further transferred to some desired
state in a high probability close to 100\% by a given unitary
transformation, then the output state of the quantum circuit could be
considered to be almost independent of any initial functional state. Of
course, it is impossible that the unitary transformation can change all
these $m_{r}$ wave-packet states to the same desired state in the
probability 100\%. It could be better to choose the desired state as the
wave-packet state $|n_{h}\rangle =|n_{h}^{\prime },CM0,R_{0}\rangle $ $%
(n_{h}^{\prime }=0$ or $1)$ of the halting-qubit atom in the left-hand
potential well as the wave-packet state is stable in the double-well
potential field. Thus, it is necessary to manipulate and control coherently
the halting-qubit atom by the state-locking pulse field after the
halting-qubit atom enters into the right-hand potential well. The coherent
manipulation process have two purposes in the quantum control process to
simulate efficiently the reversible and unitary halting protocol. The first
one is that the halting-qubit atom can stay in the right-hand potential well
for a long time till the computational process finished after the
halting-qubit atom enters into the right-hand potential well. The second is
that after the computational process finished the halting-qubit atom can
return the left-hand potential well from the right-hand one and the
returning halting-qubit atom in the left-hand potential well is in the
wave-packet state $|n_{h}^{\prime },CM0,R_{0}\rangle $ $(n_{h}^{\prime }=0$
or $1)$ with a high probability close to 100\% no matter what the
wave-packet state $|n_{h}^{\prime
}(t_{m_{r}}),CM2(t_{m_{r}}),R_{2}(t_{i}+(m_{r}-i)\Delta T)\rangle $ is for $%
i=0,1,...,m_{r}-1$. The coherent manipulation of the halting-qubit atom in
the right-hand potential well generally starts after the computational
process finished, but actually this manipulating process may start at a much
earlier time $t_{mi}$ when the halting-qubit atom enters into the right-hand
potential well rather than at the end of the computational process. \newline
\newline
\newline
{\large 5. Coherently manipulating the halting-qubit atom in time and space}

In this section the issues to discuss are focused on how the state-locking
pulse field manipulates coherently the wave-packet states of the
halting-qubit atom in the right-hand potential well of the double-well
potential field so that the halting-qubit atom can stay in the right-hand
potential well for a long time till the computational process finished and
how the quantum control process that simulates efficiently the reversible
and unitary halting protocol in the atomic physical system makes its output
state insensitive to any initial functional state of the quantum circuit $%
Q_{c}$. As shown in the previous section, when the halting-qubit atom leaves
the left-hand potential well and enters into the right-hand one at the time $%
t_{mi}=t_{0i}+\Delta t_{0}$ in the $i-$th cycle of the quantum circuit, the
conditional unitary operations $U_{h}^{c}$ and $V_{h}^{c}$ and the
state-dependent trigger pulse $P_{t}$ of the quantum circuit become
unavailable. Thus, from the time $t_{mi}$ on, the quantum computational
process really terminates in effect, although it runs continuously to the
end of the quantum circuit. Meanwhile, the halting-qubit atom evolves
continuously in the right-hand potential well under the state-locking pulse
field. Note that the state-locking pulse field consists of the double-well
potential field itself and the sequences of the time- and space-dependent
electromagnetic field pulses which are applied only to the right-hand
potential well during the computational process and then to the double-well
potential field after the computational process finished. Obviously, these
sequences of the time- and space-dependent electromagnetic field pulses of
the state-locking pulse field have a negligible effect on any state of the
halting-qubit atom when the atom is in the left-hand potential well during
the computational process. Thus, it follows from the unitary transformations
(17) and (18) that, from the time $t_{mi}$ on, the time evolution process of
the whole quantum system of the quantum computer may be reduced to the
simpler quantum control process of the halting-qubit atom in the right-hand
potential well under the state-locking pulse field, \newline
\begin{equation*}
P_{SL}(\{\varphi _{k}\},t,t_{mi})|1,CM2(t_{mi}),R_{2}(t_{mi})\rangle
\end{equation*}%
\begin{equation}
\hspace{1in}=|n_{h}^{\prime }(t),CM2(t),R_{2}(t)\rangle ,\text{ }t_{mi}\leq
t.  \label{19}
\end{equation}%
Here the wave-packet state $|n_{h}^{\prime }(t),CM2(t),R_{2}(t)\rangle $ ($%
t_{mi}\leq t$) generally could be expanded in terms of the motional basis
states of the halting-qubit atom in the right-hand potential well [38--43], 
\begin{equation*}
|n_{h}^{\prime }(t),CM2(t),R_{2}(t)\rangle =\underset{n_{h}^{\prime }}{\sum }%
\underset{CM2}{\sum }a(n_{h}^{\prime },CM2,t)|n_{h}^{\prime }\rangle
|CM2,R_{2}\rangle .
\end{equation*}%
Now the first purpose of the quantum control process is to design a unitary
sequence of the time- and space-dependent electromagnetic pulses and/or the
shaped potential fields of the state-locking pulse field $P_{SL}(\{\varphi
_{k}\},t,$ $t_{mi})$ to manipulate the halting-qubit atom so that the atom
is able to stay in the right-hand potential well till the end of the
computational process. For convenience, suppose that the lowest point of the
left-hand harmonic potential well is equal to the bottom of the right-hand
square potential well and both are set to zero. As stated before, when the
halting-qubit atom is in the unstable wave-packet state $%
|1,CM1(t_{0i}),R_{1}(t_{0i})\rangle $ the total motional energy of the atom
which includes the kinetic and potential energies in the left-hand harmonic
potential well but not the atomic internal energy is much higher than the
height of the intermediate potential barrier in the double-well potential
field. Since the quantum scattering process is an energy-conservative
process the total motional energy of the halting-qubit atom remains
unchanged when the halting-qubit atom enters into the right-hand potential
well from the left-hand one, but it is completely converted into the kinetic
energy as the potential energy of the halting-qubit atom is zero in the
right-hand square potential well. Suppose that the relativistic effect is
negligible for the motional halting-qubit atom in the double-well potential
field. Then the motional velocity of the halting-qubit atom is given by $%
v_{h}=\sqrt{2E_{h}/m_{h}}$ at the time $t_{mi}$ after the atom enters into
the right-hand potential well, where $E_{h}$ and $m_{h}$ are the total
motional energy and mass of the halting-qubit atom, respectively. Therefore,
the motional velocity $v_{h}$ of the halting-qubit atom could become very
large when the atom enters into the right-hand potential well from the
left-hand one. Since the geometric length of the right-hand square potential
well is limited it is impossible to keep the halting-qubit atom in the
right-hand potential well for a long time required by the computational
process if the motional velocity $v_{h}$ is very large. Thus, the motional
velocity of the halting-qubit atom must be decelerated greatly by the
state-locking pulse field so that the halting-qubit atom does not leave in a
short time the right-hand potential well. This decelerating process could be
achieved by the unitary sequence of the time- and space-dependent laser
light pulses of the state-locking pulse field. The decelerating process of
the halting-qubit atom is really very similar to the conventional atomic
laser cooling processes [36]. The interactions between the halting-qubit
atom in motion and the laser light pulse field become important in the
decelerating process, while it is known that the dipole force of the laser
light pulse field exerting a motional atom plays an important role in the
atomic laser cooling processes [36b]. There are a variety of the atomic
laser cooling methods and techniques which have been discovered and
developed in the past decades and used extensively to cool an atomic
ensemble to an extremely low temperature [36], but most of these atomic
laser cooling methods are non-unitary. Thus, not all these atomic laser
cooling methods are suitable for building up the unitary decelerating
sequence because the decelerating process for the halting-qubit atom must be
a unitary process. This is quite different from a conventional atomic laser
cooling process in an atomic ensemble. Only when an atomic laser cooling
method is unitary or can be made unitary can it be exploited to decelerate
the halting-qubit atom in the quantum control process. Thus, only the
coherent atomic laser cooling methods could be used to build up the unitary
decelerating sequence. Another difference between an atomic laser cooling
process and the unitary decelerating process is that the unitary
decelerating process is simpler as the current atomic physical system is the
pure quantum-state system of an atomic ion or a neutral atom in the
double-well potential field, while it is usually more complex to cool an
atomic ensemble by an atomic laser cooling technique. The third difference
is that it could not be necessary to decelerate the halting-qubit atom to
zero velocity, while the target of a conventional laser cooling technique is
to cool an atomic ensemble to an extremely low temperature as close to zero
degree as possible. Therefore, the unitary decelerating process for the
halting-qubit atom is relatively easy to be achieved by a coherent atomic
laser cooling technique.

The mechanisms for the laser cooling in an atomic ensemble have been studied
extensively and thoroughly in the past years [36]. However, in order to
investigate the mechanism of the unitary decelerating process for the
halting-qubit atom here the basic atomic-laser-cooling mechanism is
introduced briefly. A conventional atomic laser cooling method usually
consists of both the atomic optical-pumping process (or the atomic
optical-absorption process) and the atomic optical-emission process [36c].
The optical pumping process is that an atom under cooling is excited from
the ground state to the excited state by absorbing photons from the laser
light field, while the optical emission process is that the atom in the
excited state returns the ground state by emitting photons to the laser
light field or environment. The optical pumping process generally may be
made unitary easily, but the optical emission process usually is simply
chosen as a spontaneous and random process in a conventional laser cooling
technique partly due to that the lifetime of the atomic excited state
usually is very short. Hence the optical emission process usually is a
non-unitary process in an atomic ensemble for a conventional laser cooling
technique. Suppose that a free atomic ion or neutral atom with the mass $m$
in the ground state is irradiated by a laser light field and makes a
transition from the ground state to the excited state by absorbing a photon
from the laser light field. For convenience, here only nonrelativistic limit
is considered for the atomic motion. The optical pumping process generally
obeys the energy-, momentum-, and angular momentum-conservative laws. Denote
that before the transition the atom in the ground state has the
internal-state energy $E_{a}$, the kinetic energy $p_{a}^{2}/(2m)$, the
momentum $p_{a}=m\mathbf{v}_{a}$, and the angular momentum $J_{a}$,
respectively, and the photon that will be absorbed by the atom has the
photonic energy $E_{c}=\hslash \omega ,$ the momentum $p_{c}=\hslash \mathbf{%
k}$ ($|p_{c}|=\hslash \omega /c$), and the angular momentum $J_{c}$,
respectively. After the transition the photon is absorbed by the atom and
hence the photonic energy, momentum, and angular momentum are transferred to
the atom. The atom now is in the excited state. Suppose that after the
transition the atom in the excited state has the internal-state energy $%
E_{b} $, the kinetic energy $E_{b}=p_{b}^{2}/(2m),$ the momentum $p_{b}=m%
\mathbf{v}_{b},$ and the angular momentum $J_{b},$ respectively. Then the
energy conservation before and after the transition shows that there holds
the relation: 
\begin{equation}
E_{a}+p_{a}^{2}/(2m)+\hslash \omega =E_{b}+p_{b}^{2}/(2m)  \label{20}
\end{equation}%
where $\omega $ is the photonic frequency. The momentum conservation leads
to that 
\begin{equation}
m\mathbf{v}_{a}+\hslash \mathbf{k}=m\mathbf{v}_{b},  \label{21}
\end{equation}%
where $\mathbf{k}$ ($|\mathbf{k|=}\omega /c$) is the wave vector of the
photon before the transition and $\mathbf{v}_{a}$ and $\mathbf{v}_{b}$ are
the motional velocity vectors of the atom before and after the transition,
respectively. The angular momentum conservation between the angular momentum 
$J_{a}$ of the atom and the photonic angular momentum $J_{c}$ before the
transition and the angular momentum $J_{b}$ of the atom after the transition
will not be discussed in detail here. If the optical pumping process (or the
photon absorption process) for the atom is one-dimensional, then the wave
vector $\mathbf{k}$ of the photon is either co-direction to the motional
velocity $\mathbf{v}_{a}$ of the atom or opposite to the velocity $\mathbf{v}%
_{a}$ before the transition. For the first case that both the wave
propagating of the laser light field with the vector $\mathbf{k}$ and the
motion of the atom with the velocity $\mathbf{v}_{a}$ are co-direction to
each other the motional velocity of the atom after the transition is given
by 
\begin{equation}
v_{b}=v_{a}+\hslash k/m>v_{a},  \label{22}
\end{equation}%
hence the motional atom is accelerated by the copropagating laser light
field, while for the second case that the laser light field propagates in
the opposite direction to the atomic motion the atomic motional velocity
after the transition is written as 
\begin{equation}
v_{b}=v_{a}-\hslash k/m<v_{a},  \label{23}
\end{equation}%
hence the motional atom is decelerated by the opposite propagating laser
light field. Obviously, here the atom is slowed down by $\hslash k/m$ when
the atom absorbs a photon from the opposite propagating laser light field.
When the atom is in the excited state it no longer absorbs any photons from
the opposite propagating laser light field and hence can not be further
slowed down. One of the schemes to decelerate further the atom is that the
atom in the excited state first jumps back to the ground state without
changing significantly its total motional momentum, and then it absorbs a
photon from the opposite propagating laser light field again and hence is
decelerated further. Note that the process that the excited-state atom jumps
back to the ground state by emitting photons to the laser light field or its
environment is just the atomic optical-emission process. Thus, the atomic
laser cooling process consists of a number of the atomic optical
absorption-emission cycles that the atom in the ground state absorbs a
photon to make a transition to the excited state and then jumps back to the
ground state from the excited state by emitting photons. The optical
emission process is either a spontaneous emission process in a random form
or the stimulated emission process in a coherent form. In the spontaneous
emission process photons are emitted in a random form by the atoms in the
excited state so that the total momentum of the emitting photons is zero and
hence the atomic motional momentum does not change significantly after the
emission process. Therefore, the atom is slowed down basically by the
optical pumping process in every optical absorption-emission cycle if the
optical emission process is spontaneous and random. Most of the conventional
atomic laser cooling methods and techniques use the spontaneous
optical-emission process as their key component to cool an atom ensemble.
Every optical absorption-emission cycle can make the atom to be slowed down
by $\hslash k/m$ and the atom can be slowed down continuously by a sequence
of the optical absorption-emission cycles. However, the spontaneous
optical-emission process of the conventional laser cooling methods is not
allowed due to its own non-unitarity if these laser cooling methods are used
to slow down the halting-qubit atom in the quantum control process. On the
other hand, the coherent optical-emission process is different from the
spontaneous optical-emission process in that the coherent optical-emission
process may be a unitary process. The coherent optical-emission process
generally may be stimulated by an external laser light field [5a]. The
momentum of the emitting photons from the atom in the excited state in the
coherent optical-emission process is not zero on average and hence can make
a significant contribution to the motional momentum of the atom after the
emission process. The coherent atomic optical-emission process still obeys
the energy-, momentum-, and angular momentum-conservative laws. Therefore,
if the emitting photon travels along the same direction to the atomic
motion, then the atom will lose part of its motional momentum after the
emission process and hence is slowed down. In the optical-emission process
not only the atomic internal energy ($E_{b}-E_{a}$) of the excited state is
transferred to the photonic energy but also part of the motional energy of
the atom in the excited state is converted into the photonic energy.
However, the atom will receive a recoil momentum from the emitting photon
and hence is accelerated after the emission process if the emitting photon
travels along the opposite direction to the atomic motion. In this atomic
optical-emission process the atomic internal energy of the excited state is
transferred partly to the photonic energy and partly to the atomic kinetic
energy at the same time. In the quantum control process the halting-qubit
atom in the right-hand potential well must be first slowed down greatly so
that it is able to stay in the right-hand potential well for a long time
till the end of the computational process in the quantum circuit $Q_{c}$,
and then the atom is sped up in a unitary form after the computational
process finished such that the atom can return to the left-hand potential
well. The coherent atomic laser cooling methods and techniques therefore
provide a possible way to generate both the decelerating and accelerating
processes for the halting-qubit atom in the quantum control process.

A conventional laser cooling method based on the optical absorption-emission
(spontaneous) cycles usually is realized more easily than a coherent one in
an atomic ensemble. In general, the spontaneous optical-emission process
from the atomic excited state to the ground states in the atomic ensemble
occurs easily in nature as the atomic excited state usually has a much
shorter lifetime than those ground states of the atom. This in turn implies
that a coherent atomic laser cooling method could be more complex as the
non-unitary spontaneous optical emission must be avoided in the coherent
laser cooling process. The stimulated Raman adiabatic passage ($STIRAP$)
laser cooling method is one of the important atomic laser cooling methods.
The $STIRAP$ method has been used extensively to cool an atomic ensemble to
an extremely low temperature [42, 43], to cool a trapped atomic ion to the
ground state for quantum computation [30, 44], to manipulate a coherent
atomic beam in the atomic interferometry [38, 39, 40, 41], and to prepare
and manipulate a nonclassical motional state in a trapped-ion physical
system [37, 45]. In particular, the $STIRAP$ laser cooling method could be
used conveniently to cool a multi-level atomic ensemble with many internal
states to an extremely low temperature. The coherent $STIRAP$ laser cooling
(or decelerating) method could be a better candidate to avoid the
non-unitary spontaneous optical emission of the atom from the excited state
to the ground states. This is because the cooling (or decelerating) atom
does not stay in the excited state at all or could stay in the excited state
in a much shorter time than the lifetime of the excited state during the
coherent $STIRAP$ laser cooling process if the Raman adiabatic laser pulses
are detuned properly from the excited state. Since an adiabatic laser beam
usually has a much wider frequency bandwidth than a conventional $CW$ laser
beam the $STIRAP$ laser cooling method is able to take the Doppler effect
into account conveniently during the atomic laser cooling process. It is
well known that a general Raman adiabatic laser pulse consists of a pair of
the adiabatic laser beams with the specific characteristic parameters.
Generally, the characteristic parameters for the adiabatic laser beams of a
Raman adiabatic laser pulse include the carrier frequencies and detunings,
the frequency bandwidths, the amplitudes and phases of the adiabatic laser
light fields, the laser-beam durations, the propagation directions and
polarizations (e.g., $\sigma _{+}$ or $\sigma _{-}$), and the spatial action
positions and zones. Suppose now that the states $|g_{1}\rangle ,$ $%
|g_{2}\rangle ,$ and $|n_{e}\rangle $ are three different internal states of
the cooling atom and their corresponding wave-packet states of the atom are
written as $|g_{1},CM_{1},R_{1}\rangle ,$ $|g_{2},CM_{2},R_{2}\rangle ,$ and 
$|n_{e},CM_{e},R_{e}\rangle ,$ respectively. These two internal states $%
|g_{1}\rangle $ and $|g_{2}\rangle $ usually may be chosen as a pair of
ground internal states or two lowest energy-level internal states of the
atom, while the internal state $|n_{e}\rangle $ may be an excited state
whose energy level is much higher than those of the internal states $%
|g_{1}\rangle $ and $|g_{2}\rangle .$ In the coherent $STIRAP$ laser cooling
method an adiabatic laser beam $A$ may be used to pump the atom from the
ground internal state $|g_{1}\rangle $ to the excited state $|n_{e}\rangle $
and at the same time another adiabatic laser beam $B$ is applied to
stimulate the atom in the excited state $|n_{e}\rangle $ to jump back to the
internal state $|g_{2}\rangle $ [42, 43, 44]. The coherent $STIRAP$ process
may be formally expressed in terms of the unitary transition process: 
\begin{equation*}
|g_{1},CM_{1},R_{1}\rangle \overset{A}{\leftrightarrow }|n_{e},CM_{e},R_{e}%
\rangle \overset{B}{\leftrightarrow }|g_{2},CM_{2},R_{2}\rangle .
\end{equation*}%
Obviously, the carrier frequency of the adiabatic laser beam $A$ should be
close to the resonance frequency between the ground state $|g_{1}\rangle $
and the excited state $|n_{e}\rangle ,$ while the carrier frequency for the
adiabatic laser beam $B$ is close to the resonance frequency between the
ground state $|g_{2}\rangle $ and the excited state $|n_{e}\rangle .$ In
order to avoid occurring the non-unitary spontaneous optical emission for
the atom from the excited state $|n_{e}\rangle $ to the ground states both
the adiabatic laser beams $A$ and $B$ are detuned properly from the excited
state $|n_{e}\rangle .$ For example, if the unitary state transfer $%
|g_{1}\rangle \leftrightarrow |g_{2}\rangle $ is achieved by the
conventional $CW$ laser light irradiation, that is, the internal state $%
|g_{1}\rangle $ is first transferred completely to the excited state $%
|n_{e}\rangle $ and then to $|g_{2}\rangle $ by the $CW$ irradiation method,
then the decoherence effect usually affects largely the state transfer since
lifetime of the excited state $|n_{e}\rangle $ usually is very short and
also much shorter than those of the ground states $|g_{1}\rangle $ and $%
|g_{2}\rangle $, whereas the $STIRAP$ method can avoid such decoherence
effect on the state transfer. While the direct transition from the ground
state $|g_{1}\rangle $ $(|g_{2}\rangle )$ to another ground state $%
|g_{2}\rangle $ $(|g_{2}\rangle )$ is prohibited under the $CW$ laser light
irradiation, the coherent $STIRAP$ method is a better scheme to excite
indirectly the transition between these two ground states. Thus, the
coherent $STIRAP$\ method has some advantages over the conventional $CW$
irradiation method to transfer the ground state $|g_{1}\rangle $ $%
(|g_{2}\rangle )$ to another ground state $|g_{2}\rangle $ $(|g_{2}\rangle )$
in a unitary form. Obviously, the effective spatial bandwidth ($ESB$) for
the Raman adiabatic laser pulse must be greater than the spatial
displacement ($SD$) of the atom during the Raman adiabatic laser pulse. The
spatial displacement $(SD)$ is not more than the pulse duration ($t_{p}$)\
of the Raman adiabatic laser pulse times the maximum velocity ($v_{M}$) of
the atom during the Raman adiabatic laser pulse, that is, $SD<v_{M}\times
t_{p}\leq ESB.$

While the ground internal state $|g_{1}\rangle $ ($|g_{2}\rangle $) is
completely transferred to the state $|g_{2}\rangle $ ($|g_{1}\rangle $) by
the Raman adiabatic laser pulse, the corresponding motional state $%
|CM_{1},R_{1}\rangle $ ($|CM_{2},R_{2}\rangle $) of the atom is also changed
to another motional state $|CM_{2},R_{2}\rangle $ ($|CM_{1},R_{1}\rangle $).
During the coherent $STIRAP$ process the atom could be either sped up or
slowed down and this is mainly dependent upon the characteristic parameter
settings for these two adiabatic laser beams $A$ and $B$ of the Raman
adiabatic laser pulse and also the initial atomic motional velocity and
direction, as mentioned earlier. An example is given below. Suppose that the
atom is in the wave-packet state $|g_{1},CM_{1},R_{1}\rangle $ at the
initial time, the propagating directions of these two adiabatic laser beams $%
A$ and $B$ are opposite to each other, and the beam $A$ propagates in the
opposite direction to the atomic motion. Then the atom will be slowed down
by $\hslash k_{A}/m+\hslash k_{B}/m$ when the wave-packet state $%
|g_{1},CM_{1},R_{1}\rangle $ is transferred to $|g_{2},CM_{2},R_{2}\rangle $
by the Raman adiabatic laser pulse [43]. Here suppose that the initial
atomic velocity is much greater than $\hslash k_{A}/m+\hslash k_{B}/m.$ This
decelerating process could be understood intuitively: $(i)$ when the state $%
|g_{1},CM_{1},R_{1}\rangle $ is induced a transition to the excited state $%
|n_{e},CM_{e},R_{e}\rangle $ by the laser beam $A$ the atom is slowed down
by $\hslash k_{A}/m$ because the atom absorbs the photonic momentum $\hslash
k_{A}$ from the laser light field of the beam $A$ which travels along the
opposite direction to the atomic motion; $(ii)$ when the atom is stimulated
by the laser beam $B$ to jump to the state $|g_{2},CM_{2},R_{2}\rangle $
from the excited state $|n_{e},CM_{e},R_{e}\rangle $ it releases the
momentum $\hslash k_{B}$ to the laser light field of the beam $B$ and the
atom therefore is slowed down further by $\hslash k_{B}/m$ as the atomic
motional direction is the same as the propagating direction of the beam $B$.
Evidently, the atom can also be sped up when the atomic wave-packet state $%
|g_{1},CM_{1},R_{1}\rangle $ is transferred to $|g_{2},CM_{2},R_{2}\rangle $
by the Raman adiabatic laser pulse with the proper characteristic parameter
settings. Furthermore, the atom may be slowed down or sped up continuously
by many Raman adiabatic laser pulses with the proper characteristic
parameter settings. For example, suppose that one wants the atom to be
decelerated further after the atom is slowed down by $\hslash
k_{A}/m+\hslash k_{B}/m$ by the Raman adiabatic laser pulse $R(A,B)$ with
the beams $A$ and $B.$ Then one may apply another Raman adiabatic laser
pulse $R(A_{1},B_{1})$ with the beams $A_{1}$ and $B_{1}$ to the state $%
|g_{2},CM_{2},R_{2}\rangle $ to decelerate further the atom$.$ Since both
the spatial positions $R_{1}$ and $R_{2}$ are different for these two
wave-packet states $|g_{1},CM_{1},R_{1}\rangle $ and $|g_{2},CM_{2},R_{2}%
\rangle $ the applying spatial position $(R_{2})$ of the Raman adiabatic
laser pulse $R(A_{1},B_{1})$ is different from that one ($R_{1}$) of the
Raman adiabatic laser pulse $R(A,B).$ Here the adiabatic laser beam $A_{1}$
should travel along the opposite direction to the atomic motion, while the
beam $B_{1}$ propagates in the opposite direction to the beam $A_{1}$. Then
the atom is decelerated by $\hslash k_{A_{1}}/m+\hslash k_{B_{1}}/m$ again
after the wave-packet state $|g_{2},CM_{2},R_{2}\rangle $ is transferred to
another state $|g_{1},CM_{3},R_{3}\rangle $ by the Raman adiabatic laser
pulse $R(A_{1},B_{1}):|g_{2},CM_{2},R_{2}\rangle \overset{A_{1}}{%
\leftrightarrow }|n_{e},CM_{e}^{\prime },R_{e}^{\prime }\rangle \overset{%
B_{1}}{\leftrightarrow }|g_{1},CM_{3},R_{3}\rangle .$ The unitary
decelerating (or accelerating) sequence of the state-locking pulse field
used to decelerate (or accelerate) the halting-qubit atom in the right-hand
potential well is built up out of these coherent Raman adiabatic laser
pulses with the proper characteristic parameter settings. Obviously, both
the unitary decelerating and accelerating sequences are time- and
space-dependent. The conversion efficiency from one internal state ($%
|g_{1}\rangle )$ to another internal state ($|g_{2}\rangle )$ measures the
performance of a coherent Raman adiabatic laser pulse. A good-performance
Raman adiabatic laser pulse should be able to convert completely the
internal state $|g_{1}\rangle $ ($|g_{2}\rangle $) into the state $%
|g_{2}\rangle $ ($|g_{1}\rangle $). It has been shown theoretically [46, 38]
that in a three-state atomic system a ground internal state $|g_{1}\rangle $
($|g_{2}\rangle $) can be transferred completely to another ground state $%
|g_{2}\rangle $ ($|g_{1}\rangle $) in a unitary form through the excited
state $|n_{e}\rangle $ by a Raman adiabatic laser pulse with the proper
characteristic parameter settings.

The halting-qubit atom generally may be chosen as a multi-level atom with
many internal states in addition to the two internal states $%
\{|n_{h}^{\prime }\rangle ,$ $n_{h}=0,1\}$ of the halting quantum bit. Now
the coherent $STIRAP$ method is used to manipulate the halting-qubit atom
after the atom enters into the right-hand potential well. Here in the $%
STIRAP $ method these two internal states $\{|n_{h}^{\prime }\rangle ,$ $%
n_{h}^{\prime }=0,1\}$ could be conveniently set to the internal states $%
|g_{1}\rangle $ and $|g_{2}\rangle ,$ respectively, and the excited state $%
|n_{e}\rangle $ to some specific excited electronic state of the
halting-qubit atom. A unitary decelerating sequence $U_{D}(\{\varphi
_{k}\},t_{mi}+T_{D},t_{mi})$ consisting of the Raman adiabatic laser pulses
with the proper parameter settings $\{\varphi _{k}\}$ then is constructed to
decelerate the halting-qubit atom when the atom enters into the right-hand
potential well at the time $t_{mi}$, where $T_{D}$ is the total duration of
the unitary decelerating sequence. The total duration $T_{D}$ must be much
shorter than the period $\Delta T$ of each cycle of the quantum circuit.
Note that there are $m_{r}$ possible different times $\{t_{mi},$ $%
i=1,2,...,m_{r}\}$ for the quantum circuit $Q_{c}$. The halting-qubit atom
may enter into the right-hand potential well at any time $t_{mk}$ of these $%
m_{r}$ possible times $\{t_{mi}\}$ for $k=1,2,...,m_{r}.$ In order to
decelerate the halting-qubit atom the unitary decelerating sequence must be
applied at every time $t_{mi}$ of these $m_{r}$ possible times $\{t_{mi}\}$
in the quantum circuit. As known in the previous sections, the halting-qubit
atom completely enters into the right-hand potential well at the time $%
t_{mi}=t_{0i}+\Delta t_{0}$ in the $i-$th cycle of the quantum circuit and
at the time $t_{mi}$ the halting-qubit atom is in the wave-packet state $%
|c_{2}^{\prime }\rangle =|1,CM2(t_{mi}),R_{2}(t_{mi})\rangle $. Then the
unstable wave-packet state $|c_{2}^{\prime }\rangle $ will be changed to the
stable wave-packet state $|c_{2}(t_{mi}+T_{D})\rangle $ of the control state
subspace after the unitary decelerating sequence $U_{D}(\{\varphi
_{k}\},t_{mi}+T_{D},t_{mi})$ acts on the halting-qubit atom at the time $%
t_{mi}$ in the right-hand potential well, 
\begin{eqnarray*}
|c_{2}(t_{mi}+T_{D})\rangle &=&U_{D}(\{\varphi
_{k}\},t_{mi}+T_{D},t_{mi})|1,CM2(t_{mi}),R_{2}(t_{mi})\rangle \\
&=&|0,CM2(t_{mi}+T_{D}),R_{2}(t_{mi}+T_{D})\rangle ,
\end{eqnarray*}%
meanwhile the initial motional velocity $v_{h}$ of the halting-qubit atom at
the time $t_{mi}$ is slowed down to the velocity $v_{0}<<v_{h}$ by the
unitary decelerating sequence and the initial internal state $|1\rangle $ is
also changed to $|0\rangle $. Here it is important that after the
halting-qubit atom is acted on by the unitary decelerating sequence it is no
longer acted on by next unitary decelerating sequences. Then the motional
velocity $v_{0}$ of the halting-qubit atom must be greater than zero so that
the halting-qubit atom itself can leave in the velocity $v_{0}$ the
effective spatial action zone of the unitary decelerating sequence before
next decelerating sequence starts to apply at the time $t_{m,i+1}=t_{mi}+%
\Delta T$. The spatial displacement of the halting-qubit atom is given by $%
SD=v_{0}\times (\Delta T-T_{D})$ during the period $(\Delta T-T_{D})$ from
the time $t_{mi}+T_{D}$ after the atom is acted on by the unitary
decelerating sequence to the time $t_{m,i+1}$ before next decelerating
sequence is applied. The spatial displacement $SD$ must be large enough to
ensure that the entire wave-packet state $|c_{2}(t_{m,i+1})\rangle $ of the
halting-qubit atom at the time $t_{m,i+1}$ is outside the effective spatial
action zone of the unitary decelerating sequence, and hence it is also much
greater than the effective wave-packet spread of the wave-packet state $%
|c_{2}(t_{m,i+1})\rangle $. Now both the atomic motional velocity $v_{0}$
and energy $E_{0}=mv_{0}^{2}/2$ are much less than the initial velocity $%
v_{h}=\sqrt{2E_{h}/m_{h}}$ and energy $E_{h},$ respectively. The wave-packet
state $|c_{2}(t_{mi}+T_{D})\rangle $ is stable in the sense that the atomic
motional energy $E_{0}$ of the wave-packet state is much lower than the
height of the intermediate potential barrier in the double-well potential
field. If now the halting-qubit atom goes in the velocity $v_{0}$ through a
fixed distance $\Delta R$ in the right-hand potential well it spends the
time equal to $\Delta R/v_{0}.$ Because $v_{h}>>v_{0}$\ this time interval $%
\Delta R/v_{0}$ is much longer than the time period $\Delta R/v_{h}$ during
which the atom passes the same distance $\Delta R$ in the initial velocity $%
v_{h}$. Thus, one may imagine that the wave-packet state of the
halting-qubit atom is locked in the right-hand potential well for a long
time $(\Delta R/v_{0})$ after the atom is slowed down greatly by the unitary
decelerating sequence. As required by the quantum program and circuit, the
halting-qubit atom should stay in the right-hand potential well until the
end time ($t_{m_{r}}$) of the computational process. Then the time interval $%
\Delta R/v_{0}\geq t_{m_{r}}-(t_{m1}+T_{D}),$ where the time $t_{m1}$ is the
earliest one among all these $m_{r}$ possible times $\{t_{mi}\}.$ For
convenience, suppose $\delta t_{f}>T_{D}.$ Then at the end time $%
t_{m_{r}}=t_{0}+m_{r}\Delta T$ of the computational process the
center-of-mass spatial position $R_{2}(t_{m_{r}})$ of the wave-packet state $%
|c_{2}(t_{m_{r}})\rangle $ of the halting-qubit atom is given by 
\begin{equation}
R_{2,i}(t_{m_{r}})=R_{2}(t_{mi}+T_{D})+v_{0}\times (t_{m_{r}}-t_{mi}-T_{D}),%
\text{ }1\leq i\leq m_{r},  \label{24}
\end{equation}%
when the halting-qubit atom enters into the right-hand potential well at the
time $t_{mi}$ in the $i-$th cycle of the quantum circuit. Here the spatial
position $R_{2}(t_{m_{r}})$ is denoted as $R_{2,i}(t_{m_{r}})$ so as to show
explicitly that the position is dependent of the cyclic index value $%
i=1,2,...,m_{r}$. Though each possible spatial position $R_{2}(t_{mi}+T_{D})$
is the same for the index value $i=1,2,...,m_{r}$ just like $R_{2}(t_{mi}),$
the wave-packet spatial position $R_{2,i}(t_{m_{r}})$ at the end time $%
t_{m_{r}}$ of the computational process is different for a different index
value $i$. This is because the halting-qubit atom stays in the right-hand
potential well for a longer time and hence passes a longer spatial distance
before the end time $t_{m_{r}}$ of the computational process if it enters
into the right-hand potential well at an earlier time $t_{mi}$. The maximum
and minimum wave-packet spatial positions for the halting-qubit atom at the
end time $t_{m_{r}}$ correspond to the halting-qubit atom entering into the
right-side potential well in the first and the last cycle of the quantum
circuit, respectively. Evidently, these $m_{r}$ possible different
wave-packet spatial positions of the halting-qubit atom at the end time $%
t_{m_{r}}$ satisfy the following inequality: 
\begin{equation}
R_{2,m_{r}}(t_{m_{r}})<...<R_{2,2}(t_{m_{r}})<R_{2,1}(t_{m_{r}}).  \label{25}
\end{equation}%
Here as usual the $+x$ coordinate direction is defined as from the left-hand
potential well toward the right-hand one in the double-well potential field.
The inequality (25) is also correct for the case $\delta t_{f}\leq T_{D}$.
Of course, in this case the atomic wave-packet state and its spatial
position $R_{2,i}(t_{m_{r}}+T_{D}-\delta t_{f})$ at the time $%
t_{m_{r}}+T_{D}-\delta t_{f}$ correspond to the wave-packet state and its
spatial position $R_{2,i}(t_{m_{r}})$ at the end time $t_{m_{r}}$ in the
case $\delta t_{f}>T_{D},$ respectively. Actually, the time $%
t_{m_{r}}+T_{D}-\delta t_{f}$ is the end time of the total quantum circuit
consisting of the quantum circuit $Q_{c}$ and the unitary decelerating
sequence for the case $\delta t_{f}\leq T_{D}$, while if $\delta t_{f}>T_{D}$
the end time of the total quantum circuit is really just the end time $%
t_{m_{r}}$ of the single quantum circuit $Q_{c}$. Hereafter only the
situation $\delta t_{f}>T_{D}$ is considered. Obviously, the halting-qubit
atom at the time $t_{m_{r}}$ is always in the spatial region $%
[R_{2,m_{r}}(t_{m_{r}})-\delta R(t_{m_{r}})/2,$ $R_{2,1}(t_{m_{r}})+\delta
R(t_{m_{r}})/2]$ of the right-hand potential well, that is, any
center-of-mass spatial position $R_{2,i}(t_{m_{r}})$ of the wave packet
state $|c_{2}(t_{m_{r}})\rangle $ for $i=1,2,...,m_{r}$ satisfies: $%
R_{2,i}(t_{m_{r}})\in \lbrack R_{2,m_{r}}(t_{m_{r}}),$ $R_{2,1}(t_{m_{r}})].$
Here for convenience suppose that the spatial shape of the wave-packet state 
$|c_{2}(t_{m_{r}})\rangle $ is symmetrical and $\delta R(t_{m_{r}})$ is the
effective spatial spread of the wave-packet state at the time $t_{m_{r}}$.
Since the spatial distance between any two nearest wave-packet states takes
the same value: $\Delta
R_{2,i,i+1}(t_{m_{r}})=R_{2,i}(t_{m_{r}})-R_{2,i+1}(t_{m_{r}})=v_{0}\times
\Delta T$ for $i=1,2,...,m_{r}-1,$ as shown in Eq. (24), these $m_{r}$
possible wave-packet states distribute uniformly in the spatial region $%
[R_{2,m_{r}}(t_{m_{r}})-\delta R(t_{m_{r}})/2,$ $R_{2,1}(t_{m_{r}})+\delta
R(t_{m_{r}})/2]$ of the right-hand potential well. In general, it follows
from Eq. (24) that the spatial distance between the spatial positions $%
R_{2,i}(t_{m_{r}})$ and $R_{2,j}(t_{m_{r}})$ $(1\leq i<j\leq m_{r})$ of the
halting-qubit atom at the time $t_{m_{r}}$ can be calculated by 
\begin{equation}
\Delta
R_{2,i,j}(t_{m_{r}})=R_{2,i}(t_{m_{r}})-R_{2,j}(t_{m_{r}})=v_{0}(j-i)\Delta
T.  \label{26}
\end{equation}%
Here the time difference $(j-i)\Delta T$ is the duration between the
halting-qubit atom entering into the right-hand potential well in the $j-$th
cycle and the $i-$th cycle ($j>i)$ in the quantum circuit. Obviously, the
maximum spatial distance which is the dimensional size of the spatial region 
$[R_{2,m_{r}}(t_{m_{r}}),$ $R_{2,1}(t_{m_{r}})]$ is given by $\Delta
R_{2,1,m_{r}}(t_{m_{r}})=v_{0}(m_{r}-1)\Delta T.$ From the end time $%
t_{m_{r}}$ on, there are no longer any unitary operation of the quantum
circuit and any unitary decelerating sequence applying to the whole quantum
system of the quantum computer. However, in order that the output state of
the reversible and unitary halting protocol is not dependent sensitively
upon any initial functional state in the quantum circuit the wave-packet
state $|c_{2}(t_{m_{r}})\rangle =|0,CM2(t_{m_{r}}),R_{2,i}(t_{m_{r}})\rangle 
$ of the halting-qubit atom must be changed back to the stable halting state
such as the state $|1,CM0,R_{0}\rangle $ in a high probability. Thus, the
halting-qubit atom must ultimately return the left-hand potential well from
the right-hand one after the computational process finished. Here the
control state subspace $S(C)$ in the atomic physical system consists of a
series of wave-packet states and is not a two-state subspace. Then the state 
$|c_{2}\rangle $ of the control state subspace $S(C)$ in the quantum program 
$Q_{c}$ really corresponds to these wave-packet states of the halting-qubit
atom in the right-hand potential well and also the stable halting state $%
|1,CM0,R_{0}\rangle $ finally.

One possible scheme to force the halting-qubit atom in the right-hand
potential well to return to the left-hand potential well is to increase the
atomic motional energy and invert the motional direction of the atom. Now a
unitary accelerating sequence which consists of the Raman adiabatic laser
pulses with the proper characteristic parameter settings is constructed to
speed up in a unitary form the halting-qubit atom in the right-hand
potential well after the computational process and the unitary decelerating
sequence finished. The characteristic parameter settings for the Raman
adiabatic laser pulses of the unitary accelerating sequence are clearly
different from those of the unitary decelerating sequence. A unitary
accelerating process could be thought of as the inverse process of a unitary
decelerating process except the atomic motional direction and the spatial
action zone. There are two purposes for the unitary accelerating process to
speed up the halting-qubit atom. The first purpose is simply that after the
halting-qubit atom is sped up by the unitary accelerating sequence it hits
the right-hand wall of the right-hand potential well to change its motional
direction and then returns to the left-hand potential well in a higher
velocity so that the halting-qubit atom can pass the intermediate potential
barrier to arrive in the left-hand potential well in a shorter time. Here
define the arriving time ($T_{i},$ $i=1,2,...,m_{r};$ see below) as the
instant of time at which the entire effective wave-packet state of the
halting-qubit atom enters into the left-hand potential well from the
right-hand one and moreover the center-of-mass position of the wave-packet
state is some given spatial position (e.g., $R_{1}(t_{0i})$) within the
left-hand potential well. For a heavy atom the wave-packet picture in
quantum mechanics is very similar to the classical particle picture [5a].
Then from viewpoint of the particle picture it could be better to choose the
given spatial position such that the mean motional speed and kinetic energy
of the halting-qubit atom is zero at the given spatial position within the
left-hand potential well, that is, at the given spatial position the total
motional energy of the halting-qubit atom is pure potential energy.
Evidently, when the halting-qubit atom arrives in the left-hand potential
well there are $m_{r}$ possible different arriving times for the
halting-qubit atom with the $m_{r}$ possible wave-packet states $%
\{|0,CM2(t_{m_{r}}),R_{2,i}(t_{m_{r}})\rangle \}$ ($i=1,2,...,m_{r}$) at the
end time $t_{m_{r}}$ of the computational process. Each arriving time
corresponds one-to-one to a possible wave-packet state ($%
|0,CM2(t_{m_{r}}),R_{2,i}(t_{m_{r}})\rangle $) which locates at a different
spatial position ($R_{2,i}(t_{m_{r}})$) in the right-hand potential well.
Generally, the first wave-packet state $|0,CM2(t_{m_{r}}),R_{2,1}(t_{m_{r}})%
\rangle $ will arrive in the left-hand potential well at the earliest time,
while the last one $|0,CM2(t_{m_{r}}),R_{2,m_{r}}(t_{m_{r}})\rangle $ enters
into the left-hand potential well at the latest time. Then the second
purpose is particularly important for the quantum control process in that
the unitary accelerating sequence is really used to shorten greatly any time
differences among the $m_{r}$ possible different arriving times for the
halting-qubit atom. As pointed out earlier, all these $m_{r}$ possible
wave-packet states $\{|0,CM2(t_{m_{r}}),R_{2,i}(t_{m_{r}})\rangle \}$ of the
halting-qubit atom at the end time $t_{m_{r}}$ are within the spatial region 
$[R_{2,m_{r}}(t_{m_{r}})-\delta R(t_{m_{r}})/2,$ $R_{2,1}(t_{m_{r}})+\delta
R(t_{m_{r}})/2]$ in the right-hand potential well. In order that any one of
these $m_{r}$ possible wave-packet states can be changed back to the stable
halting state $|n_{h}^{\prime },CM0,R_{0}\rangle $ $(n_{h}^{\prime }=0$ or $%
1)$ in a high probability each of these $m_{r}$ possible wave-packet states
may be acted on by the same unitary accelerating sequence such that any time
differences among these $m_{r}$ possible arriving times can be shorten
greatly. Here the effective width of the spatial action zone of every Raman
adiabatic laser pulse of the unitary accelerating sequence must be greater
than the dimensional size of the spatial region $[R_{2,m_{r}}(t_{m_{r}})-%
\delta R(t_{m_{r}})/2,$ $R_{2,1}(t_{m_{r}})+\delta R(t_{m_{r}})/2].$ Since
the halting-qubit atom moves also a spatial displacement during the Raman
adiabatic laser pulse the effective spatial-action-zone width of the Raman
adiabatic laser pulse must also take the spatial displacement into account
in addition to the dimensional size of the spatial region. In technique it
could be better to choose those spatially uniform ultra-broadband adiabatic
laser pulses [47] as the adiabatic laser beams of the Raman adiabatic laser
pulses of the unitary accelerating sequence. Denote such an ultra-broadband
unitary accelerating sequence as $U_{A}(\{\varphi _{k}\},t+T_{A},t),$ where $%
T_{A}$ is the total duration of the accelerating sequence. When the
ultra-broadband unitary accelerating sequence acts on the halting-qubit atom
at the end time $t_{m_{r}}$ of the computational process any one of these $%
m_{r}$ possible wave-packet states $\{|0,CM2(t_{m_{r}}),R_{2,i}(t_{m_{r}})%
\rangle \}$ of the halting-qubit atom is transferred to the corresponding
unstable wave-packet state: 
\begin{eqnarray*}
|c_{2,i}^{\prime }(t_{m_{r}}+T_{A})\rangle &=&U_{A}(\{\varphi
_{k}\},t_{m_{r}}+T_{A},t_{m_{r}})|0,CM2(t_{m_{r}}),R_{2,i}(t_{m_{r}})\rangle
\\
&=&|1,CM2(t_{m_{r}}+T_{A}),R_{2,i}(t_{m_{r}}+T_{A})\rangle
\end{eqnarray*}%
where the internal state $|0\rangle $ of the halting-qubit atom is changed
to the state $|1\rangle $ after the unitary accelerating sequence, meanwhile
the halting-qubit atom is sped up from the initial motional velocity $v_{0}$
to a great velocity $v>>v_{0}$. The motional velocity $v$ usually may be
greater than the motional velocity $v_{h}=\sqrt{2E_{h}/m_{h}}$, that is, $%
v\geq v_{h}>>v_{0},$ so that the halting-qubit atom has an enough high
motional energy to pass the intermediate potential barrier to enter into the
left-hand potential well. Note that the motional velocity $v$ of the
halting-qubit atom has an upper-bound value $c,$ where $c$ is the light
speed in vacuum, and usually $v<<c.$ The important point for the unitary
accelerating process is that the center-of-mass spatial distance between any
pair of the wave-packet states among these $m_{r}$ possible wave-packet
states $\{|0,CM2(t_{m_{r}}),R_{2,i}(t_{m_{r}})\rangle \}$ of the
halting-qubit atom is kept unchanged before and after the unitary
accelerating process, although the halting-qubit atom is accelerated under
the action of the unitary accelerating sequence and its wave-packet spatial
position has been changing along the $+x$ direction. Therefore, after the
unitary accelerating sequence these $m_{r}$ possible wave-packet spatial
positions $\{R_{2,i}(t_{m_{r}}+T_{A})\}$ of the halting-qubit atom still
satisfy the inequality (25) and their possible spatial distances remain also
unchanged and are still given by Eq. (26), that is, $\Delta
R_{2,i,j}(t_{m_{r}}+T_{A})=v_{0}(j-i)\Delta T$ for $1\leq i<j\leq m_{r}.$
After the action of the accelerating sequence the halting-qubit atom moves
in the velocity $v$ along the direction $+x$ toward the right-hand potential
wall of the double-well potential field and hits ultimately the potential
wall in an elastic form. Then the halting-qubit atom is bounced off the
right-hand potential wall and its motional direction therefore is reversed
and hence changed to the direction $-x$. Evidently, this elastic bouncing
process is unitary [5a, 48]. Now the halting-qubit atom moves in the
velocity $v$ ($v\geq v_{h}$) along the direction $-x$ toward the left-hand
potential well. It first goes across the right-hand potential well, then
passes the intermediate potential barrier, and finally enters into the
left-hand harmonic potential well. Note that the spatial position $%
R_{2,1}(t_{m_{r}}+T_{A})$ is nearest the right-hand potential wall among all
these $m_{r}$ possible wave-packet spatial positions $%
\{R_{2,i}(t_{m_{r}}+T_{A})\}.$ Evidently, the shortest spatial distance
between the spatial position $R_{2,1}(t_{m_{r}}+T_{A})$ and the right-hand
potential wall must be much greater than half the wave-packet spread: $%
\delta R(t_{m_{r}}+T_{A})/2.$ The halting-qubit atom needs to spend a short
time period when the atom moves from the spatial position $%
R_{2,1}(t_{m_{r}}+T_{A})$ to the right-hand potential wall, bounces off the
potential wall, and then returns the original spatial position $%
R_{2,1}(t_{m_{r}}+T_{A})$. Denote this short period as the atomic bouncing
dead time $t_{d}$. Suppose that the time period is denoted as $t_{a}$ when
the halting-qubit atom arrives in the left-hand potential well from the
spatial position $R_{2,1}(t_{m_{r}}+T_{A})$ after the atom bounces off the
potential wall. Then the longest time period of the quantum control process
from the starting time $(t_{0i})$ of the quantum scattering process to the
time when the halting-qubit atom arrives in the left-hand potential well is
not longer than $(t_{m_{r}}-t_{01})+T_{A}+t_{d}+t_{a}.$ It follows from the
inequality (25) that if the halting-qubit atom enters into the right-hand
potential well from the left-hand one in the first cycle of the quantum
circuit, then it will first return the left-hand potential well from the
right-hand one after the unitary decelerating and accelerating sequences,
whereas the halting-qubit atom returns the left-hand potential well at the
latest time if it enters into the right-hand potential well in the latest
cycle of the quantum circuit. Suppose that the halting-qubit atom returns to
the left-hand potential well from the right-hand one and arrives at some
given spatial position within the left-hand potential well at the arriving
time $T_{i}$ ($T_{i}>t_{m_{r}}+T_{A}$) for $i=1,2,...,m_{r}$ if the atom
enters into the right-hand potential well from the left-hand one in the $i-$%
th cycle of the quantum circuit. It follows from the inequality (25) that
these $m_{r}$ possible arriving times $\{T_{i},$ $i=1,2,...,m_{r}\}$ satisfy
the following inequality: 
\begin{equation}
T_{1}<T_{2}<...<T_{m_{r}}  \label{27}
\end{equation}%
and the equation (26) shows that the arriving-time difference $\Delta
T_{j,i}=T_{j}-T_{i}$ for $1\leq i<j\leq m_{r}$ is given by 
\begin{equation}
\Delta T_{j,i}=\Delta R_{2,i,j}(t_{m_{r}}+T_{A})/v=(j-i)\Delta Tv_{0}/v,
\label{28}
\end{equation}%
and the maximum arriving-time difference equals 
\begin{equation}
\Delta T_{m_{r},1}=(m_{r}-1)\Delta Tv_{0}/v.  \label{29}
\end{equation}%
It is known that the time difference between the halting-qubit atom entering
into the right-hand potential well in the $j-$th cycle and the $i-$th cycle $%
(j>i)$ of the quantum circuit is given by $(j-i)\Delta T.$ But after the
halting-qubit atom is acted on by the unitary decelerating and accelerating
sequences in the right-hand potential well the corresponding arriving-time
difference becomes $\Delta T_{j,i}=(j-i)\Delta Tv_{0}/v.$ Since the motional
velocity $v$ is much greater than the velocity $v_{0},$ that is, the
time-compressing factor $v_{0}/v<<1,$ the arriving-time difference $\Delta
T_{j,i}$ is much shorter than the original time difference $(j-i)\Delta T,$
indicating that the original time difference is greatly compressed after the
time- and space-dependent quantum control process which contains the unitary
decelerating and accelerating processes.

During the quantum control process the halting-qubit atom carries out
consecutively the quantum scattering process in which the atom goes from the
left-hand potential well to the right-hand one, decelerating process,
accelerating process, elastic bouncing process, the second decelerating
process (see below), and finally the second quantum scattering process in
which the atom returns from the right-hand potential well to the left-hand
one. But the internal state of the halting-qubit atom could be changed only
in the unitary decelerating and accelerating processes among these
processes. For simplicity, here the second decelerating process is not
considered temporarily. For convenience, for the case that the halting-qubit
atom enters into the right-hand potential well in the $i-$th cycle of the
quantum circuit and then returns and arrives in the left-hand potential well
at the arriving time $T_{i}$ the wave-packet state of the halting-qubit atom
in the left-hand potential well at the arriving time $T_{i}$ is denoted as $%
|1,CM1(T_{i}),R_{1,i}(T_{i})\rangle $ for $i=1,2,...,m_{r}$. Here the
wave-packet state $|1,CM1(T_{i}),R_{1,i}(T_{i})\rangle $ has the same
internal state as the state $%
|1,CM2(t_{m_{r}}+T_{A}),R_{2,i}(t_{m_{r}}+T_{A})\rangle $ of the
halting-qubit atom in the right-hand potential well after the unitary
accelerating process. Evidently, in an ideal case all these wave-packet
states $\{|1,CM1(T_{i}),R_{1,i}(T_{i})\rangle \}$ are really the same$:$%
\begin{eqnarray*}
|1,CM1(T_{1}),R_{1,1}(T_{1})\rangle &=&|1,CM1(T_{2}),R_{1,2}(T_{2})\rangle \\
&=&...=|1,CM1(T_{m_{r}}),R_{1,m_{r}}(T_{m_{r}})\rangle ,
\end{eqnarray*}%
although the arriving time $T_{i}$ for the halting-qubit atom is different
for a different cycle index value $i=1,2,...,m_{r}$, as can be seen in (27).
However, the wave-packet motional state $|CM1(T_{i}),R_{1,i}(T_{i})\rangle $
for $i=1,2,...,m_{r}$ is generally different from the desired motional state 
$|CM0,R_{0}\rangle $. Since the motional energy of the halting-qubit atom
with the wave-packet state $|1,CM1(T_{i}),R_{1,i}(T_{i})\rangle $ is much
higher than the height of the intermediate potential barrier in the
double-well potential field the wave-packet state $|1,CM1(T_{i}),$ $%
R_{1,i}(T_{i})\rangle $ is unstable and hence it must be transferred to the
stable halting state $|1,CM0,R_{0}\rangle $. In general, there is a unitary
transformation that transfers completely the wave-packet state $%
|1,CM1(T_{1}),R_{1,1}(T_{1})\rangle $ to the halting state $%
|1,CM0,R_{0}\rangle .$ This unitary transformation is defined as 
\begin{equation}
U(\{\varphi _{k}\},T_{t}+T_{1},T_{1})|1,CM1(T_{1}),R_{1,1}(T_{1})\rangle
=|1,CM0,R_{0}\rangle .  \label{30}
\end{equation}%
Here the duration $T_{t}$ of the unitary operation $U(\{\varphi
_{k}\},T_{t}+T_{1},T_{1})$ usually could be much longer than the maximum
arriving-time difference $\Delta T_{m_{r},1}=T_{m_{r}}-T_{1}$ (see Eq. (29))
and $\{\varphi _{k}\}$ are the control parameters of the unitary operation.
The unitary operation $U(\{\varphi _{k}\},T_{t}+T_{1},T_{1})$ starts to act
on the wave-packet state $|1,CM1(T_{1}),R_{1,1}(T_{1})\rangle $ of the
halting-qubit atom in the left-hand potential well at the arriving time $%
T_{1}$ and it does not change any internal state of the halting-qubit atom.
It could be generated by applying the coherent Raman adiabatic laser pulse
and the time-dependent potential field to the left-hand harmonic potential
well. The coherent Raman adiabatic laser trigger pulse $P_{t}$ transfers the
lower motional-energy state $|1,CM0,R_{0}\rangle $ to the higher
motional-energy wave-packet state $|1,CM1(t_{0i}),R_{1}(t_{0i})\rangle $,
while here the unitary operation $U(\{\varphi _{k}\},T_{t}+T_{1},T_{1})$
converts the higher motional-energy wave-packet state $%
|1,CM1(T_{1}),R_{1,1}(T_{1})\rangle $ into the lower motional-energy state $%
|1,CM0,R_{0}\rangle $ in the left-hand potential well. However, the unitary
operation $U(\{\varphi _{k}\},T_{t}+T_{1},T_{1})$ is more complex than the
coherent Raman adiabatic laser trigger pulse $P_{t}$. There is a difference
between these two motional states $|CM1(T_{1}),R_{1,1}(T_{1})\rangle $ and $%
|CM0,R_{0}\rangle .$ The difference is in the atomic motional energy,
momentum, spatial position, wave-packet shape (e.g., the effective spread)
and phase, and so forth. Thus, in order to convert the wave-packet motional
state $|CM1(T_{1}),R_{1,1}(T_{1})\rangle $ into the ground motional state $%
|CM0,R_{0}\rangle $ one could apply the coherent Raman adiabatic laser
pulses and also an external time-dependent potential field to the
halting-qubit atom in the left-hand potential well. Here the external
time-dependent potential field may be used to modulate the left-hand
harmonic potential field or even the whole double-well potential field. Both
the external time-dependent potential field and the coherent Raman adiabatic
laser pulses could also be thought of as the components of the state-locking
pulse field.

As shown in the inequality (27), the arriving time $T_{i}$ is different for $%
i=1,2,...,m_{r}$ for the halting-qubit atom returning and arriving in the
left-hand potential well after the atom enters into the right-hand potential
well in a different cycle ($i-$th cycle) of the quantum circuit. The
earliest and latest arriving times are $T_{1}$ and $T_{m_{r}},$
respectively. It is certain that the halting-qubit atom is within the
left-hand potential well at the arriving time $T_{i}$ if the atom enters
into the right-hand potential well in the $i-$th cycle of the quantum
circuit. However, the halting-qubit atom could not be in the left-hand
potential well at the arriving time $T_{i}$ if the atom enters into the
right-hand potential well in the $j-$th cycle of the quantum circuit with
the cyclic index $j\neq i$ rather than in the $i-$th cycle. If the
halting-qubit atom enters into the right-hand potential well in the $j-$th
cycle ($m_{r}\geq j>1$) rather than in the first cycle of the quantum
circuit, then at the arriving time $T_{1}$ the wave-packet state of the
halting-qubit atom could not be the state $|1,CM1(T_{1}),R_{1,1}(T_{1})%
\rangle $ or the state $|1,CM1(T_{j}),R_{1,j}(T_{j})\rangle $, but it could
be another wave-packet state $|1,CM1(T_{1}),R_{1,j}(T_{1})\rangle $
different from the state $|1,CM1(T_{1}),R_{1,1}(T_{1})\rangle .$ Here the
spatial position $R_{1,j}(T_{1})$ for $m_{r}\geq j>1$ is also different from 
$R_{1,1}(T_{1})$ or $R_{1,j}(T_{j})$ and will not be constrained to be
within the left-hand potential well. Then the unitary operation $U(\{\varphi
_{k}\},T_{t}+T_{1},T_{1})$ of Eq. (30) transfers the wave-packet state $%
|1,CM1(T_{1}),R_{1,j}(T_{1})\rangle $ with the cyclic index $m_{r}\geq j>1$
to the state $|1,CM0,R_{0}\rangle $ in a probability less than 100\%. The
unitary state transformation may be generally written as 
\begin{equation*}
|1,CM0(j),R_{0}(j)\rangle \equiv U(\{\varphi
_{k}\},T_{t}+T_{1},T_{1})|1,CM1(T_{1}),R_{1,j}(T_{1})\rangle
\end{equation*}%
\begin{equation*}
=A_{j}(CM0,R_{0},T_{t},T_{1})|1,CM0,R_{0}\rangle
\end{equation*}%
\begin{equation}
\quad \quad \quad \quad +\underset{CM}{\sum }%
A_{j}(CM,R_{1,j}(T_{1}),T_{t},T_{1})|1,CM,R\rangle ,  \label{31}
\end{equation}%
where the first term on the right-hand side is the desired state $%
|1,CM0,R_{0}\rangle $ and the second term a superposition state which is
orthogonal to the desired state $|1,CM0,R_{0}\rangle $. The absolute
amplitude $|A_{j}(CM0,R_{0},T_{t},T_{1})|$ measures the conversion
efficiency from the state $|1,CM1(T_{1}),R_{1,j}(T_{1})\rangle $ ($m_{r}\geq
j\geq 1$) to the state $|1,CM0,R_{0}\rangle $ under the action of the
unitary operation $U(\{\varphi _{k}\},T_{t}+T_{1},T_{1})$. By comparing Eq.
(30) with Eq. (31) one sees that the amplitude $A_{1}(CM0,$ $%
R_{0},T_{t},T_{1})=1$ and $A_{1}(CM,R_{1,1}(T_{1}),T_{t},T_{1})=0$ for any
index value $CM.$ A theoretical calculation for the amplitude $%
A_{j}(CM0,R_{0},T_{t},$ $T_{1})$ ($m_{r}\geq j>1$) usually could be more
complex.

Denote that $H_{0}$ and $U_{0}(t,t_{0})$ are the Hamiltonian and time
evolution propagator of the halting-qubit atom in the time-independent
double-well potential field without the Raman adiabatic laser pulses and the
time-dependent external potential field, respectively. The propagator $%
U_{0}(t,t_{0})$ does not change any internal state of the halting-qubit
atom. Then there is the relation between both the wave-packet states $%
|1,CM1(T_{1}),R_{1,j}(T_{1})\rangle $ and $|1,CM1(T_{j}),$ $%
R_{1,j}(T_{j})\rangle $ for $j=1,2,...,m_{r},$%
\begin{equation}
U_{0}(T_{j},T_{1})|1,CM1(T_{1}),R_{1,j}(T_{1})\rangle
=|1,CM1(T_{j}),R_{1,j}(T_{j})\rangle .  \label{32}
\end{equation}%
This is because the halting-qubit atom in the wave-packet state $%
|1,CM1(T_{1}),$ $R_{1,j}(T_{1})\rangle $ at the time $T_{1}$ will arrive in
the left-hand potential well at the arriving time $T_{j}\geq T_{1}$ and
moreover it is in the wave-packet state $|1,CM1(T_{j}),$ $%
R_{1,j}(T_{j})\rangle $ at the arriving time $T_{j}$. By the equation (32)
and the relation $|1,CM1(T_{j}),$ $R_{1,j}(T_{j})\rangle =|1,CM1(T_{1}),$ $%
R_{1,1}(T_{1})\rangle $ the unitary state transformation (31) may be reduced
to the form 
\begin{equation*}
U(\{\varphi _{k}\},T_{t}+T_{1},T_{1})|1,CM1(T_{1}),R_{1,j}(T_{1})\rangle
\qquad \qquad \qquad \qquad
\end{equation*}%
\begin{equation}
=U(\{\varphi
_{k}\},T_{t}+T_{1},T_{1})U_{0}(T_{j},T_{1})^{+}|1,CM1(T_{1}),R_{1,1}(T_{1})%
\rangle .  \label{33}
\end{equation}%
The wave-packet motional state $|CM1(T_{1}),R_{1,1}(T_{1})\rangle $
generally is not an exact eigenstate of the Hamiltonian $H_{0}$ of the
halting-qubit atom in the double-well potential field, although the motional
state $|CM0,R_{0}\rangle $ could be approximately an eigenstate of the
Hamiltonian $H_{0}$ with the motional energy eigenvalue $E_{0}=\hslash
\omega _{0}/2$ since it is approximately the ground motional state of the
halting-qubit atom in the left-hand harmonic potential well. Thus, equation
(33) shows that not every state $|1,CM1(T_{1}),R_{1,j}(T_{1})\rangle $ for
the cyclic index $j=1,2,....,m_{r}$ can be converted completely into the
same state $|1,CM0,R_{0}\rangle $ by the same unitary operation $U(\{\varphi
_{k}\},T_{t}+T_{1},T_{1}).$ It follows from Eqs. (30), (31), and (33) that
the amplitude $A_{j}(CM0,R_{0},T_{t},T_{1})$ of the state $%
|1,CM0,R_{0}\rangle $ on the right-hand side of Eq. (31)\ can be written as 
\begin{equation*}
A_{j}(CM0,R_{0},T_{t},T_{1})=
\end{equation*}%
\begin{equation}
\langle
1,CM1(T_{1}),R_{1,1}(T_{1})|U_{0}(T_{j},T_{1})^{+}|1,CM1(T_{1}),R_{1,1}(T_{1})\rangle
\label{34}
\end{equation}%
This equation may be used to calculate the amplitude $%
A_{j}(CM0,R_{0},T_{t},T_{1})$ if the wave-packet state $%
|1,CM1(T_{1}),R_{1,1}(T_{1})\rangle $ and the unitary operation $%
U_{0}(T_{j},T_{1})$ are explicitly given for $j=1,2,...,m_{r}$. It follows
from Eqs. (32) and (34) that the probability $%
|A_{j}(CM0,R_{0},T_{t},T_{1})|^{2}$ is really the project probability of the
wave-packet state $|1,CM1(T_{1}),R_{1,j}(T_{1})\rangle $ for $%
j=1,2,...,m_{r} $ to the wave-packet state $|1,CM1(T_{1}),R_{1,1}(T_{1})%
\rangle .$

The quantum control process really starts at the time $t_{0i}$ (here for
convenience the excitation process of the trigger pulse $P_{t}$ is not
considered) and its initial state is the unstable wave-packet state $%
|1,CM1(t_{0i}),$ $R_{1}(t_{0i})\rangle $ of the halting-qubit atom in the
left-hand potential well. For convenience the mean motional velocity and
kinetic energy of the halting-qubit atom may be prepared to be zero when the
atom is prepared to be in the wave-packet state $|1,CM1(t_{0i}),$ $%
R_{1}(t_{0i})\rangle $ at the time $t_{0i}.$ This can be achieved by the
suitable state-dependent coherent Raman adiabatic laser trigger pulse $P_{t}$%
, as shown in the quantum control unit $Q_{h}$. From the viewpoint of the
particle picture the total motional energy $E_{h}$ of the halting-qubit atom
in the left-hand harmonic potential well is really pure potential energy at
the time $t_{0i}$ due to zero motional velocity of the atom when the atom is
in the wave-packet state $|1,CM1(t_{0i}),R_{1}(t_{0i})\rangle $. Thus, the
center-of-mass position $R_{1}(t_{0i})$ of the halting-qubit atom in the
left-hand harmonic potential well is really determined uniquely by the total
motional energy $E_{h}$. The wave-packet motional state $%
|CM1(t_{0i}),R_{1}(t_{0i})\rangle $ may be considered as a coherent state
[49] for the halting-qubit atom in the left-hand harmonic potential well as
it is generated from the ground motional state $|CM0,R_{0}\rangle $ by the
coherent Raman adiabatic laser trigger pulse $P_{t}$, as shown in Ref. [37].
The wave-packet state $|1,CM1(t_{0i}),R_{1}(t_{0i})\rangle $ is transferred
to the wave-packet state $|1,CM1(T_{i}),R_{1,i}(T_{i})\rangle $ ($%
i=1,2,...,m_{r}$) by the quantum control process that the halting-qubit atom
starts at the position $R_{1}(t_{0i})$ and at the time $t_{0i}$ from the
left-hand potential well to enter into the right-hand potential well by the
quantum scattering process, then it is manipulated by the decelerating
process, the accelerating process, and the elastic bouncing process in the
right-hand potential well, and finally it returns to the left-hand potential
well by another quantum scattering process and arrives in the left-hand
potential well at the arriving time $T_{i}$. For convenience, consider the
specific case that the halting-qubit atom arrives at the original position $%
R_{1}(t_{0i})$ in the left-hand potential well at the arriving time $T_{i}$
after it goes a cycle along the double-well potential field in the quantum
control process. Then in an ideal condition the wave-packet state $%
|1,CM1(T_{i}),R_{1,i}(T_{i})\rangle $ is just the original wave-packet state 
$|1,CM1(t_{0i}),R_{1}(t_{0i})\rangle $ and hence it is also a coherent
state. In order to achieve this point the total motional energy of the
halting-qubit atom in the wave-packet state $|1,CM1(T_{i}),R_{1,i}(T_{i})%
\rangle $ must be equal to that one $E_{h}$ of the atom in the wave-packet
state $|1,CM1(t_{0i}),R_{1}(t_{0i})\rangle .$ Then this requires that the
motional velocity $v$ of the halting-qubit atom after the unitary
accelerating process be equal to the motional velocity $v_{h}$ of the atom
just before the unitary decelerating process. Actually, this means that for
the halting-qubit atom the quantum scattering process from the left-hand
potential well to the right-hand one is the inverse process of the second
quantum scattering process from the right-hand potential well to the
left-hand one. On the other hand, both the wave-packet shapes for the
motional states $|CM1(t_{0i}),R_{1}(t_{0i})\rangle $ and $%
|CM1(T_{i}),R_{1,i}(T_{i})\rangle $ could be different from each other in
practice. The wave-packet spread of the motional state $%
|CM1(T_{i}),R_{1,i}(T_{i})\rangle $ usually is greater than that of the
motional state $|CM1(t_{0i}),R_{1}(t_{0i})\rangle $ [5a, 48, 53] even if the
halting-qubit atom returns the original position $R_{1}(t_{0i})$ at the
arriving time $T_{i}$ in the left-hand potential well after it goes a cycle
along the double-well potential field. This is because those processes
including the quantum scattering processes, the decelerating and
accelerating processes, and the elastic bouncing process as well as the
freely atomic motional processes in the quantum control process can change
the wave-packet shape of motional state of the halting-qubit atom and
usually could broaden the wave-packet spread of the motional state. In order
to minimize the effect of these processes on the wave-packet spread of the
motional state in the quantum control process the unitary accelerating
process may be constructed as the inverse process of the unitary
decelerating process except the atomic motional direction and spatial
position. This means that the motional velocity $v_{h}$ of the halting-qubit
atom before the unitary decelerating process equals the velocity $v$ of the
atom after the unitary accelerating process. Then in the case $v=v_{h}$ the
effect of the decelerating process could cancel that effect of the
accelerating process on the wave-packet spread of the motional state and
likely the effect of the first quantum scattering process could cancel that
of the last quantum scattering process in the quantum control process. As a
result, the net effect for both the decelerating and accelerating processes
and both the first and the last quantum scattering processes on the
wave-packet spread of the motional state could become so small that it can
be neglected. Thus, in the quantum control process the effect on the
wave-packet spread of the motional state could mainly come from the elastic
bouncing process and the freely atomic motional processes. However, this
effect could also be small and negligible if the halting-qubit atom has a
large mass $m_{h}$ and there is a short period for the quantum control
process [53]. Therefore, in the case $v=v_{h}$ the wave-packet motional
state $|CM1(T_{i}),R_{1,i}(T_{i})\rangle $ ($i=1,2,...,m_{r}$) still could
be considered approximately as the original motional state $%
|CM1(t_{0i}),R_{1}(t_{0i})\rangle $ in practice, although the wave-packet
spread of a motional state of the halting-qubit atom could change slightly
after the quantum control process. Since the motional state $%
|CM1(t_{0i}),R_{1}(t_{0i})\rangle $ ($i=1$) is a coherent state the
wave-packet motional state $|CM1(T_{1}),R_{1,1}(T_{1})\rangle $ may be
approximately expressed in a coherent-state form [49, 37, 5c], 
\begin{equation}
|CM1(T_{1}),R_{1,1}(T_{1})\rangle =|\alpha \rangle \equiv \exp (-\frac{1}{2}%
|\alpha |^{2})\overset{\infty }{\underset{k=0}{\sum }}\frac{\alpha ^{k}}{%
\sqrt{k!}}|k\rangle  \label{35}
\end{equation}%
where the relevant global phase factor is omitted and the state $|k\rangle $
is the energy eigenstate of the Hamiltonian of the halting-qubit atom in the
left-hand harmonic potential well and $\alpha $ a complex parameter$.$ The
absolute value $|\alpha |^{2}$ is the mean motional energy of the
halting-qubit atom in the wave-packet motional state $%
|CM1(T_{1}),R_{1,1}(T_{1})\rangle $ in the left-hand harmonic potential well
[5c]. Now consider other wave-packet motional states $%
\{|CM1(T_{1}),R_{1,j}(T_{1})\rangle \}$ ($m_{r}\geq j>1)$. If the entire
effective wave-packet motional state $|CM1(T_{1}),$ $R_{1,j}(T_{1})\rangle $
is within the left-hand harmonic potential well at the time $T_{1},$ it is
also a coherent state approximately just like the coherent state $%
|CM1(T_{1}),$ $R_{1,1}(T_{1})\rangle $ or $|CM1(T_{j}),R_{1,j}(T_{j})\rangle
.$ Here these two coherent states $|CM1(T_{1}),$ $R_{1,j}(T_{1})\rangle $
and $|CM1(T_{1}),R_{1,1}(T_{1})\rangle $ are connected by the unitary
transformation (32). If now the right-hand potential wall of the left-hand
potential well is sufficiently high, then that the time evolution process of
the halting-qubit atom in the left-hand potential well is governed by the
Hamiltonian of the double-well potential field is really reduced to the
simple one that the time evolution process is governed by the Hamiltonian of
the single left-hand harmonic potential well. Here the Hamiltonian $H_{0}$
and the propagator $U_{0}(t,t_{0})=\exp [-iH_{0}(t-t_{0})/\hslash ]$ of the
double-well potential field are also reduced to those of the single
left-hand harmonic potential field, respectively. Since the state $|k\rangle 
$ in the coherent state of Eq. (35) is an eigenstate of the Hamiltonian of
the single left-hand harmonic potential well, there holds the
eigen-equation: $U_{0}(T_{j},T_{1})^{+}|k\rangle =\exp [i(k+1/2)\omega
_{0}(T_{j}-T_{1})]|k\rangle .$ It follows from Eqs. (32) and (35) that the
coherent state $|CM1(T_{1}),R_{1,j}(T_{1})\rangle $ for $j=1,2,...,m_{r}$
can be written as 
\begin{equation*}
|CM1(T_{1}),R_{1,j}(T_{1})\rangle =|\alpha \exp [i\omega
_{0}(T_{j}-T_{1})]\rangle \quad
\end{equation*}%
\begin{equation}
=\exp (-\frac{1}{2}|\alpha |^{2})\overset{\infty }{\underset{k=0}{\sum }}%
\frac{\{\alpha \exp [i\omega _{0}(T_{j}-T_{1})]\}^{k}}{\sqrt{k!}}|k\rangle 
\tag{35a}
\end{equation}%
\newline
where the global phase factor is also omitted and $\omega _{0}=\sqrt{K/m_{h}}
$ is the basic oscillatory frequency of the harmonic oscillator, i.e., the
halting-qubit atom in the left-hand harmonic potential well, and $K$ a force
constant of the harmonic oscillator.

There are two different situations to be considered below. For the first
situation the maximum arriving-time difference $T_{m_{r}}-T_{1}$ for the
halting-qubit atom is much shorter than the basic oscillatory period $2\pi
/\omega _{0}$ of the halting-qubit atom in the left-hand harmonic potential
well, that is, $T_{m_{r}}-T_{1}<<\pi /\omega _{0}.$ In the first situation
there is not the second unitary decelerating process in the quantum control
process. From the viewpoint of the particle picture the halting-qubit atom
at the position $R_{1}(t_{0i})$ leaving the left-hand potential well needs
to spend roughly the time $\pi /\omega _{0}.$ Then this really means that in
the case $T_{m_{r}}-T_{1}<<\pi /\omega _{0}$ each of these $m_{r}$ possible
wave-packet states $\{|1,CM1(T_{1}),R_{1,j}(T_{1})\rangle \}$ for the
halting-qubit atom is able to enter completely into the left-hand potential
well at the time $T_{1}$. Once each of these $m_{r}$ possible wave-packet
states $\{|1,CM1(T_{1}),R_{1,j}(T_{1})\rangle \}$ of the halting-qubit atom
enters completely into the left-hand potential well at the time $T_{1}$, one
may suddenly change the double-well potential field to the single left-hand
harmonic potential field. Then in this case it is much easier to calculate
theoretically the amplitude $A_{j}(CM0,R_{0},T_{t},T_{1})$ in Eq. (31) as
the quantum dynamics can be exactly solved in a single harmonic oscillator
even when the Hamiltonian of the harmonic oscillator is time-dependent. Now
all these $m_{r}$ possible wave-packet states $%
\{|1,CM1(T_{1}),R_{1,j}(T_{1})\rangle \}$ are the coherent states which are
given explicitly by Eqs. (35) and (35a). With the help of Eq. (32) and by
inserting the coherent states of Eqs. (35) and (35a) into Eq. (34) the
absolute amplitude $|A_{j}(CM0,R_{0},T_{t},T_{1})|$ of Eq. (34) is reduced
to the simple form 
\begin{equation*}
|A_{j}(CM0,R_{0},T_{t},T_{1})|=|\langle
1,CM1(T_{1}),R_{1,1}(T_{1})|1,CM1(T_{1}),R_{1,j}(T_{1})\rangle |
\end{equation*}%
\begin{equation}
\qquad \qquad =\exp \{-|\alpha |^{2}(1-\cos [\omega _{0}(T_{j}-T_{1})])\}
\label{36}
\end{equation}%
where the relation for a pair of coherent states $|\alpha \rangle $ and $%
|\beta \rangle :\langle \alpha |\beta \rangle =\exp \{-|\alpha
|^{2}/2-|\beta |^{2}/2+\alpha ^{\ast }\beta \}$ [49, 51b, 52] is used. When
the arriving time $T_{j}=T_{1}$ it follows from Eq. (36)\ that the absolute
amplitude $|A_{1}(CM0,R_{0},T_{t},T_{1})|$ $=1.$ This result is consistent
with Eq. (30). However, if the arriving time $T_{j}>T_{1}$ ($m_{r}\geq j>1$)$%
,$ then equation (36) shows that the absolute amplitude $|A_{j}(CM0,R_{0},$ $%
T_{t},T_{1})|$ decays exponentially as the mean motional energy $|\alpha
|^{2}$ of the halting-qubit atom and the term $(1-\cos [\omega
_{0}(T_{j}-T_{1})])$. In order that the absolute amplitude $%
|A_{j}(CM0,R_{0},T_{t},T_{1})|$ is close to unity one should decrease
greatly either the mean motional energy $|\alpha |^{2}$ and the basic
oscillatory frequency $\omega _{0}$ or the arriving-time differences $%
\{(T_{j}-T_{1})\}$. Since for the current situation the basic oscillatory
frequency $\omega _{0}$ is fixed and the mean motional energy $|\alpha |^{2}$
of the halting-qubit atom must be much greater than the height of the
intermediate potential barrier one can only decrease greatly the
arriving-time differences $\{(T_{j}-T_{1})\}$ to make the absolute amplitude 
$|A_{j}(CM0,R_{0},T_{t},T_{1})|$ close to unity. Because the arriving-time
difference $T_{j}-T_{i}=(j-i)\Delta Tv_{0}/v$ for $1\leq i<j\leq m_{r},$ as
shown in Eq. (28), and the atomic motional velocity $v=v_{h}$ is determined
by the mean motional energy $|\alpha |^{2}$ in the current situation, one
has to slow down greatly the atomic motional velocity $v_{h}$ to a much
smaller velocity $v_{0}$ which could be close to zero by the unitary
decelerating process, so that the arriving-time differences can be shortened
greatly and thus the probability $|A_{j}(CM0,R_{0},T_{t},T_{1})|^{2}$ of Eq.
(36) can be close to 100\%. Though a larger atomic motional velocity $v$
leads to a smaller arriving-time difference $(T_{j}-T_{1}),$ it also leads
to a larger mean motional energy $|\alpha |^{2}\varpropto v^{2},$ and hence
the absolute amplitude $|A_{j}(CM0,R_{0},T_{t},T_{1})|$ does not become
larger and closer to unity significantly for a larger atomic motional
velocity $v=v_{h}$.

Now consider the second situation that the arriving-time differences are not
always much shorter than the basic oscillatory period $2\pi /\omega _{0}$ of
the halting-qubit atom in the left-hand harmonic potential well, that is,
the condition $T_{j}-T_{1}<\pi /\omega _{0}$ may not hold for some index
values $j=1,2,...,m_{r}$. In this case not every of these $m_{r}$ possible
wave-packet states $\{|1,CM1(T_{1}),$ $R_{1,j}(T_{1})\rangle \}$ for the
halting-qubit atom is able to enter completely into the left-hand potential
well at the time $T_{1}$. Since in this case some arriving-time differences $%
\{(T_{j}-T_{1})\}$ are larger the corresponding amplitudes $%
\{|A_{j}(CM0,R_{0},$ $T_{t},T_{1})|\}$ of Eq. (36) are not close to unity.
One scheme to solve the problem is that the basic oscillatory frequency $%
\omega _{0}$ is switched to a smaller value such that the new basic
oscillatory period can be much longer than the maximum arriving-time
difference and meanwhile the mean motional energy $|\alpha |^{2}$ of the
halting-qubit atom is lowed down accordingly before the halting-qubit atom
enters into the left-hand harmonic potential well. In this case the
amplitudes $\{|A_{j}(CM0,R_{0},$ $T_{t},T_{1})|\}$ become larger and may be
close to unity, as can be seen from Eq. (36). This scheme may be described
as follows. Since in the current situation the shortest arriving-time
difference $\Delta T_{i+1,i}=\Delta Tv_{0}/v$ ($1\leq i\leq m_{r}-1$)
generally is long and it is also much longer than the pulse duration of the
Raman adiabatic laser pulse, one may use a unitary decelerating sequence
consisting of the Raman adiabatic laser pulses to decelerate the
halting-qubit atom so that the atom has a low motional energy and speed
before the atom enters into the left-hand potential well from the right-hand
one by the second quantum scattering process. Here every arriving-time
difference $\Delta T_{j,i}$ for $1\leq i<j\leq m_{r}$ must be kept unchanged
before and after the unitary decelerating process. This decelerating process
is the second unitary decelerating process of the quantum control process.
It is really a space-compressing process for the $m_{r}$ possible
wave-packet states of the halting-qubit atom. The unitary decelerating
sequence may be applied after the atom bounces off the right-hand potential
wall of the double-well potential field and its spatial action zone could be
the same as that one of the first unitary decelerating sequence of the
quantum control process. After the quantum computational process finished
and long before the halting-qubit atom returns the left-hand potential well
the original oscillatory frequency $\omega _{0}$ of the halting-qubit atom
in the left-hand potential well is switched to a much smaller oscillatory
frequency and the height of the intermediate potential barrier is changed to
be much lower than the motional energy of the halting-qubit atom such that
the atom with a low motional energy can also enter completely into the
left-hand potential well even after it is decelerated by the second unitary
decelerating sequence. Of course, the intermediate potential barrier may
also be cancelled (i.e., switched off) and in the case the height of the
right-hand potential wall of the left-hand potential well has to be lowed
down correspondingly. These operations can be achieved by applying external
potential fields to modulate the left-hand harmonic potential well and the
intermediate potential barrier. Denote that $\omega _{c}$ is the basic
oscillatory frequency for the halting-qubit atom in the new left-hand
harmonic potential well and $|\alpha _{c}|^{2}$ the mean motional energy of
the halting-qubit atom after the atom is decelerated by the second unitary
decelerating sequence. Then $\omega _{c}<<\omega _{0}$ and $|\alpha
_{c}|^{2}<<$ $|\alpha |^{2}.$ Now the basic oscillatory period $2\pi /\omega
_{c}$ is longer for a smaller basic oscillatory frequency $\omega _{c}$.
Since the time-compressing factor $v_{0}/v=v_{0}/v_{h}<<1$ and $2\pi /\omega
<<2\pi /\omega _{c}$ now the condition: $T_{m_{r}}-T_{1}=(m_{r}-1)\Delta
T(v_{0}/v_{h})<<\pi /\omega _{c}$ may be met easily by setting an enough
small basic oscillatory frequency $\omega _{c}$. Then in this case each of
these $m_{r}$ possible wave-packet states $\{|1,CM1(T_{1}),R_{1,j}(T_{1})%
\rangle \}$ is able to enter completely into the new left-hand potential
well at the time $T_{1}$ after the second decelerating process. After each
of these $m_{r}$ possible states $\{|1,CM1(T_{1}),R_{1,j}(T_{1})\rangle \}$
enters completely into the new left-hand potential well one may change
suddenly the double-well potential field to the single left-hand harmonic
potential field by increasing greatly and quickly the height of the
right-hand potential wall of the left-hand potential well. Here the spatial
position $R_{1,1}^{c}(T_{1})$ of the wave-packet state $%
|1,CM1(T_{1}),R_{1,1}^{c}(T_{1})\rangle $ in the new left-hand potential
well is determined by the mean motional energy $|\alpha _{c}|^{2}$ instead
of $|\alpha |^{2}.$ Obviously, the wave-packet state $|\alpha _{c}\rangle
=|1,CM1(T_{1}),R_{1,1}^{c}(T_{1})\rangle $ of the new left-hand potential
well is fully different from the original state $|\alpha \rangle
=|1,CM1(T_{1}),R_{1,1}(T_{1})\rangle $ and the state $%
|1,CM1(t_{0i}),R_{1}(t_{0i})\rangle $ of the original left-hand potential
well, but both the mean motional velocity and kinetic energy are zero for
the halting-qubit atom with any one of the three wave-packet states.
Evidently, these wave-packet states $\{|1,CM1(T_{1}),R_{1,j}^{c}(T_{1})%
\rangle \}$ are still approximately coherent states of the new left-hand
harmonic potential well and the equation (36) holds also for them, where the
oscillatory frequency $\omega _{0}$ and the mean motional energy $|\alpha
|^{2}$ are replaced with $\omega _{c}$ and $|\alpha _{c}|^{2},$
respectively. Now the time-compressing factor $v_{0}/v_{h}<<1$ leads to that 
$\omega _{c}(T_{j}-T_{1})<<1$ and $|\alpha _{c}|^{2}(1-\cos [\omega
_{c}(T_{j}-T_{1})])<<1$ for $j=1,2,...,m_{r}.$ Thus, it follows from Eq.
(36) that the project probability of the wave-packet state $%
|1,CM1(T_{1}),R_{1,j}^{c}(T_{1})\rangle $ to the wave-packet state $%
|1,CM1(T_{1}),R_{1,1}^{c}(T_{1})\rangle $ for the halting-qubit atom in the
new left-hand potential well may be given by 
\begin{equation*}
|\langle
1,CM1(T_{1}),R_{1,j}^{c}(T_{1})|1,CM1(T_{1}),R_{1,1}^{c}(T_{1})\rangle |^{2}
\end{equation*}%
\begin{equation}
=\exp \{-2|\alpha _{c}|^{2}(1-\cos [\omega _{c}(T_{j}-T_{1})])\}.  \label{37}
\end{equation}%
Here one must pay attention to that unlike in Eq. (36) the project
probability in Eq. (37) generally is not equal to the probability $%
|A_{j}(CM0,R_{0},T_{t},T_{1})|^{2}$ of the desired state $%
|1,CM0,R_{0}\rangle $ in Eq. (31), because the current left-hand potential
well is not the original one.

In the definition (30) of the unitary transformation $U(\{\varphi
_{k}\},T_{t}+T_{1},T_{1})$ the wave-packet state $%
|1,CM1(T_{1}),R_{1,1}(T_{1})\rangle $ now is replaced with the state $%
|1,CM1(T_{1}),R_{1,1}^{c}(T_{1})\rangle $ of the halting-qubit atom in the
new left-hand potential well, while the state $|1,CM0,R_{0}\rangle $ is
still of the atom in the original left-hand potential well. If now there is
a unitary transformation $V_{1}(\{\varphi _{k}\},\tau +T_{1},T_{1})$ such
that the motional state $|\alpha _{c}\rangle
=|CM1(T_{1}),R_{1,1}^{c}(T_{1})\rangle $ of the new left-hand potential well
is completely converted into another coherent state $|\beta \rangle $ of the
original left-hand potential well and there is also another unitary
transformation $V_{2}(\{\varphi _{k}\},T_{t}+T_{1},\tau +T_{1})$ such that
the coherent state $|\beta \rangle $ is further converted completely into
the desired motional state $|CM0,R_{0}\rangle $, then it can turn out that
the project probability of Eq. (37) is really just the probability $%
|A_{j}(CM0,R_{0},T_{t},T_{1})|^{2}$ of the desired state $%
|1,CM0,R_{0}\rangle $ in Eq. (31), and the probability $%
|A_{j}(CM0,R_{0},T_{t},T_{1})|^{2}$ now can be expanded as the power series
of the arriving-time difference $(T_{j}-T_{1})$, up to the second order
approximation, 
\begin{equation}
|A_{j}(CM0,R_{0},T_{t},T_{1})|^{2}=1-|\alpha _{c}|^{2}(\omega
_{c})^{2}(\Delta T)^{2}(j-1)^{2}(v_{0}/v_{h})^{2}+....  \label{38}
\end{equation}%
Therefore, the motional state $|CM1(T_{1}),R_{1,j}^{c}(T_{1})\rangle $ of
the new left-hand potential well for $j=1,2,...,m_{r}$ is converted into the
desired motional state $|CM0,R_{0}\rangle $ in the probability $%
|A_{j}(CM0,R_{0},T_{t},T_{1})|^{2}$ of Eq. (38) by the unitary
transformation $U(\{\varphi _{k}\},T_{t}+T_{1},T_{1})$ of Eq. (30) which now
is a product of the unitary transformations $V_{1}$ and $V_{2}$. The
equation (38) shows that the lower bound of the probability $|A_{j}(CM0,$ $%
R_{0},T_{t},T_{1})|^{2}$ of the desired state $|1,CM0,R_{0}\rangle $ in Eq.
(31) is dependent upon the time-compressing factor $v_{0}/v_{h}$ in a
quadratic form when the time-compressing factor $v_{0}/v_{h}<<1$. One sees
from Eq. (38) that the probability $|A_{j}(CM0,$ $R_{0},T_{t},T_{1})|^{2}$
becomes closer to unity when the initial motional velocity ($v_{h}=v)$
increases. This point is quite different from that one in the first
situation (see Eq. (36)) and it may make the scheme to increase the
probability $|A_{j}(CM0,$ $R_{0},T_{t},T_{1})|^{2}$ in the second situation
better than that one in the first situation.

Obviously, the unitary transformation $U(\{\varphi _{k}\},T_{t}+T_{1},T_{1})$
of Eq. (30) now may be explicitly expressed as 
\begin{equation*}
U(\{\varphi _{k}\},T_{t}+T_{1},T_{1})=V_{2}(\{\varphi
_{k}\},T_{t}+T_{1},\tau +T_{1})V_{1}(\{\varphi _{k}\},\tau +T_{1},T_{1}).
\end{equation*}%
Here the unitary transformations $V_{1}(\{\varphi _{k}\},\tau +T_{1},T_{1})$
and $V_{2}(\{\varphi _{k}\},T_{t}+T_{1},\tau +T_{1})$ are respectively
defined as 
\begin{equation}
V_{1}(\{\varphi _{k}\},\tau +T_{1},T_{1})|\alpha _{c}\rangle =|\beta \rangle
,
\end{equation}%
\begin{equation}
V_{2}(\{\varphi _{k}\},T_{t}+T_{1},\tau +T_{1})|\beta \rangle
=|CM0,R_{0}\rangle .  \label{40}
\end{equation}%
The unitary transformation $V_{2}(\{\varphi _{k}\},T_{t}+T_{1},\tau +T_{1})$
can be generated simply by the coherent Raman adiabatic laser pulse applying
to the original left-hand harmonic potential well [37] as both the motional
states $|\beta \rangle $ and $|CM0,R_{0}\rangle $ are coherent states of the
halting-qubit atom in the original left-hand potential well. Of course, the
unitary operation $V_{2}(\{\varphi _{k}\},T_{t}+T_{1},\tau +T_{1})$ is the
unity operation if the motional state $|\beta \rangle $ is just $%
|CM0,R_{0}\rangle .$ Now one needs to construct the unitary transformation $%
V_{1}(\{\varphi _{k}\},\tau +T_{1},T_{1})$. Firstly, one must pay attention
to that both the desired motional state $|CM0,R_{0}\rangle $ and the
coherent motional state $|\beta \rangle $ belong to the halting-qubit atom
in the original left-hand harmonic potential well whose basic oscillatory
frequency is $\omega _{0},$ while the coherent motional state $|\alpha
_{c}\rangle =|CM1(T_{1}),R_{1,1}^{c}(T_{1})\rangle $ is of the atom in the
new left-hand harmonic potential well whose basic oscillatory frequency is $%
\omega _{c}.$ Therefore, the unitary transformation $V_{1}(\{\varphi
_{k}\},\tau +T_{1},T_{1})$ is involved in the time-dependent
oscillatory-frequency-varying evolution process of the halting-qubit atom in
the left-hand potential well and in the process the initial and final
oscillatory frequencies are $\omega _{c}$ and $\omega _{0}$, respectively.
In order to construct the unitary transformation $V_{1}(\{\varphi
_{k}\},\tau +T_{1},T_{1})$ one may apply the time-dependent potential field
to modulate the left-hand harmonic potential well with the initial and final
basic oscillatory frequencies $\omega _{c}$ and $\omega _{0}$, respectively,
after the halting-qubit atom enters into the new left-hand harmonic
potential well. Note that the time-dependent potential field does not change
any internal state of the halting-qubit atom in the left-hand potential
well. Denote that $H(t)$ and $U(t,t_{0})$ are the Hamiltonian and evolution
propagator of the halting-qubit atom in the time-dependent frequency-varying
left-hand harmonic potential well, respectively. Then the time evolution
propagator $U(t,t_{0})$ of the atomic physical system which can be
considered as a conventional harmonic oscillator may be written as 
\begin{equation}
U(t,t_{0})=T\exp \{-\frac{i}{\hslash }\overset{t}{\underset{t_{0}}{\int }}%
H(t^{\prime })dt^{\prime }\}  \label{41}
\end{equation}%
where the operator $T$ is the Dyson time-ordering operator and the
time-dependent frequency-modulation Hamiltonian $H(t)$ of the harmonic
oscillator is simply written as 
\begin{equation}
H(t)=\frac{1}{2m_{h}}p_{x}^{2}+\frac{1}{2}m_{h}\omega (t)^{2}x^{2}
\label{42}
\end{equation}%
where the frequency-modulation function $\omega (t)$ satisfies $\omega
(t_{0})=\omega _{c}$ for $t_{0}=T_{1}$ and $\omega (t)=\omega _{0}$ for $%
t=\tau +T_{1}.$ The unitary transformation $V_{1}(\{\varphi _{k}\},\tau
+T_{1},T_{1})$ is just the propagator $U(\tau +T_{1},T_{1})$. The unitary
propagator $U(\tau +T_{1},T_{1})$ may be generally expressed as [50g, 50h,
50i, 51, 52] 
\begin{eqnarray}
U(\tau +T_{1},T_{1}) &=&\exp [\frac{1}{2}z(\tau ,T_{1})a^{2}-\frac{1}{2}%
z(\tau ,T_{1})^{\ast }(a^{+})^{2}]  \notag \\
&&\times \exp [-i\phi (\tau ,T_{1})a^{+}a]  \label{43}
\end{eqnarray}%
where the operators $a$ and $a^{+}$ are creation and destruction operators
of the conventional harmonic oscillator, respectively, the complex parameter 
$z(\tau ,T_{1})=|z(\tau ,T_{1})|\exp [i\varphi (\tau ,T_{1})],$ and $\phi
=\phi (\tau ,T_{1})$ is the real angular frequency. The time-dependent
frequency-modulation function $\omega (t)$ of the Hamiltonian (42) must be
designed suitably. The unitary operation $U(\tau +T_{1},T_{1})$ may be
generally determined from Eq. (41) and the Hamiltonian of Eq. (42) by the
Magnus expansion method [50a, 50c, 50d, 50e] or by the Lie algebra approach
[50b, 50c, 50f, 50g, 50h, 50j]. The most convenient method to calculate
approximately the parameters $|z(\tau ,T_{1})|,$ $\varphi (\tau ,T_{1}),$
and $\phi (\tau ,T_{1})$ in Eq. (43)\ from the Hamiltonian $H(t)$ of Eq.
(42) could be the Magnus expansion method [50a, 50c] or the average
Hamiltonian theory [50d, 50e, 50j].

The unitary operation $V_{1}(\{\varphi _{k}\},\tau +T_{1},T_{1})=U(\tau
+T_{1},T_{1})$ of Eq. (43)\ can be determined from the Hamiltonian of Eq.
(42) with the time-dependent frequency-modulation function $\omega (t)$, but
here the important thing is how to generate a proper time-dependent
frequency-modulation function $\omega (t)$ so that the coherent state $%
|\alpha _{c}\rangle =|CM1(T_{1}),R_{1,1}^{c}(T_{1})\rangle $ can be
completely transferred to another coherent state $|\beta \rangle $ by the
unitary operation $V_{1}(\{\varphi _{k}\},\tau +T_{1},T_{1})$ in Eq. (39).
Firstly, the second term $\exp [-i\phi (\tau ,T_{1})a^{+}a]$ of the unitary
operation $U(\tau +T_{1},T_{1})$\ can transfer the coherent state $|\alpha
_{c}\rangle $ to another coherent state [49, 5c, 51, 52], 
\begin{equation}
\exp [-i\phi (\tau ,T_{1})a^{+}a]|\alpha _{c}\rangle =|\alpha _{c}\exp
[-i\phi (\tau ,T_{1})]\rangle .  \label{44}
\end{equation}%
Note that the first term of the unitary operation $U(\tau +T_{1},T_{1})$ is
a double-photon unitary operator $S(z)\equiv \exp [\frac{1}{2}%
(za^{2}-z^{\ast }(a^{+})^{2})]$ of the conventional harmonic oscillator [51,
52]. Then, the double-photon unitary operator $S(z(\tau ,T_{1}))$ acting on
the coherent state of Eq. (44) will transfer the coherent state to other
coherent states. There is a general formula to calculate the transition
amplitude between a pair of coherent states $|\alpha \rangle $ and $|\beta
\rangle $ induced by the double-photon unitary operator $S(z)$ [51]: 
\begin{eqnarray}
\langle \alpha _{1}|S(z)|\beta _{1}\rangle &=&C_{r}^{-1/2}\exp \{-\frac{1}{2}%
(|\alpha _{1}|^{2}+|\beta _{1}|^{2})+C_{r}^{-1}\alpha _{1}^{\ast }\beta
_{1}\}  \notag \\
&&\times \exp \{\frac{1}{2}S_{r}C_{r}^{-1}[\beta _{1}^{2}\exp (i\varphi
)-(\alpha _{1}^{\ast })^{2}\exp (-i\varphi )]\}  \label{45}
\end{eqnarray}%
where the parameters $z=r\exp (i\varphi ),$ $C_{r}=\cosh r,$ and $%
S_{r}=\sinh r.$ Now inserting the unitary operation $V_{1}(\{\varphi
_{k}\},\tau +T_{1},T_{1})=U(\tau +T_{1},T_{1})$ of Eq. (43) into the
equation (39), then using the state transformation (44) and the formula
(45), and setting $\beta _{1}=\alpha _{c}\exp [-i\phi (\tau ,T_{1})]$ $%
=|\alpha _{c}|\exp (i\gamma _{c})\exp [-i\phi (\tau ,T_{1})],$ $\alpha
_{1}=\beta =|\beta |\exp (i\gamma ),$ $r=|z(\tau ,T_{1})|,$ and $\varphi
=\varphi (\tau ,T_{1}),$ one obtains 
\begin{equation*}
|\langle \beta |V_{1}(\{\varphi _{k}\},\tau +T_{1},T_{1})|\alpha _{c}\rangle
|=
\end{equation*}%
\begin{equation*}
C_{r}^{-1/2}\exp \{-\frac{1}{2}(|\alpha _{c}|^{2}+|\beta
|^{2})+C_{r}^{-1}|\alpha _{c}\beta |\cos [-\gamma +\gamma _{c}-\phi (\tau
,T_{1})]\}
\end{equation*}%
\begin{equation*}
\times \exp \{\frac{1}{2}S_{r}C_{r}^{-1}|\alpha _{c}|^{2}\cos [2\gamma
_{c}-2\phi (\tau ,T_{1})+\varphi (\tau ,T_{1})]\}
\end{equation*}%
\begin{equation}
\times \exp \{-\frac{1}{2}S_{r}C_{r}^{-1}|\beta |^{2}\cos [2\gamma +\varphi
(\tau ,T_{1})]\}.  \label{46}
\end{equation}%
In order that the coherent state $|\alpha _{c}\rangle
=|CM1(T_{1}),R_{1,1}^{c}(T_{1})\rangle $ can be transferred completely to
the coherent state $|\beta \rangle $ by the unitary operation $%
V_{1}(\{\varphi _{k}\},$ $\tau +T_{1},T_{1})$ the absolute amplitude $%
|\langle \beta |V_{1}(\{\varphi _{k}\},\tau +T_{1},T_{1})|\alpha _{c}\rangle
|$ must be equal to unity. Then it follows from Eq. (46) that there holds
the relation when $|\langle \beta |V_{1}(\{\varphi _{k}\},\tau
+T_{1},T_{1})|\alpha _{c}\rangle |^{2}=1:$ 
\begin{equation*}
\ln C_{r}=-|\alpha _{c}|^{2}(1-S_{r}C_{r}^{-1}\cos [2\gamma _{c}-2\phi (\tau
,T_{1})+\varphi (\tau ,T_{1})])
\end{equation*}%
\begin{equation*}
+2|\alpha _{c}\beta |C_{r}^{-1}\cos [-\gamma +\gamma _{c}-\phi (\tau ,T_{1})]
\end{equation*}%
\begin{equation}
-|\beta |^{2}(1+S_{r}C_{r}^{-1}\cos [2\gamma +\varphi (\tau ,T_{1})]).
\label{47}
\end{equation}%
Equation (47)\ is used to construct the time-dependent frequency-modulation
function $\omega (t)$ in the Hamiltonian $H(t)$\ of Eq. (42), because the
equation (47) has to be met if there exists the Hamiltonian $H(t)$\ of Eq.
(42) such that the coherent state $|\alpha _{c}\rangle $ can be completely
converted into the coherent state $|\beta \rangle $ by the unitary
transformation $U(\tau +T_{1},T_{1}).$ Here the time-dependent
frequency-modulation function $\omega (t)$ must satisfy the initial and
final conditions: $\omega (T_{1})=\omega _{c}$ and $\omega (\tau
+T_{1})=\omega _{0}.$ The amplitude $|\alpha _{c}|$ and phase $\gamma _{c}$
are given by the coherent state $|\alpha _{c}\rangle $ and hence there are
only four independent variables $|\beta |,$ $\gamma ,$ and $r=|z(\tau
,T_{1})|,$ $\phi (\tau ,T_{1})$ in Eq. (47). In particular, when $r=0$ such
that $C_{r}=1,$ $\ln C_{r}=0,$ and $S_{r}=0,$ the equation (47) is reduced
to the form 
\begin{equation*}
(|\alpha _{c}|-|\beta |)^{2}+2|\alpha _{c}\beta |(1-\cos [-\gamma +\gamma
_{c}-\phi (\tau ,T_{1})])=0.
\end{equation*}%
This equation has the unique solution: $|\beta |=|\alpha _{c}|$ and $\gamma
=2k\pi +\gamma _{c}-\phi (\tau ,T_{1}),$ where $k$ is an integer and usually
set to zero. This result is obvious because the unitary transformation $%
V_{1}(\{\varphi _{k}\},\tau +T_{1},T_{1})$ becomes the unitary operation $%
\exp [-i\phi (\tau ,T_{1})a^{+}a]$ if $r=0$. For a general case $r=|z(\tau
,T_{1})|\neq 0$ the solutions to the equation (47) are more complex. For
simplicity, suppose that the phase $\gamma =const.$ The equation (47) is
reduced to the form 
\begin{equation}
a_{1}|\beta |^{2}+b_{1}|\beta |+c_{1}=0  \label{48}
\end{equation}%
where the three coefficients $a_{1}$, $b_{1}$, and $c_{1}$ are given by 
\begin{equation*}
a_{1}=1+S_{r}C_{r}^{-1}\cos [2\gamma +\varphi (\tau ,T_{1})],
\end{equation*}%
\begin{equation*}
b_{1}=-2|\alpha _{c}|C_{r}^{-1}\cos [-\gamma +\gamma _{c}-\phi (\tau
,T_{1})],
\end{equation*}%
\begin{equation*}
c_{1}=\ln C_{r}+|\alpha _{c}|^{2}(1-S_{r}C_{r}^{-1}\cos [2\gamma _{c}-2\phi
(\tau ,T_{1})+\varphi (\tau ,T_{1})]).
\end{equation*}%
Here the parameters $r=|z(\tau ,T_{1})|,$ $\varphi (\tau ,T_{1}),$ and $\phi
(\tau ,T_{1})$ are determined from the Hamiltonian $H(t)$ of Eq. (42). The
equation (48) has the solutions when the coefficient $a_{1}\neq 0:$ 
\begin{equation*}
|\beta |=\frac{1}{2a_{1}}\{-b_{1}\pm \sqrt{b_{1}^{2}-4a_{1}c_{1}}\}.
\end{equation*}%
The equation (48) has a real solution at least only when $%
b_{1}^{2}-4a_{1}c_{1}\geq 0.$ However, the reasonable solutions to the
equation (48) are those positive real solutions. Therefore, the
time-dependent frequency-modulation function $\omega (t)$ of the Hamiltonian 
$H(t)$ of Eq. (42) must be constructed such that there exists at least one
positive real solution to the equation (48).

After finishing both the quantum computational and quantum control processes
the initial state $|0,CM0,R_{0}\rangle |0\rangle |f_{r}(x_{0})\rangle $ with
the initial functional state $|f_{r}(x_{0})\rangle $ is transferred by the
quantum circuit $Q_{c}$ to the output state $|1,CM0,R_{0}\rangle |1\rangle
|0\rangle $ in the probability $|A_{i}(CM0,R_{0},T_{t},T_{1})|^{2}$ given by
Eq. (36) or (38), where the indices $i$ and $x_{0}$ satisfy $%
x_{0}=(x_{f}-i+1)\func{mod}m_{r}$ or $i=(x_{f}-x_{0}+1)\func{mod}m_{r}$ for $%
i=1,2,...,m_{r}.$ Obviously, the halting state $|1,CM0,R_{0}\rangle $ and
the branch-control state $|b_{h}\rangle =|1\rangle $ in the output state can
also be further converted completely into the halting state $%
|0,CM0,R_{0}\rangle $ and the state $|0\rangle $ by the unitary
transformations, respectively. Now the quantum circuit $Q_{c}$ with the
initial state $|0,CM0,R_{0}\rangle |0\rangle |f_{r}(x_{0})\rangle $ outputs
the desired state $|0,CM0,R_{0}\rangle |0\rangle |0\rangle $ in the
probability $|A_{i}(CM0,R_{0},$ $T_{t},T_{1})|^{2}$ given by Eq. (36) or
(38). As shown in Eqs. (36) and (38), the probability $%
|A_{i}(CM0,R_{0},T_{t},T_{1})|^{2}$ for $i=(x_{f}-x_{0}+1)\func{mod}%
m_{r}=1,2,...,m_{r}$ is close to 100\% when the time-compressing factor $%
v_{0}/v_{h}<<1$ no matter what the initial functional state $%
|f_{r}(x_{0})\rangle $ is in the initial state $|0,CM0,R_{0}\rangle
|0\rangle |f_{r}(x_{0})\rangle $ of the quantum circuit. This result shows
that the desired output state (or the state of Eq. (31)) of the quantum
circuit $Q_{c}$ is almost independent upon any initial functional state $%
|f_{r}(x_{0})\rangle $ with $x_{0}=0,1,...,m_{r}-1.$ This is just the
characteristic feature of the reversible and unitary halting protocol based
on the state-locking pulse field. Evidently, such a quantum circuit whose
output state is dependent insensitively upon any initial state could be used
to build up an efficient quantum search process just like the quantum
program and circuit proposed in the previous paper [24]. However, the lower
bound for the probability $|A_{i}(CM0,R_{0},T_{t},T_{1})|^{2}$ of Eq. (36)
or (38) for any index value $i=1,2,...,m_{r}$ must be greater than some
threshold value if the quantum circuit $Q_{c}$ is used to solve efficiently
the unsorted quantum search problem in the Hilbert space of an $n-$qubit
quantum system. Here the Hilbert space should be the quantum search space.
Note that the $2^{n}-$dimensional Hilbert state space $S(Z_{2^{n}})$ of an $%
n-$qubit quantum system is really a direct product state space of $n$
two-dimensional additive-cycle-group state subspaces $%
S(Z_{2}):S(Z_{2^{n}})=S(Z_{2})\bigotimes S(Z_{2})\bigotimes ...\bigotimes
S(Z_{2}).$ It can turn out that the lower-bound probability of Eq. (36) or
(38) must be greater than $(1-\ln p(n)/n)$, here $p(n)$ is a polynomial of
the qubit number $n$, if the minimum transfer probability from the entire
initial state $\dbigotimes\limits_{k=1}^{n}\{|0,CM0,R_{0}\rangle |0\rangle
|f_{r}(x_{0})\rangle \}_{k}$ to the final state $\dbigotimes%
\limits_{k=1}^{n}\{|0,CM0,R_{0}\rangle |0\rangle |0\rangle \}_{k}$ is equal
to $1/p(n)$ in the $2^{n}-$dimensional quantum search space $S(Z_{2^{n}})$.
Thus, if the time-compressing factor $v_{0}/v$ is small enough, that is, $%
v_{0}/v<<1,$ such that the lower-bound probability of $%
\{|A_{i}(CM0,R_{0},T_{t},T_{1})|^{2}\}$ of Eq. (36) or (38) is greater than $%
(1-\ln p(n)/n),$ then the quantum search process based on the quantum
circuit $Q_{c}$ is enough powerful to solve efficiently the unsorted quantum
search problem in the Hilbert space of the $n-$qubit quantum system. Here
the quantum parallel principle may be applied in the way that quantum
computation is performed simultaneously in all factor state subspaces of the
whole Hilbert space, and in this case the quantum parallel principle is
compatible with both the unitary quantum dynamical principle and the
mathematical logic principles used in the quantum search process.

The amplitudes $\{A_{i}(CM0,R_{0},T_{t},T_{1})\}$ of the state $%
|1,CM0,R_{0}\rangle $ in Eq. (31) corresponding to different initial states
in the quantum circuit $Q_{c}$ are dependent upon the time-compressing
factor $v_{0}/v$ and the atomic mean motional energy $|\alpha _{c}|^{2},$ as
shown in Eqs. (36) and (38). Of course, each of these amplitudes is also
dependent on a global phase factor which may be different for a different
index $i$. Obviously, the global phase factor has not a net contribution to
the probability $|A_{i}(CM0,R_{0},T_{t},T_{1})|^{2}$. Therefore, the basic
principle to make the reversible and unitary halting protocol insensitive to
any initial state includes two parts, one is by the time-compressing process
to decrease the difference among the dynamical phase factors which determine
the probabilities $\{|A_{i}(CM0,R_{0},T_{t},T_{1})|^{2}\}$ of the state $%
|1,CM0,R_{0}\rangle $ in Eq. (31), another is by manipulating the
Hamiltonian that governs the quantum control process to decrease the
difference among the dynamical phase factors. Take a simple and intuitive
physical model to illustrate the principle. Consider the simple physical
model: a simple rotation pulse $\exp (-i\omega _{p}I_{x}\Delta t)$ acting
upon the state $|0\rangle $ to convert it into the state $|1\rangle :$ 
\begin{equation*}
\exp (-i\omega _{p}I_{x}\Delta t)|0\rangle =\cos \theta |0\rangle -i\sin
\theta |1\rangle .
\end{equation*}%
Here the pulse field strength $\omega _{p}$ is the control parameter of the
Hamiltonian $H_{p}=\omega _{p}I_{x}=\omega _{p}\sigma _{x}/2$ ($\sigma _{x}$
is the Pauli operator) that governs the state transfer process from the
state $|0\rangle $ to the state $|1\rangle .$ The Hamiltonian $H_{p}$ is
time- and space-independent. Note that the eigenvalue of the Hamiltonian $%
H_{p}$ is $\omega _{p}m_{x}$, here $m_{x}=1/2$ or $-1/2$ is the eigenvalue ($%
\hslash =1$) of the operator $I_{x}$. Then the dynamical phase factor $\exp
[-im_{x}\omega _{p}\Delta t]$ is determined by the dynamical phase angle,
i.e., the rotation pulse angle $\theta =\theta (\Delta t)=\omega _{p}\Delta
t/2.$ The dynamical phase angle is proportional to the pulse duration $%
\Delta t$ and the pulse field strength $\omega _{p}.$ The amplitude $\sin
\theta $ of the state $|1\rangle $ and the transfer efficiency ($\sin
^{2}\theta $) from the state $|0\rangle $ to the state $|1\rangle $ induced
by the rotation pulse are determined by the dynamical phase angle $\theta $
up to a global phase factor $-i.$ If the rotation pulse angle $\theta
=\omega _{p}\Delta t/2=\pi /2,$ then the state $|0\rangle $ is transferred
completely to the state $|1\rangle $ by the rotation pulse and in this case
the pulse duration $\Delta t$ is equal to $\Delta t_{p}=\pi /\omega _{p}.$
Now suppose that the pulse duration $\Delta t$ is slightly different from $%
\Delta t_{p},$ for example, $\Delta t=\Delta t_{p}-\delta t$ with $|\delta
t|<<\Delta t_{p}.$ Then there is a difference between the dynamical phase
angles $\theta (\Delta t)$ and $\theta (\Delta t_{p})=\pi /2$, which is
given by $\Delta \theta (\Delta t)=\theta (\Delta t)-\theta (\Delta
t_{p})=-\omega _{p}\delta t/2.$ Therefore, the transfer efficiency is given
by 
\begin{equation*}
\sin ^{2}\theta (\Delta t)=\cos ^{2}[\Delta \theta (\Delta t)]=1-[\omega
_{p}\delta t/2]^{2}+....
\end{equation*}%
One sees that the transfer efficiency $\sin ^{2}\theta $ is determined by
the dynamical-phase-angle difference $\Delta \theta (\Delta t).$ Now
consider the case $|\delta t|<\pi /\omega _{p}$ such that $|\Delta \theta
(\Delta t)|<\pi /2.$ Then the smaller the dynamical-phase-angle difference $%
|\Delta \theta (\Delta t)|$, the higher the transfer efficiency $\sin
^{2}\theta $. Note that the dynamical-phase-angle difference $|\Delta \theta
(\Delta t)|$ is proportional to the time difference $\delta t$ and the pulse
field strength $\omega _{p}$ which controls the Hamiltonian $H_{p}.$ Thus,
the transfer efficiency may be increased and is closer to unity by
decreasing either the time difference or the pulse field strength or both.
By comparing this formula with Eq. (38) one sees that there is a
correspondence $\delta t\leftrightarrow (j-1)\Delta Tv_{0}/v.$ Now suppose
that this simple rotation-pulse model is available approximately for the
quantum control process of Eq. (2). Then the probability $1-|\varepsilon
(t,t_{0j})|^{2}$ of the state $|c_{2}\rangle $ of the control state subspace
in Eq. (2) could be evaluated roughly by setting $\delta t=(j-1)\Delta
Tv_{0}/v$ for $j=1,2,...,m_{r}$ in the transfer efficiency $\sin ^{2}\theta
(\Delta t)$ above$,$%
\begin{equation*}
1-|\varepsilon (t,t_{0j})|^{2}\thickapprox 1-\frac{1}{4}[\omega
_{p}(j-1)\Delta T(v_{0}/v)]^{2}.
\end{equation*}%
The probability of the state $|c_{2}\rangle $ is dependent on the
time-compressing factor $v_{0}/v$ in a quadratic form when $v_{0}/v<<1$.
Actually, the current atomic physical model is much more complicated than
the simple rotation-pulse model and especially the Hamiltonian that governs
the quantum control process is time- and space-dependent in the atomic
physical model so that this simple rotation-pulse model is not suited for
describing the quantum control process in the atomic physical model, but the
basic principle to make the reversible and unitary halting protocol
insensitive to the initial states is the same for both the models! \newline
\newline
\newline
{\large 6. Discussion}

In the paper the reversible and unitary halting protocol and the
state-locking pulse field have been investigated in detail and extensively.
Though the halting protocol for the universal quantum computational models
usually is irreversible and non-unitary, it can be made reversible and
unitary in the specific case that any initial state is limited to a single
basis state in the halting protocol, and in this case it can also be
simulated efficiently and faithfully (in a probability close to 100\%)
within the universal quantum computational models. From the viewpoint of the
conventional halting-operation property that the output state of the halting
operation may be completely independent of any initial state the reversible
and unitary halting protocol is generally different from the classical
irreversible one. In quantum computation the reversible and unitary halting
protocol can only achieve such a unitary halting-operation property that the
output state of the reversible and unitary halting protocol may be almost
independent of any initial state (in a probability approaching to 100\%) but
it can not be completely independent of any initial state due to the
limitation of unitarity, while the classical halting protocol may have
completely the conventional halting-operation property in classical
computation. The unitary halting-operation property is very important for
the reversible and unitary halting protocol because whether or not the
quantum search process built up out of the reversible and unitary halting
protocol can solve efficiently the quantum search problem is mainly
dependent upon this property. The state-locking pulse field plays a key role
in constructing such a reversible and unitary halting protocol that has the
unitary halting-operation property in quantum computation. It is shown in
the paper that a reversible and unitary halting protocol may be simulated
efficiently in a real quantum physical system. A simple atomic physical
system which consists of an atomic ion or a neutral atom in the double-well
potential field therefore is proposed to implement the reversible and
unitary halting protocol. The correctness for the atomic physical model to
simulate efficiently the reversible and unitary halting protocol is based on
these facts that $(i)$ for a heavy atom the atomic wave-packet picture
quantum mechanically is very close to the classical particle picture, $(ii)$%
\ the quantum motional behavior of an atom is much like the classical
motional behavior of a particle as an atom is generally much heavier than an
electron, and $(iii)$ the time evolution processes for the atomic physical
system in a variety of atomic motions still obey the Schr\"{o}dinger
equation and hence are governed by the unitary quantum dynamics. Perhaps
this simple physical model could not be the best one, but it does provide
one with a much intuitive physical picture to understand the mechanism of
the reversible and unitary halting protocol and show how the state-locking
pulse field works in the quantum control process to simulate efficiently the
reversible and unitary halting protocol. From this simple physical model one
can see clearly the reason why unitarity in quantum computation and hence
the unitary quantum dynamics are so important.

The unitarity of the quantum control process is of crucial importance if the
quantum circuit $Q_{c}$ is used to build up an efficient quantum search
process based on the unitary quantum dynamics. It follows from the quantum
control process that each possible wave-packet state $|1,CM0(j),R_{0}(j)%
\rangle $ of Eq. (31)\ for $j=1,2,...,m_{r}$ corresponds one-to-one to a
unique arriving time $T_{j}$ (see (27)) and a wave-packet spatial position $%
R_{2,j}(t_{m_{r}})$ at the end time $t_{m_{r}}$ of the computational process
(see Eq. (24)). On the other hand, it follows from the quantum circuit $%
Q_{c} $ that a different initial functional state $|f_{r}(x_{0})\rangle $
with the index $x_{0}=1,2,...,m_{r}$ corresponds one-to-one to a different
wave-packet spatial position $R_{2,j}(t_{m_{r}})$ and also a different
arriving time $T_{j}$ for the index $j=(x_{f}-x_{0}+1)\func{mod}m_{r}$.
Therefore, an initial functional state $|f_{r}(x_{0})\rangle $ for $%
x_{0}=1,2,...,m_{r}$ corresponds one-to-one to an output wave-packet state $%
|1,CM0(j),R_{0}(j)\rangle $ of Eq. (31) with the index $j=(x_{f}-x_{0}+1)%
\func{mod}m_{r}$. The one-to-one correspondence is ensured by the
wave-packet spatial-position order such as the inequality (25): $%
R_{2,m_{r}}(t_{m_{r}})<...<R_{2,2}(t_{m_{r}})<R_{2,1}(t_{m_{r}})$ and also
the arriving-time order such as the inequality (27): $%
T_{1}<T_{2}<...<T_{m_{r}}$. The one-to-one correspondence between the
initial functional states $\{|f_{r}(x_{0})\rangle \}$ and the output
wave-packet states $\{|1,CM0(j),R_{0}(j)\rangle \}$ is not only necessary
for the reversible and unitary halting protocol itself, but also it is of
crucial importance if the quantum circuit $Q_{c}$ containing the quantum
control unit that simulates efficiently the reversible and unitary halting
protocol is used to construct the quantum search process. The quantum search
process requires that the probabilities of the desired state $%
|1,CM0,R_{0}\rangle $ in the wave-packet states $\{|1,CM0(j),R_{0}(j)\rangle
\}$ (see Eqs. (31), (36), and (38)) become closer and closer to unity as the
qubit number increases. This requires further that the time-compressing
factor $(v_{0}/v)$ or the atomic motional velocity $v_{0}$ become smaller
and smaller, as shown in Eqs. (36) and (38). Note that the atomic motional
velocity $v$ is limited by $v<<c$ (the speed of light in vacuum). Therefore,
as the qubit number increases the difference among these $m_{r}$ possible
arriving times $\{T_{j}\}$ (see Eq. (28)) becomes smaller and smaller and so
does the difference among these $m_{r}$ possible wave-packet spatial
positions $\{R_{2,j}(t_{m_{r}})\}$ $(1\leq j\leq m_{r})$ (see Eq. (26)). If
now the unitarity of the quantum control process is destroyed slightly due
to some factors such as imperfect parameter settings, then the wave-packet
spatial-position order such as the inequality (25) and/or the arriving-time
order such as the inequality (27) could be easily destroyed in the quantum
control process when the qubit number is large due to the small differences
among these wave-packet spatial positions $\{R_{2,j}(t_{m_{r}})\}$ and among
these arriving times $\{T_{j}\}$. Then the one-to-one correspondence between
the initial functional states $\{|f_{r}(x_{0})\rangle \}$ and the output
states $\{|1,CM0(j),R_{0}(j)\rangle \}$ could be destroyed easily and this
will directly result in that the quantum search process based upon the
quantum circuit $Q_{c}$ could not work normally.

In particular, the importance of the unitarity of the quantum control
process can be seen more clearly from the inverse process of the quantum
control process. It is known in the quantum circuit $Q_{c}$ that these $%
m_{r} $ possible initial functional states $\{|f_{r}(x_{0})\rangle \}$ are
different from each other and also orthogonal to one another and so are
these $m_{r}$ possible initial states $\{|0,CM0,R_{0}\rangle |0\rangle
|f_{r}(x_{0})\rangle \}$ of the quantum circuit. By the quantum circuit $%
Q_{c}$ these $m_{r}$ well-distinguished initial states are converted
one-to-one into these $m_{r}$ output states $\{|1,CM0(j),$ $R_{0}(j)\rangle
|1\rangle |0\rangle \}$ for $j=(x_{f}-x_{0}+1)\func{mod}m_{r}=1,2,...,m_{r}$%
. Now for the $2^{n}-$ dimensional Hilbert state space $S(Z_{2^{n}})$ with a
large qubit number $n$ which is taken as the quantum search space it is
required by the quantum circuit $Q_{c}$ that all these $m_{r}$ possible
probabilities $\{|A_{j}(CM0,R_{0},T_{t},T_{1})|^{2}\}$ of the state $%
|1,CM0,R_{0}\rangle $ in the wave-packet states $\{|1,CM0(j),R_{0}(j)\rangle
\}$ of Eq. (31) be close to unity, that is, $%
|A_{j}(CM0,R_{0},T_{t},T_{1})|^{2}>(1-\ln p(n)/n)$ for $j=1,2,...,m_{r},$ so
that the quantum search process based on the quantum circuit can be made
efficient. Thus, this means that all these $m_{r}$ wave-packet states $%
\{|1,CM0(j),$ $R_{0}(j)\rangle \}$ are very similar to the same state $%
|1,CM0,R_{0}\rangle $, indicating that all these states $%
\{|1,CM0(j),R_{0}(j)\rangle \}$ are very similar to each other and not yet
orthogonal to one another. Thus, all these $m_{r}$ possible output states $%
\{|1,CM0(j),R_{0}(j)\rangle |1\rangle |0\rangle \}$ of the quantum circuit
are very close to the same state $|1,CM0,R_{0}\rangle |1\rangle |0\rangle .$
Hence they are very similar to each other and not yet orthogonal to one
another. The difference among these output states becomes also smaller and
smaller as the qubit number increases. This property for these output states
is completely different from that one for their corresponding initial
states. When these output states $\{|1,CM0(j),R_{0}(j)\rangle |1\rangle
|0\rangle \}$ are transferred one-to-one back to the initial states $%
\{|0,CM0,R_{0}\rangle |0\rangle |f_{r}(x_{0})\rangle \}$ by the inverse
processes of both the quantum control process and the quantum computational
process of the quantum circuit the small difference among these output
states is amplified to be a large one among those initial states, that is,
almost indistinguishable output states are changed one-to-one back to
well-distinguished initial states. Then some output states could not be
converted correctly into their corresponding initial states by the inverse
processes if the unitarity of the inverse processes is slightly destroyed.
Therefore, the unitarity of the inverse processes is of crucial importance
particularly in the first time stage of the whole inverse process that these
output states are converted one-to-one back into those initial states. Here
the first time stage of the whole inverse process is really the starting
time stage (between $T_{t}+T_{1}$ and $t_{m_{r}}$) of the inverse process of
the quantum control process in the quantum circuit.

Evidently, these $m_{r}$ possible initial states $\{|0,CM0,R_{0}\rangle
|0\rangle |f_{r}(x_{0})\rangle \}$ of the quantum circuit have the same
global phase factor and are orthogonal to one another. On the other hand,
their corresponding $m_{r}$ possible output states $\{|1,CM0(j),R_{0}(j)%
\rangle |1\rangle |0\rangle \}$ are very similar to each other and not yet
orthogonal to one another, but the global phase factors for these output
states may be very different from each other. This shows that the difference
among these discrete and orthogonal initial states is mainly transformed to
the global-phase difference among these similar output states by the quantum
program or circuit $Q_{c}$ which consists of the time- and space-dependent
unitary evolution processes. The conventional unitary operations which are
generally space-independent are not generally suitable for constructing such
a state unitary transformation. Only the time- and space-dependent unitary
evolution processes can provide one with the best way to generate such a
state unitary transformation. It is well known that there exists the square
speedup limit for the standard quantum search algorithm whose unitary
evolution process is space-independent and can be thought of as a single
time-dependent evolution process. Then a quantum search process could be
able to break through the square speedup limit if it is constructed with the
time- and space-dependent unitary evolution processes. The cost to break
through the square speedup limit is that the quantum search space is
enlarged and the single time-dependent unitary evolution process is extended
to a time- and space-dependent unitary evolution process. It is also known
[23] that an unknown quantum state can be transferred efficiently and
completely to a large state subspace from a small subspace in the $2^{n}-$%
dimensional Hilbert state space of an $n-$qubit spin system whose unitary
evolution process is space-independent, while the inverse unitary
state-transfer process is relatively hard in the same Hilbert space. Indeed,
when the Hilbert state space is added extra space dimensions and hence the
unitary evolution processes are extended to be in a multi-dimensional
Hilbert state space of time and coordinate spaces, the inverse
state-transfer process that an unknown quantum state is transferred to a
small subspace from a large one now could become more efficient than before,
as shown in the paper. However, there still exists a difference between the
unitary state-transfer process and its inverse process in the time and space
multi-dimensional Hilbert state space, although the difference may be much
smaller than that one in the $2^{n}-$dimensional Hilbert space in which any
unitary evolution process is space-independent. The results in the paper
show that an unknown quantum state of a large state subspace may be
transferred to a given state of a small state subspace in a probability
close to 100\%, but it is impossible due to the limitation of unitarity that
an unknown quantum state is transferred completely (in the probability
100\%) into a small state subspace from a large one by any given unitary
quantum dynamical process. On the other hand, it has been shown that an
unknown quantum state always can be transferred completely (in the
probability 100\%) into a large state subspace from a small one [23]. One
therefore obtains an important theorem: \textit{it is impossible to transfer
completely (in the probability 100\%) every quantum state of a large state
subspace into a small state subspace in the Hilbert space of a quantum
system by any given unitary dynamical process quantum mechanically}. This
theorem could play an important role in understanding deeply the time
evolution process of a quantum ensemble from a non-equilibrium state to the
equilibrium state from viewpoint of the unitary quantum dynamics. It is well
known in statistical physics that such a non-equilibrium evolution process
in a quantum ensemble is generally described through the stochastic
probability theory. A direct result of the theorem is that there may exist a
computational-power difference [23, 24] between a time- and space-dependent
unitary dynamical process, which transfers an unknown quantum state from one
state subspace to another, and its inverse process in a quantum system
consisting of many quantum bits. The computational-power difference for the
quantum search problem is very large for the discrete $2^{n}-$dimensional
Hilbert space of an $n-$qubit quantum system whose unitary evolution
processes are space-independent, but it could become much smaller when the
quantum system permits the time- and space-dependent unitary evolution
processes in the quantum search process. Finally, it can be believed that a
quantum system allowing to have time- and space-dependent unitary evolution
processes will have a deep effect on the conventional quantum cloning
theorem which usually works in a quantum system with single time-dependent
unitary evolution processes.

The paper has not answered the question how high it is the mean motional
energy of the unstable wave-packet state $|1,CM1(t_{0i}),R_{1}(t_{0i})%
\rangle $ of the halting-qubit atom in the left-hand harmonic potential well
so that the halting-qubit atom can overcome the intermediate potential
barrier and enter almost completely into the right-hand potential well by
the quantum scattering process. One may determine the lower bound of the
mean motional energy by solving the quantum scattering problem of the atomic
physical system. One needs also to consider the effect of the wave-packet
spread on the quantum control process as the wave-packet spread of the
motional state of the halting-qubit atom usually broadens during the atomic
motional processes in the quantum control process. The effect may be
investigated by solving the Schr\"{o}dinger equation of the atomic physical
system for the quantum control process. However, it can be expected that for
a heavy halting-qubit atom the wave-packet spread should not have a
significant effect on the lower-bound probability of Eq. (36) or (38). 
\newline
\newline
\newline
{\large Acknowledge}

Many references in the paper are found from the Boston University Science
and Engineering library and the Harvard Wolbach library. Author thanks both
the libraries for helps. Author would like to thank Charles H. Bennett to
point out the reversible halting protocol of the reversible computational
model. \newline
\newline
\newline
{\large References}\newline
1. (a)\ C. H. Bennett, \textit{Logical reversibility of computation}, IBM J.
Res. Develop. 17, 525 (1973); \textit{Time/space trade-offs for reversible
computation}, SIAM J. Comput. 18, 766 (1989)

(b) Y. Lecerf, \textit{Machines de Turing r\'{e}versibles}, C. R. Acad. Sci.
257, 2597 (1963) \newline
2. P. Benioff, \textit{The computer as a physical system: A microscopic
quantum mechanical Hamiltonian model of computers as represented by Turing
machines}, J. Statist. Phys. 22, 563 (1980) \newline
3. D. Deutsch, \textit{Quantum theory, the Church-Turing principle and the
universal quantum computer}, Proc. Roy. Soc. Lond. A 400, 96 (1985)\newline
4. D. Deutsch, \textit{Quantum computational networks}, Proc. Roy. Soc.
Lond. A 425, 73 (1989)\newline
5. (a) L. I. Schiff, \textit{Quantum mechanics}, 3rd, McGraw-Hill book
company, New York, 1968; (b) J. von Neumann, \textit{Mathematical
foundations of quantum mechanics}, (translated from the German ed. by R. T.
Beyer), Princeton University Press, 1955; (c) K. Gottfried and T. M. Yan, 
\textit{Quantum mechanics: Fundamentals}, 2nd., Springer, 2003 \newline
6. P. W. Shor, \textit{Polynomial-time algorithms for prime factorization
and discrete logarithms on a quantum computer}, SIAM J. Comput. 26, 1484
(1997); also see: Proc. 35th Annual Symposium on Foundations of Computer
Science, IEEE Computer Society, Los Alamitos, CA, pp.124 (1994)\newline
7. L. K. Grover, \textit{Quantum mechanics helps in searching for a needle
in a haystack}, Phys. Rev. Lett. 79, 325 (1997)\newline
8. J. M. Myers, \textit{Can a universal quantum computer be fully quantum?},
Phys. Rev. Lett. 78, 1823 (1997)\newline
9. E. Bernstein and U. Vazirani, \textit{Quantum computation complexity},
SIAM J. Comput. 26, 1411 (1997)\newline
10. M. Ozawa, \textit{Quantum Turing machines: local transition,
preparation, measurement, and halting},
http://arxiv.org/abs/quant-ph/9809038 (1998)\newline
11. N. Linden and S. Popescu, \textit{The halting problem for quantum
computers}, http://arxiv.org/abs/quant-ph/9806054 (1998)\newline
12. Y. Shi, \textit{Remarks on universal quantum computer},
http://arxiv.org/abs /quant-ph/9908074 (1999) \newline
13. R. Jozsa, \textit{Quantum algorithms and the Fourier transform}, Proc.
Roy. Soc. Lond. A 454, 323 (1998); \textit{Quantum factoring, discrete
logarithms and the hidden subgroup problem},
http://arxiv.org/abs/quant-ph/0012084 (2000)\newline
14. (a) R. Cleve, A. Ekert, C. Macchiavello, and M. Mosca, \textit{Quantum
algorithms revisited}, Proc. Roy. Soc. Lond. A 454, 339 (1998)

(b)\ M. Mosca and A. Ekert, \textit{The hidden subgroup problems and
eigenvalue estimation on a quantum computer},
http://arxiv.org/abs/quant-ph/9903071 (1999)\newline
15. C. H. Bennett, E. Bernstein, G. Brassard, and U. Vazirani, \textit{%
Strengths and weaknesses of quantum computing},
http://arxiv.org/abs/quant-ph/9701001 (1997)\newline
16. R. Br\"{u}schweiler, \textit{Novel strategy for database searching in
spin Liouville space by NMR ensemble computing}, Phys. Rev. Lett. 85, 4815
(2000)\newline
17. (a)\ X. Miao, \textit{Universal construction for the unsorted quantum
search algorithms}, http://arxiv.org/abs/quant-ph/0101126 (2001)

(b) X. Miao, \textit{Solving the quantum search problem in polynomial time
on an NMR quantum computer}, http://arxiv.org/abs/quant-ph/0206102 (2002)%
\newline
18. X. Miao, \textit{A polynomial-time solution to the parity problem on an
NMR quantum computer}, http://arxiv.org/abs/quant-ph/0108116 (2001)\newline
19. R. Beals, H. Buhrman, R. Cleve, M. Mosca, and R. De Wolf, \textit{%
Quantum lower bounds by polynomials}, Proceedings of 39th Annual Symposium
on Foundations of Computer Science, pp. 352 (1998); also see:
http://arxiv.org/ abs/quant-ph/9802049 (1998)\newline
20. (a) X. Miao, \textit{A prime factorization based on quantum dynamics on
a spin ensemble (I)}, http://arxiv.org/abs/quant-ph/0302153 (2003)

(b) X. Miao, unpublished\newline
21. (a)\ Y. S. Yen and A. Pines, \textit{Multiple-quantum NMR in solids}, J.
Chem. Phys. 78, 3579 (1983)

(b) D. P. Weitekamp, \textit{Time-domain multiple-quantum NMR}, Adv. Magn.
Reson. 11, 111 (1983)\newline
22. X. Miao, \textit{Multiple-quantum operator algebra spaces and
description for the unitary time evolution of multilevel spin systems},
Molec. Phys. 98, 625 (2000)\newline
23. X. Miao, \textit{Efficient multiple-quantum transition processes in an }$%
n-$\textit{qubit spin system}, http://arxiv.org/abs/quant-ph/0411046 (2004)%
\newline
24. X. Miao, \textit{Quantum search processes in the cyclic group state
spaces}, http://arxiv.org/abs/quant-ph/0507236 (2005)\newline
25. A. Barenco, C. H. Bennett, R. Cleve, D. DiVincenzo, N. Margolus, P. W.
Shor, T. Sleator, J. Smolin, and H. Weinfurter, \textit{Elementary gates for
quantum computation}, Phys. Rev. A 52, 3457 (1995)\newline
26. X. Miao, \textit{Universal construction of unitary transformation of the
quantum computation with one- and two-body interactions},
http://arxiv.org/abs/ quant-ph/0003068 (2000)\newline
27. A. Yao, \textit{Quantum circuit complexity}, Proc. 34th Annual Symposium
on Foundations of Computer Science, IEEE Computer Society Press, Los
Alamitos, CA, pp. 352\newline
28. R. Jozsa, \textit{An introduction to measurement based quantum
computation}, http://arxiv.org/abs/quant-ph/0508124 (2005)\newline
29. A. Ekert and R. Jozsa, \textit{Quantum algorithms: entanglement enhanced
information processing}, http://arxiv.org/abs/quant-ph/9803072 (1998)\newline
30. D. J. Wineland, C. Monroe, W. M. Itano, D. Leibfried, B. E. King, D. M.
Meekhof, \textit{Experimental issues in coherent quantum-state manipulation
of trapped atomic ions}, J. Res. NIST, 103, 259 (1998) \newline
31. (a) A. S$\phi $rensen and K. M$\phi $lmer, \textit{Quantum computation
with ions in thermal motion}, Phys. Rev. Lett. 82, 1971 (1999); K. M$\phi $%
lmer and A. S$\phi $rensen, \textit{Multi-particle entanglement of hot
trapped ions}, Phys. Rev. Lett. 82, 1835 (1999)

(b) G. J. Milburn, \textit{Simulating nonlinear spin models in an ion trap},
http:// arxiv.org/abs/quant-ph/9908037 (1999)

(c) J. J. Garcia-Ripoll, P. Zoller, and J. I. Cirac, \ \textit{Coherent \
control \ of \ trapped \ ions \ using \ off-resonant \ lasers},
http://arxiv.org/ abs/ quant-ph/

0411103 (2004)

(d) P. J. Lee, K. A. Brickman, L. Deslauriers, P. C. Haljan, L. M. Duan, and
C. Monroe, \textit{Phase control of trapped ion quantum gates},
http://arxiv.org/ abs/quant-ph/0505203 (2005)\newline
32. (a) G. K. Brennen, C. M. Caves, P. S. Jessen, and I. H. Deutsch, \textit{%
Quantum logic gates in optical lattices},
http://arxiv.org/abs/quant-ph/9806021 (1998), also see: Phys. Rev. Lett. 82,
1060 (1999); G. K. Brennen, I. H. Deutsch, and C. J. Williams, \textit{%
Quantum logic for trapped atoms via molecular hyperfine interactions}, Phys.
Rev. A 65, 022313 (2002)

(b) A. Beige, S. F. Huelga, P. L. Knight, M. B. Plenio, and R. C. Thompson, 
\textit{Coherent manipulation of two dipole-dipole interacting ions},
http:// arxiv.org/abs/quant-ph/9903059 (1999), also see: J. Mod. Opt. 47,
401 (2000)

(c) D. Jaksch, J. I. Cirac, P. Zoller, S. L. Rolstom, R. Cote, and M. D.
Lukin, \textit{Fast quantum gates for neutral atoms}, http://arxiv.org/abs
/quant-ph/0004038 (2000); M. Cozzini, T. Calarco, A. Recati, and P. Zoller, 
\textit{Fast Rydberg gates without dipole-dipole blockade via quantum control%
}, http:// arxiv.org/abs/quant-ph/0511118 (2005) \newline
33. J. I. Cirac and P. Zoller, \textit{Quantum computations with cold
trapped ions}, Phys. Rev. Lett. 74, 4091 (1995)\newline
34. M. A. Rowe, A. Ben-Kish, B. Demarco, D. Leibfried, V. Meyer, J. Beall,
J. Britton, J. Hughes, W. M. Itano, B. Jelenkovic, C. Langer, T. Rosenband,
and D. J. Wineland, \textit{Transport of quantum states and separation of
ions in a dual RF ion trap}, http://arxiv.org/abs/quant-ph/0205094 (2002)%
\newline
35. W. Paul, \textit{Electromagnetic traps for charged and neutral particles}%
, Rev. Mod. Phys. 62, 531 (1990); and references therein\newline
36. (a)\ S. Chu, \textit{Nobel Lecture: The manipulation of neutral particles%
}, Rev. Mod. Phys. 70, 685 (1998), and references therein;

(b) C. N. Cohen-Tannoudji, \textit{Nobel Lecture: Manipulating atoms with
photons}, Rev. Mod. Phys. 70, 707 (1998), and references therein;

(c)\ W. D. Phillips, \textit{Nobel Lecture: Laser cooling and trapping of
neutral atoms}, Rev. Mod. Phys. 70, 721 (1998), and references therein 
\newline
37. (a) C. Monroe, D. M. Meekhof, B. E. King, and D. J. Wineland, \textit{A }%
$^{\prime \prime }$\textit{Schr\"{o}dinger cat}$^{\prime \prime }$\textit{\
superposition state of an atom,} Science 272, 1131 (1996); (b) D. M.
Meekhof, C. Monroe, B. E. King, W. M. Itano, and D. J. Wineland, \textit{%
Generation of nonclassical motional states of a trapped atom}, Phys. Rev.
Lett. 76, 1796 (1996); Erratum, 77, 2346 (1996)\newline
38. P. Marte, P. Zoller, and J. L. Hall, \textit{Coherent atomic mirrors and
beam splitters by adiabatic passage in multilevel systems}, Phys. Rev. A 44,
R4118 (1991)\newline
39. L. S. Goldner, C. Gerz, R. J. C. Spreeuw, S. L. Rolston, C. I.
Westbrook, W. D. Phillips, P. Marte, and P. Zoller, \textit{Momentum
transfer in laser-cooled Cesium by adiabatic passage in a light field},
Phys. Rev. Lett. 72, 997 (1994) \newline
40. J. Lawall and M. Prentiss, \textit{Demonstration of a novel atomic beam
splitter}, Phys. Rev. Lett. 72, 993 (1994)\newline
41. M. Weitz, B. C. Young, and S. Chu, \textit{Atomic interferometer based
on adiabatic population transfer}, Phys. Rev. Lett. 73, 2563 (1994) \newline
42. (a) A. Aspect, E. Arimondo, R. Kaiser, N. Vansteenkiste, and C.
Cohen-Tannoudji, \textit{Laser cooling below the one-photon recoil energy by
velocity selective coherent population trapping}, Phys. Rev. Lett. 61, 826
(1988); (b) J. Lawall, S. Kulin, B. Saubamea, N. Bigelow, M. Leduc, and C.
Cohen-Tannoudji, \textit{Three- dimensional \ laser \ cooling \ of \ Helium
\ beyond the single-photon recoil limit}, Phys. Rev. Lett. 75, 4194 (1995);
(c) S. Kulin, B. Saubamea, E. Peik, J. Lawall, T. W. Hijmans, M. Leduc, and
C. Cohen-Tannoudji, \textit{Coherent manipulation of atomic wave-packets by
adiabatic transfer}, Phys. Rev. Lett. 78, 4185 (1997) \newline
43. M. Kasevich and S. Chu, \textit{Laser cooling below a photon recoil with
three-level atoms}, Phys. Rev. Lett. 69, 1741 (1992)\newline
44. C. Monroe, D. M. Meekhof, B. E. King, S. R. Jefferts, W. M. Itano, D. J.
Wineland, and P. Gould, \textit{Ressolved-sideband Raman cooling of a bound
atom to the 3D zero-point energy}, Phys. Rev. Lett. 75, 4011 (1995)\newline
45. J. I. Cirac, R. Blatt, and P. Zoller, \textit{Nonclassical states of
motion in a three-dimensional ion trap by addiabatic passage}, Phys. Rev. A
49, R3174 (1994)\newline
46. (a) J. Oreg, F. T. Hioe, and J. H. Eberly, \textit{Adiabatic following
in multilevel systems}, Phys. Rev. A 29, 690 (1984); (b) J. R. Kuklinski, U.
Gaubatz, F. T. Hioe, and K. Bergmann, \textit{Adiabatic population transfer
in a three-level system driven by delayed laser pulses}, Phys. Rev. A 40,
6741 (1989); (c) C. E. Carroll and F. T. Hioe, \textit{Analytic solutions
for three-state systems with overlapping pulses}, Phys. Rev. A 42, 1522
(1990)\newline
47. B. B. Blinov, R. N. Kohn Jr., M. J. Madsen, P. Maunz, D. L. Moehring,
and C. Monroe, \textit{Broadband laser cooling of trapped atoms with
ultrafast pulses}, http://arxiv.org/abs/quant-ph/0507074 (2005)\newline
48. (a) M. Goodman, \textit{Path integral solution to the infinite square
well}, Am. J. Phys. 49, 843 (1981). For an intutitive physical picture for
the Gaussian wave packets bouncing off walls see: (b)\ M. Andrews, \textit{%
Wave packets bouncing off walls}, Am. J. Phys. 66, 252 (1998)\newline
49. R. J. Glauber, \textit{Coherent and incoherent states of the radiation
field}, Phys. Rev. 131, 2766 (1963)\newline
50. (a) W. Magnus, \textit{On the exponential solution of differential
equations for a linear operator}, Commun. Pure Appl. Math. 7, 649 (1954);
(b) J. Wei and E. Norman, \textit{Lie algebraic solution of linear
differential equations}, J. Math. Phys. 4, 575 (1963); \textit{On global
representations of the solutions of linear differential equations as a
product of exponentials}, Proc. Am. Math. Soc. 15, 327 (1964); (c) R. M.
Wilcox, \textit{Exponential operators and parameter differentiation in
quantum physics}, J. Math. Phys. 8, 962 (1967); (d) U. Haeberlen and J.
Waugh, \textit{Coherent averaging effects in magnetic resonance}, Phys. Rev.
175, 453 (1968); (e) M. Matti Maricq, \textit{Application of average
Hamiltonian theory to the NMR of solids}, Phys. Rev. B 25, 6622 (1982); also
see a review for the Magnus expansion and its convergence: M. Matti Maricq, 
\textit{Long-time limitations of the Average Hamiltonian Theory: A
dressed-states viewpoint}, Adv. Magn. Reson. 14, 151 (1990); (f) G.
Campolieti and B. C. Sanctuary, \textit{The Wei-Norman Lie-algebraic
technique applied to field modulation in nuclear magnetic resonance}, J.
Chem. Phys. 91, 2108 (1989); (g)\ C. M. Cheng and P. C. W. Fung, \textit{The
evolution operator technique in solving the Schr\"{o}dinger equation, and
its application to disentangling exponential operators and solving the
problem of a mass-varying harmonic oscillator}, J. Phys. A 21, 4115 (1988);
(h) C. F. Lo, \textit{Generating displaced and squeezed number states by a
general driven time-dependent oscillator}, Phys. Rev. A 43, 404 (1991); 
\textit{Time evolution of a charged oscillator with a time-dependent mass
and frequency in a time-dependent electromagnetic field}, Phys. Rev. A 45,
5262 (1992); (i) J. Y. Ji, J. K. Kim, and S. P. Kim, \textit{%
Heisenberg-picture approach to the exact quantum motion of a time-dependent
harmonic oscillator}, Phys. Rev. A 51, 4268 (1995); H. C. Kim, M. H. Lee, J.
Y. Ji, and J. K. Kim, \textit{Heisenberg-picture approach to the exact
quantum motion of a time-dependent forced harmonic oscillator}, Phys. Rev. A
53, 3767 (1997); (j) X. Miao, \textit{An explicit criterion for existence of
the Magnus solution for a coupled spin system under a time-dependent
radiofrequency pulse}, Phys. Lett. A 271, 296 (2000) \newline
51. (a)\ D. Stoler, \textit{Equivalence classes of minimum uncertainty
packets}, Phys. Rev. D 1, 3217 (1970); \textit{Equivalence classes of
minimum-uncertainty packets. II}, Phys. Rev. D 4, 1925 (1971)

(b) R. F. Bishop and A. Vourdas, \textit{Generalised coherent states and
Bogoliubov transformations}, J. Phys. A 19, 2525 (1986)\newline
52. (a) J. R. Klauder and E. C. G. Sudarshan, \textit{Foundamentals of
quantum optics}, Chapt. 7, W. A. Benjamin, New York, 1968

(b) R. Loudon and P. L. Knight, \textit{Squeezed light}, J. Mod. Opt. 34,
709 (1987)\newline
53. M. L. Goldberger and K. M. Watson, \textit{Collision theory}, Chapt. 3,
Wiley, New York, 1964\newline
54. C. H. Bennett, private communication\newline
55. The density operator may be written as $\rho _{eq}=\alpha E+\varepsilon
\sum_{k=1}^{n}I_{kz}$ in the high temperature approximation for a
homonuclear $n-$spin ($I=1/2$) system in the thermal equilibrium state in a
high magnetic field ($B_{z}$) and at room temperature. The total
magnetization $M_{0}$ of the spin system in the thermal equilibrium state is
determined by $M_{0}=Tr\{\rho _{eq}\sum_{l=1}^{n}I_{lz}\}=\varepsilon
n2^{n}/4.$ Since all the multiple-quantum coherences in the spin system are
created from the total spin magnetization $M_{0}$ by the suitable pulse
sequences in the inphase NMR multiple-quantum experiments [21] and the
factoring multiple-quantum experiment [20] the total spectral intensity of
the multiple-quantum-transition spectrum is not more than the total spin
magnetization $M_{0}$ up to a scale factor. In the factoring
multiple-quantum experiment one part of the total spin magnetization $M_{0}$
are converted into the multiple-quantum coherences and the rest into the
longitudinal magnetization and spin order components. Thus, the inphase NMR
multiple-quantum spectrum for the factoring algorithm consists of those
spectral lines of both the multiple-quantum coherences and the longitudinal
magnetization and spin order components. Then the total spectral intensity
for the NMR multiple-quantum spectrum is kept constant in experiment and
equals really the total spin magnetization $M_{0}$ up to a scale factor
(also see [21b]). Here the spectral line of the longitudinal magnetization
and spin order components really overlaps with the zero-order quantum
transition spectral line in the NMR multiple-quantum spectrum. \newline
\newline

\end{document}